\documentclass[twocolumn]{aastex62}

\graphicspath{{./}{figures/}}

\received{February 20, 2026}
\revised{May 26, 2026}
\accepted{June 4, 2026}
\submitjournal{ApJ}

\shorttitle{Positron Annihilation from X-Ray Binaries}
\shortauthors{Detoito \& Krawczynski}

\usepackage{enumitem}

\usepackage{natbib}

\usepackage{amsmath}

\usepackage[x11names]{xcolor}

\begin{document}

\title{Galactic Electron-Positron Annihilation Emission from X-Ray Binaries with Transient and Persistent Jets}

\correspondingauthor{Argen Gian Detoito}
\email{a.detoito@wustl.edu}

\author[0000-0003-0354-4268]{Argen Gian Detoito}
\affil{Washington University in St. Louis \\
1 Brookings Dr. \\
Saint Louis, MO 63130, USA}

\author{Henric Krawczynski}
\affiliation{Washington University in St. Louis \\
1 Brookings Dr. \\
Saint Louis, MO 63130, USA}



\begin{abstract}
The Galactic Center exhibits a persistent 0.511 MeV emission characteristic of electron-positron annihilation.
Various astrophysical sources may be responsible for supplying the positrons, yet their contributions with respect to the total annihilation rate is uncertain.
X-ray binaries have historically been a primary candidate, possibly contributing a significant fraction of the Galactic positrons. 
X-ray binaries can produce positrons through the Blandford-Znajek mechanism, two-photon pair production in the corona, and/or one-photon pair production for strongly-magnetized compact objects.
These positrons may then be channeled into the Interstellar Medium through jets or winds.
In this work, we argue that X-ray binaries cannot contribute to the electron-positron annihilation flux alone.
We present 0.511 MeV line fluxes and all-sky maps produced by X-ray binaries and stellar flares from the Galactic stellar population, along with the gamma-ray continuum from positrons produced by X-ray binaries.
To do so, we solve the particle transfer equation using Green’s functions.
The total morphology is slightly asymmetric towards negative longitudes and positive latitudes.
We find five 0.511 MeV flux hotspots in the generated all sky maps and scrutinize the contribution of individual sources to these hotspots.
Observations of these hotspots can reveal the contribution of X-ray binaries to the Galactic positrons.
The upcoming Compton Spectrometer and Imager (COSI) mission could image a hotspot near Cygnus X-1. The detection of the other hotspots requires a telescope with improved arcminute angular resolution.
\end{abstract}



\section{Introduction} \label{sec:intro}

%
%
In the late 1960s and 1970s, several balloon-borne experiments detected a strong $\gamma$-ray signal at from the Galactic Center \citep{haymes_observation_1969, johnson_spectrum_1972, leventhal_detection_1978}.
Shortly after, the $\gamma$-ray emission was constrained to originate from electron-positron ($e^-$--$e^+$) annihilation and a low-energy continuum attributed to the three-photon decay of ortho-Positronium (o-Ps), an exotic bound state between an electron and positron with a spin of $1$.
Over the past few decades, we have investigated where the Galactic positrons preferentially annihilate in the Interstellar Medium (ISM) through the relevant annihilation processes \citep{guessoum_lives_2005, churazov_2005_emission, jean_spectral_2006}, the variability of the $e^+$ annihilation flux over year-to-decade timescales \citep{churazov_2011_emission}, the morphology of the 0.511 MeV flux through imaging studies \citep{purcell_1997_imaging, knodlseder_2005_imaging, bouchet_2010_imaging}, and the in-flight annihilation spectrum \citep{beacom_in_flight_2006, knodlseder_in_flight_2025}.
Recent measurements of the 0.511 MeV fluxes come from SPI-INTEGRAL \citep{skinner_galactic_2015, siegert_flux_2016} and the COSI balloon flight \citep{kierans_cosi_2020, siegert_cosi_2020}, reporting line fluxes on the order of $10^{-3}\,\text{ph}\,\text{cm}^{-2}\,\text{s}^{-1}$.

Despite theoretical and observational advancements on this problem, we have yet to constrain the origin of these Galactic positrons due to a variety of issues.
Positrons may propagate a significant distance away from their sources, obscuring the candidate positron sources in all-sky maps.
Positron propagation, cooling, and annihilation in the ISM has been studied in a variety of theoretical works \citep{higdon_SNe_2009, jean_positron_2009, martin_nucleosynthesis_2012, alexis_MC_2014, panther_propagation_2018, siegert_2022_propagation}.
However, positrons injected with particularly low energies, about $\mathcal{O}(1\,\text{MeV})$, could annihilate reasonably ``close'' their production sites.
Some sources could then produce a diffuse and point-like component to the annihilation morphology.
Unfortunately, instruments operating within this energy range are hampered by strong instrumental background, making the distinction between candidate sources difficult.
As a result, several sources can replicate the same observed flux and morphology.
The Galactic positrons may originate from nucleosynthesis in stellar sources like massive stars \citep{diehl_massive_stars_1995, prantzos_massive_stars_1996}, hypernovae and gamma-ray bursts \citep{casse_hypernovae_2004, parizot_GRB_2005, bertone_GRB_2006}, and supernovae \citep{clayton_SNe_1973, milne_SNe_1999, mazzali_SNe_2007, seitenzahl_SNe_2009}; cosmic ray interactions with the ISM \citep{moskalenko_CRs_1998, porter_CRs_2008, adriani_CRs_2009}; exotic sources like dark matter interactions \citep{boehm_DM_2004, hooper_DM_2004, picciotto_DM_2005, gunion_DM_2006, sizun_DM_2006, finkbeiner_DM_2007}; or sources with compact objects like the supermassive black hole in the Galactic Center \citep{riegler_sag_A_1981, fatuzzo_sag_A_2001, cheng_sag_A_2006, totani_sag_A_2006}, pulsars \citep{harding_pulsars_1981, zhang_pulsars_1997}, and X-ray binaries.

Microquasars, a subset of luminous X-ray binaries, were first proposed as a positron source because early SPI/INTEGRAL measurements showed an asymmetric flux morphology towards negative longitudes \citep{weidenspointner_asymmetric_2008}.
\citet{guessoum_microquasars_2006} showed that microquasars with injection rates of $\sim10^{41}\,e^+\,\text{s}^{-1}$ on average could contribute a notable fraction to the observed SPI/INTEGRAL annihilation flux.
These positrons would be produced by photon-photon processes in the inner accretion disk, then funneled out of the system through discrete ejections (transient jets).
X-ray binaries that are not microquasars were initially ignored because of their lower X-ray luminosities than microquasars.
In addition, the collective X-ray emissivity of low-mass X-ray binaries is stronger in the Galactic disk than the bulge, where the latter would be incompatible with the spatial morphology of positron annihilation.
\citet{bandyopadhyay_emission_2009} later investigated how low-mass X-ray binaries with steady jets would affect the Galactic positron annihilation flux.
The authors found that low-luminosity X-ray binaries with black holes could reasonably contribute most of the Galactic 0.511 MeV emission, but emphasized the unknown nature of the jet positron and proton content.
About a decade later, \citet{bartels_galactic_2018} explored how low-mass X-ray binaries with neutron stars may be responsible for the Galactic Center positron annihilation rate, motivated by morphological similarities between the 0.511 MeV excess and Fermi-LAT GeV excess \citep{goodenough_2009_GeV_excess}.

In this work, we attempt to unify the contribution of all X-ray binaries to the Galactic positron annihilation rate through ejections by transient and persistent jets.
We argue that X-ray binaries injecting positrons through both transient and persistent jets cannot reproduce the Galactic Center $e^+$ annihilation flux on their own.
However, when combined with a disk-dominated positron source, X-ray binaries contribute the bulge emission component that a disk-dominated source might not reproduce.
For our disk positron source, we consider stellar flares from all stellar classes capable of flaring activity.
We use best-fit models for the stellar mass distribution and up-to-date catalogues on X-ray binaries in order to determine the total flux and the flux morphology.
We describe the formulas we use to calculate fluxes for stellar flares in Section~\ref{sec:SFs} and X-ray binaries in Section~\ref{sec:XRBs}.
In Section~\ref{sec:calculations}, we describe how we create all-sky maps of the 0.511 MeV flux and compute the gamma-ray continuum from X-ray binaries in the Galaxy.
We interpret and discuss our results in Section~\ref{sec:results}, then provide conclusions and suggestions for future work in Section~\ref{sec:conclusion}.

\section{Stellar Flares} \label{sec:SFs}

%
%
In the standard flare model, flaring occurs due to strong, disordered magnetic fields sustained by differential rotation and outer convection zones.
The stellar dynamo twists and deforms field lines in the atmosphere.
Opposing magnetic field lines then reconnect, releasing magnetic energy in the form of particle kinetic energy \citep{shibata_hot-plasma_1995, chintzoglou_origin_2019}.
The reconnection energy accelerates electrons, protons, and ions in the stellar corona, which can interact to form heavier isotopes and mesons.
For the former, some isotopes can $\beta^+$ decay into a lighter isotope, emitting a positron in the process (${}^{\text{A}}_{\text{Z}}\text{X} \rightarrow {}^{\text{A}}_{\text{Z-1}}\text{Y} + \nu_e + e^+$).
The relevant processes for the latter are proton-proton ($p+p\rightarrow\pi^{\pm,0}+\cdots$) and photopion ($p+\gamma\rightarrow\pi^{\pm,0}+\cdots$) interactions.
The pions then undergo decay processes that produce positrons \citep{murphy_high-energy_1987}.

%
%
Previous 0.511 MeV studies of stellar flares have centered on M-type stars because they flare the most \citep{bisnovatyi-kogan_SFs_2017, mittal_flares_2024}.
F, G, K, and M-type stars have convective zones and stellar dynamo powerful enough to produce flares.
O and B-type stars do not exhibit disordered magnetic fields because they have radiative envelopes instead of outer convection zones.
Magnetic reconnection is not expected to occur for these stars, thus inhibiting flare activity.
A-type stars have been observed to flare, and yet their flares likely do not originate from magnetic reconnection \citep{kowalski_stellar_2024}.
Here, we attempt to calculate the energy released by stellar flares from F, G, K, and M stars.
In our calculations, we will neglect O, B, and A stars.

%
%
We calculate the 0.511 MeV flux from stellar flares using the following formula,
\begin{equation} \label{eq:F_SFs}
    F_{511,\text{F}} =
    \eta \sum_{\text{C}} f_{\text{C}} \dot{E}_{\text{C}} I_{\text{C}},
\end{equation}
where the factor $\eta$ represents the amount of positrons produced per bolometric flare energy.
\citet{bisnovatyi-kogan_SFs_2017} considers both $\eta = 120\,e^+\,\text{erg}^{-1}$ and $\eta = 340\,e^+\,\text{erg}^{-1}$.
The authors calculate the smaller $\eta$ value with an X-class solar flare measurement from RHESSI \citep{share_high-resolution_2003} while the authors argue that the larger $\eta$ value is required in order to match the $e^+$ annihilation from the bulge.
For our calculations, we use $\eta = 120\,e^+\,\text{erg}^{-1}$ since it comes from a flare measurement.
While \citet{bisnovatyi-kogan_SFs_2017} states that the larger $\eta$ value is required, we demonstrate later that $\eta = 120\,e^+\,\text{erg}^{-1}$ can still reproduce observed 0.511 MeV fluxes.

We assume in our calculations that the positrons produced by stellar flares annihilates close to their production sites.
This is a reasonable assumption because the $e^+$ annihilation timescale is on the order of $\mathcal{O}(20\,\text{s})$ \citep{murphy_radioactive_2014}.
In addition, \citet{bisnovatyi-kogan_SFs_2017} expects that a negligible fraction of $e^+$, about $\mathcal{O}(10^{-6})$, would escape from the stellar corona.
Therefore, we expect that modeling positron propagation would be trivial to the flux morphology.

%
%
The factor $f_{\text{C}}\dot{E}_{\text{C}}$ represents the total energy released by stellar flares per unit time for a specific stellar class $\text{C}$.
$\dot{E}_{\text{C}}$ is dependent on the flare frequency distribution (FFD), $\nu(E)$, which is commonly expressed as a power law \citep{gershberg_results_1972, lacy_uv_1976},
\begin{equation} \label{eq:FFD}
    \nu(E) = \nu_0 \left( \frac{E}{E_0} \right)^{-\alpha}
\end{equation}
Then the total released energy per unit time can be calculated by integrating over the FFD.
It is important to note that while all stars generally follow $\alpha\sim2$, measurements from Kepler have found different $\alpha$ values depending on the stellar class \citep{audard_extreme-ultraviolet_2000, barnes_rotational_2003}.
Therefore we will adopt class-dependent FFD parameters from \citet{yang_flare_2019}, then integrate over the FFD in Equation (\ref{eq:FFD}) (see Table~\ref{tab:SF_FFD_table}).
To determine the contribution of a stellar class $\text{C}$, we use a stellar initial mass function (IMF) to determine the fraction of stars in our galaxy belonging to
class $\text{C}$, which we denote as $f_{\text{C}}$ in Equation (\ref{eq:F_SFs}).
We use the IMF from \citet{chabrier_galactic_2003}, normalize the IMF, then integrate over the mass range of each stellar class in order to calculate their fractions.
\begin{equation}
    \dot{E}_{\text{C}} =
    \int_{E_i}^{E_f} E \nu_{\text{C}}(E) dE =
    \left| \nu_{\text{0,C}} E_{\text{0,C}}^{\alpha_{\text{C}}} \right| \int_{E_i}^{E_f} E^{1-\alpha_{\text{C}}} dE
\end{equation}

%
%
\startlongtable
\begin{deluxetable*}{c|c|c|c|c|c|c|c}
\tablecaption{Average stellar masses, fraction of Galactic stellar mass, and FFD parameters for the \\ stellar classes expected to contribute to the 0.511 MeV emission. \label{tab:SF_FFD_table}}
\tablehead{
\colhead{Stellar Class} & \colhead{Mass Range} & \colhead{$M_{\text{avg,C}}$\tablenotemark{a}} & \colhead{$f_{\text{C}}$\tablenotemark{a}} & \colhead{$\nu_{0,\text{C}}$\tablenotemark{b}} & \colhead{$E_{0,\text{C}}$\tablenotemark{b}} & \colhead{$\alpha_{\text{C}}$\tablenotemark{b}} & \colhead{$f_{\text{C}}\dot{E}_{\text{C}}$} \\
\colhead{} & \colhead{($M_{\odot}$)} & \colhead{($M_{\odot}$)} & \colhead{} & \colhead{($\text{erg}^{-1}\,\text{yr}^{-1}$)} & \colhead{($\text{erg}$)\tablenotemark{b}} & \colhead{} & \colhead{($\text{erg}\,\text{s}^{-1}$)}
}
\startdata
F & $1.04$ to $1.40$ & $1.169$ & $0.010$ & $2.0\times10^{-33}$ & $10^{35}$ & $2.11 \pm 0.09$ & $(9.027 \pm 1.408)\times10^{29}$ \\
G & $0.80$ to $1.04$ & $0.906$ & $0.021$ & $1.0\times10^{-32}$ & $10^{35}$ & $1.96 \pm 0.04$ & $(1.453 \pm 0.041)\times10^{30}$ \\
K & $0.45$ to $0.80$ & $0.587$ & $0.067$ & $2.0\times10^{-31}$ & $10^{34}$ & $1.78 \pm 0.02$ & $(1.027 \pm 0.006)\times10^{30}$ \\
M & $0.08$ to $0.45$ & $0.175$ & $0.375$ & $2.0\times10^{-32}$ & $10^{34}$ & $2.13 \pm 0.05$ & $(3.582 \pm 0.285)\times10^{30}$
\enddata
\tablenotetext{a}{Calculated from the IMF in \citet{chabrier_galactic_2003}.}
\tablenotetext{b}{Obtained from \citet{yang_flare_2019}.}
\end{deluxetable*}

\begin{figure*}
\plotone{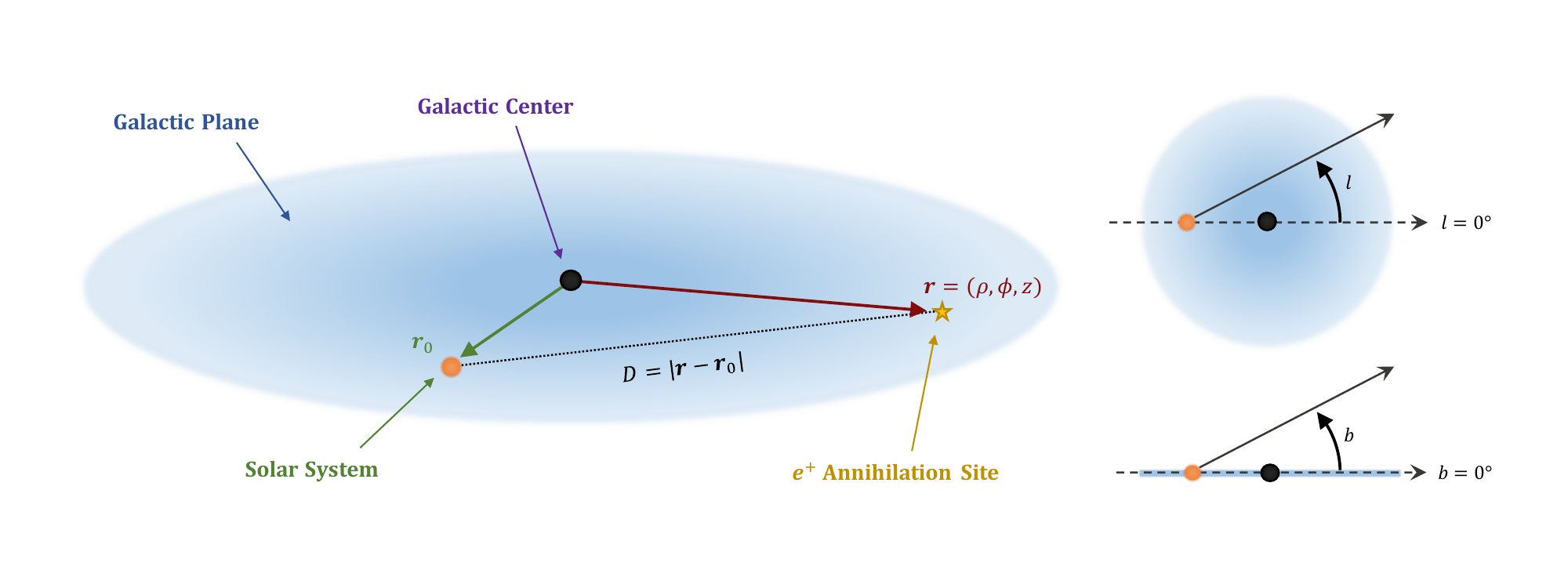}
\caption{
The coordinate systems used for flux calculations with a cartoon Galactic Plane for reference.
The stellar mass models are given in cylindrical coordinates $\mathbf{r}=(\rho,\phi,z)$ centered on the Galactic Center.
However, the fluxes are calculated using Galactic coordinates.
To calculate fluxes, the stellar mass models are converted to Galactic coordinates by shifting the origin to the Solar System (at $\mathbf{r}_0)$ and converting from cylindrical to spherical coordinates.
}
\label{fig:coordinates}
\end{figure*}

%
%
The $I_{\text{C}}$ term in Equation (\ref{eq:F_SFs}) encapsulates the position-dependent information of $e^+$ annihilation.
\begin{equation}
    I_{\text{C}} = \int
    \frac{d\mathbf{r}}{4\pi \mathbf{r}^2}
    \left[2 \left( 1 - \frac{3}{4} f_{\text{Ps}}(\mathbf{r}) \right) \right]
    \left( \frac{n_{\text{stars}} (\mathbf{r})}{M_{\text{avg,C}}} \right)
\end{equation}
Here, we let $\mathbf{r} = (D,l,b)$ denote Galactic coordinates with line-of-sight distance $D = \left| \mathbf{r} \right|$. 
In this coordinate system, $l=0^\circ$ points towards the Galactic Center and $b=0^\circ$ is the Galactic Plane (see Figure \ref{fig:coordinates}).
The first integration factor, $\propto \mathbf{r}^{-2}$, relates the $e^+$ annihilation rate at location $\mathbf{r}$ to the observed flux.

%
%
The second integration factor, $\propto f_{\text{Ps}} (\mathbf{r})$, encodes information about the fraction of positrons which annihilate into two 0.511 MeV photons, where $f_{\text{Ps}}$ is the fraction of positrons which form Ps (the ``Positronium fraction'').
Past analyses of the 0.511 MeV emission indicate with confidence that the Galactic positrons annihilate primarily in the Warm Ionized Medium (WIM) and Warm Neutral Medium (WNM) of our galaxy -- primarily due to charge exchange with ISM gas particles \citep{guessoum_lives_2005}.
Spectral fits of the $e^+$ annihilation spectrum from SPI/INTEGRAL give a value of $f_{\text{Ps}} = 0.97\pm0.02$ \citep{jean_spectral_2006}.
We adopt this Ps fraction so that $f_{\text{Ps}}$ is independent of position.

%
%
The third integration factor estimates the number of stars belonging to stellar class $\text{C}$, where $M_{\text{avg,C}}$ is the average mass of a star in class $\text{C}$ and $n_{\text{stars}} (\mathbf{r})$ is the stellar mass density.
To calculate $M_{\text{avg,C}}$, we take a weighted average of each class's mass range using the same IMF from \citet{chabrier_galactic_2003}.
For $n_{\text{stars}} (\mathbf{r})$, we use the stellar mass models given in \citet{mcmillan_mass_2017}.
\begin{equation} \label{eq:n_B}
    n_{\text{B}} (\tilde{\mathbf{r}}) =
    \frac{n_{0,\text{B}}}{(1 + \rho'/\rho_0)^{\beta}}
    \, \exp \left[ -\left( \frac{\rho'}{\rho_{\text{cut}}} \right) \right]
\end{equation}
\begin{equation} \label{eq:n_rho_prime}
    \rho' = \left[ \rho^2 + \left( \frac{z}{q} \right)^2 \right]^{0.5}
\end{equation}
\begin{equation} \label{eq:n_D_thin}
    n_{\text{D,thin}} (\tilde{\mathbf{r}}) =
    \frac{\Sigma_0}{2z_{\text{D,thin}}}
    \, \exp \left( -\frac{\left| z \right|}{z_{\text{D,thin}}} -\frac{\rho}{\rho_{\text{D,thin}}} \right)
\end{equation}
\begin{equation} \label{eq:n_D_thick}
    n_{\text{D,thick}} (\tilde{\mathbf{r}}) =
    \frac{\Sigma_0}{2z_{\text{D,thick}}}
    \, \exp \left( -\frac{\left| z \right|}{z_{\text{D,thick}}} -\frac{\rho}{\rho_{\text{D,thick}}} \right)
\end{equation}
Here we denote $\tilde{\mathbf{r}} = (\rho, \phi, z)$ as right-handed coordinates centered on the Galactic Center (or ``Galactocentric coordinates'').
The stellar mass is decomposed into a stellar bulge, thin disk, and thick disk so that $n_{\text{stars}} = n_{\text{B}} + n_{\text{D,thin}} + n_{\text{D,thick}}$.

\section{X-Ray Binaries} \label{sec:XRBs}

%
%
X-ray binaries are binary systems containing a compact object (a black hole or neutron star) and a stellar companion.
Matter from the stellar companion accretes onto the compact object into an accretion disk.
The infalling matter loses gravitational energy in the process, radiating some of its energy away in the form of thermal photons \citep{shakura_BHs_1973}.
The inner accretion disk emits a thermal X-ray blackbody spectrum.
X-ray binaries also exhibit a non-thermal, high-energy tail that is attributed to a hot, radiatively compact region known as the corona.
Much is still unknown about the corona, but it is believed that it has a characteristic electron temperature of $k_B T_e \approx 100\,\text{keV}$.
The corona heats and cools down as X-ray binaries cycle between the low-hard and high-soft states respectively.

%
%
It is believed that X-ray binaries produce positrons through photopair processes.
Thermal disk photons can upscatter up to MeV energies in the corona.
These photons then collide with other photons to produce pairs.
Electrons can also be accelerated by unscreened electric fields in the black hole's magnetosphere \citep{blandford_electromagnetic_1977}.
These electrons can emit photons that pair produce with other photons, resulting in a electron-positron pair cascade.
These positrons are produced primarily through photopair processes ($\gamma+\gamma\rightarrow e^+ + e^-$).
Any positrons which annihilate in the X-ray binary corona are expected to produce a thermally broadened spectrum instead of a clean annihilation line \citep{coppi_hybrid_plasma_1999}.
Therefore, we do not expect to observe an electron-positron annihilation line directly from X-ray binary spectra unless the positrons annihilate in a predominantly non-thermal plasma.
Instead, we may find annihilation lines from X-ray binaries after they have expelled positrons from the system into the ISM.
In this scenario, positrons can be expelled from the system through jets or winds.
Some X-ray binaries may host pulsars, where their intense radiation and magnetic fields may also produce electron-positron pair cascades ($\gamma+B\rightarrow e^+ + e^-$) \citep{harding_pulsars_1981, zhang_pulsars_1997}.
Due to their high magnetic field strength, pulsars suppress the formation of jets and instead, electron-positron pairs can escape through the pulsar wind.

X-ray binaries have been observed to launch two types of jets.
X-ray binaries can launch steady (also known as ``compact'' or ``persistent'') jets which, as the name implies, are continuously replenished outflows that are mildly-relativistic ($\Gamma\sim1.4$).
Persistent jets are mostly associated with the low/hard state, where the jet is sustained through steadily maintained photopair interactions in the inner accretion disk.
On the other hand, X-ray binaries can also launch discrete ejections with high radio luminosities -- these are transient jets.
Unlike persistent jets, transient jets expel material at more relativistic speeds ($\Gamma \gtrsim 2$).
The most commonly invoked origin for transient jets involves instabilities in the accretion disk.
Low-mass X-ray binaries experience thermoviscous disk instabilities which can suddenly ramp up accretion rate, leading to discrete ejections at highly relativistic speeds \citep{dubus_2001_disk_instability}.
However, transient jets may also be launched due to fast magnetic reconnection near the jet base \citep{mckinney_reconnection_2012}.
As such, transient jets are typically associated with outbursting low-mass X-ray binaries, launched during their transition from the hard/intermediate state to the soft/intermediate state.
However some high-mass X-ray binaries, like supergiant fast X-ray transients (SFXTs) and Be-star X-ray binaries, have been observed to launch transient jets for similar reasons \citep{romero_transient_jet_2007, garcia_SFXTs_2014}.
Meanwhile, persistent jets are associated more with high-mass X-ray binaries because these X-ray binaries typically exhibit persistent jets most of the time (duty cycles of about 50\% to 90\%).
Whereas jets are associated with accretion onto the compact object, pulsar winds are driven by the rotation power of the neutron star.

For our calculations, we will ignore pulsar winds and focus only on positrons ejected by jets from non-pulsar X-ray binaries.
We expect a significant fraction of positrons channeled into the ISM by X-ray binaries to have energies of $\mathcal{O}(1\,\text{MeV})$.
This fact contrasts the pulsar scenario, where ejected positrons typically have very high Lorentz factors of $\gamma\sim\mathcal{O}(10^6)$.
As such, pulsar positrons may propagate very large distances away from their injection sites.
The low-end positron energies ejected by pulsar winds are also not compatible with the methods we want to use to estimate the flux morphology of X-ray binaries, which we explain and justify in Section~\ref{ssec:XRB_511_methods}.
For the same reasons, we will not look at the contribution of pulsars to the 0.511 MeV emission and instead only focus on X-ray binaries with black holes and weakly-magnetized neutron stars.

\subsection{Positron Injection Rates} \label{ssec:XRB_inj_rates}

\begin{figure*}
\plotone{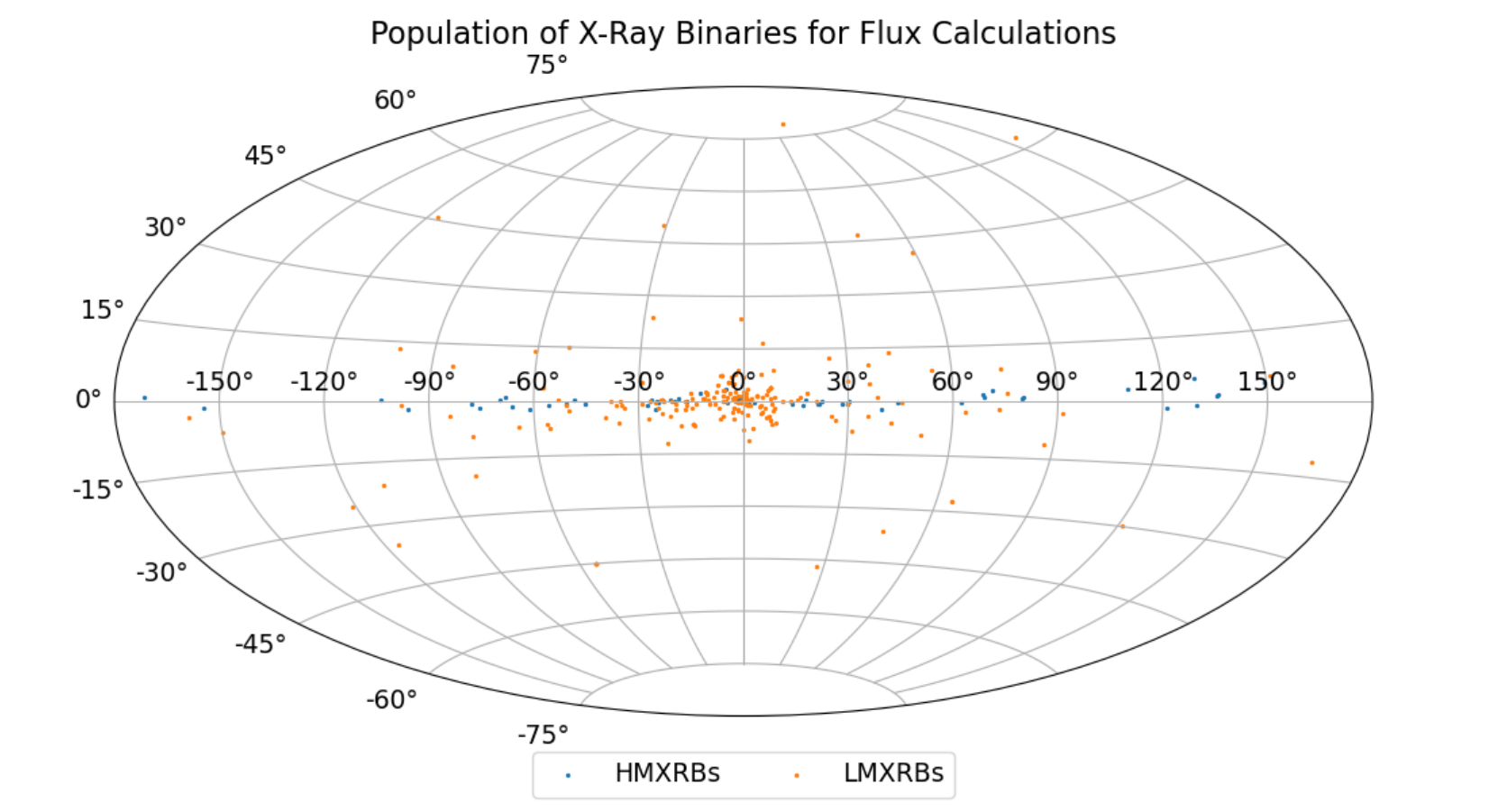}
\caption{
The population of high-mass and low-mass X-ray binaries used in our calculation of the Galactic $e^+$ annihilation flux and continuum \citep{avakyan_xrbcats_2023, neumann_xrbcats_2023}.
Most of the low-mass X-ray binaries reside in the Galactic bulge while the high-mass X-ray binaries primarily populate the disk.
}
\label{fig:XRBs_used_in_calcs}
\end{figure*}

The particle content of jets is still mostly unknown (see \citet{romero_jet_2017} for a review).
The jet content is either hadronic (electrons and protons), leptonic (electrons and positrons), or a mix of hadrons and leptons.
Our best estimates for the pair-to-proton ratio originate from observations of active galactic nuclei (AGN).
\citet{reynolds_jet_1996} reports observations of M87, where they hint at jets being likely dominated by cold electron-positron pairs.
However, it would not be surprising if jets are also comprised of, or dominated by, protons \citep{romero_hadrons_2005}.
For the sake of calculations, we assume that jets from X-ray binaries are pair-dominated.
To start, the observed jet luminosity correlates to the bolometric luminosity of the accretion disk \citep{fender_jet_model_2004},
\begin{equation} \label{eq:XRB_empirical}
    L_{\text{jet}} \approx
    \begin{cases}
        A_{\text{BH}} L_{\text{bol}}^{0.5}, & \text{BH X-ray binary} \\
        A_{\text{NS}} L_{\text{bol}}^{0.5}, & \text{NS X-ray binary}
    \end{cases},
\end{equation}
where the luminosities in Equation (\ref{eq:XRB_empirical}) are given in Eddington units.
For persistent jets, we use proportionality factors of $A_{\text{BH}}=0.1$, $A_{\text{NS}}=0.3A_{\text{BH}}$, and $L_{\text{bol}} = 0.02$.
This choice of proportionality factors corresponds to a soft-to-hard state transition at $0.01\dot{M}_{\text{Edd}}$ in black hole X-ray binaries and a NS-to-BH radio luminosity ratio of $0.3$ (\citet{bartels_galactic_2018} and references therein).
Transient jets follow the same empirical relation, albeit more loosely than their persistent counterparts \citep{fender_jet_model_2004}.
For our purposes, we will exploit this fact and follow the arguments found in \citet{fender_2003_jets}.
Since transient jets are typically associated with $L_{\text{bol}}\sim 1$ \citep{fender_jet_model_2004}, we find and use $A_{\text{BH}}=0.7$ for transient jets launched by black hole X-ray binaries.
From here, we take $A_{\text{NS}}=0.3A_{\text{BH}}$ since the NS-to-BH luminosity relation remains roughly invariant regardless of jet type.
We refer to the XRBcats catalogue \citep{avakyan_xrbcats_2023, neumann_xrbcats_2023} to get the population of X-ray binaries while filtering out X-ray binaries that do not have definitive measurements of $(D,l,b)$ and soft X-ray fluxes.
Then we use the appropriate proportonality factor depending on the compact object.
Figure~\ref{fig:XRBs_used_in_calcs} shows the X-ray binaries used in our flux calculations.
We use the Swift-XRT fluxes reported in this catalogue as an estimate of $L_{\text{bol}}$.
For jets with leptonic content, the jet luminosity is related to the positron injection rate into the ISM through the following formula from \citet{bartels_galactic_2018},
\begin{equation} \label{eq:injection_rate}
    \dot{N}_{+,i} =
    \frac{L_{\text{jet}}}{2 \langle\gamma\rangle \Gamma_{\text{jet}} m_e c^2}
\end{equation}
We assume that the jet particle content is cold ($\langle\gamma\rangle \approx 1$) and mildly-relativistic ($\Gamma_{\text{jet}} \gtrsim 1$), estimating $\Gamma_{\text{jet}} \approx 1.4$ for persistent jets and $\Gamma_{\text{jet}} \approx 2$ for transient jets.
We will denote the source-specific injection rate of persistent jets as $\dot{N}_{+,\text{p},i}$, and $\dot{N}_{+,\text{t},i}$ for transient jets.

\subsection{Electron-Positron Annihilation Line Flux} \label{ssec:XRB_511_methods}
%
%
Unlike the stellar flare scenario, we expect the positrons to propagate far (with respect to Galactic distances) from their production sites.
This problem is typically approached with Monte Carlo methods \citep{jean_positron_2009}.
These simulations instantiate a large number of positrons, then propagate them according to the transport equations.
The differential cross sections for ionization, scattering, and other energy loss processes are used to propagate positrons before they annihilate in-flight or at their rest energies.
To get an estimate of the $e^+$ annihilation flux and its morphology in Galactic coordinates without relying on Monte Carlo simulations, we convert the relevant differential equations into integrals using Green's functions.
We first determine how $e^+$ propagation affects the distribution of $e^+$ annihilation sites by considering the one-particle transfer equation \citep{ginzburg_cosmic_rays_1964}.
The one-particle transfer equation describes particles whose distribution evolves due to diffusion and energy losses, given by,
\begin{equation} \label{eq:one_transfer_eq}
    \frac{\partial N}{\partial t} - D\nabla^2 N + \frac{\partial}{\partial E} (bN) - \frac{N}{T} =
    \mathcal{Q}(E,\mathbf{r},t)
\end{equation}
In the transfer equation, $T(E) = (n\sigma_{\text{ann}} v)^{-1}$ is the mean $e^+$ lifetime, where $n = n_{\text{H}} + n_{\text{H}_2} + n_{\text{He}}$.
$D(E)$ is the energy-dependent diffusion coefficient of the particles.
It is important to mention that the transport of low-energy positrons is not yet understood.
Here, propagation may be dominated by collisions off of gas particles in the ISM (collisional transport) or scattering off of magnetic turbulence in the ISM magnetic field \citep{prantzos_511_2011}.
The latter mechanism, collisionless transport, is often invoked in cosmic ray propagation for energies of GeV or higher.
Collisionless transport is usually described with a spatial diffusion coefficient taken as a power law or broken power law in rigidity, where $D(E)\propto \rho^{\delta}$ and $\delta=1/3$ for Kolmogorov turbulence.
The parameters for the collisionless diffusion coefficient can vary (e.g., see \citet{strong_GALPROP_1998}).
To make simple estimates, we assume the transport of positrons from X-ray binaries is dominated by collisions \citep{jean_positron_2009}.
We make this assumption because most of the positrons in our calculations have initial energies of $\mathcal{O}(1\,\text{MeV})$, a topic we discuss later.
Collisional diffusion depends on the ionization cross sections with atomic hydrogen, molecular hydrogen, and helium,
\begin{equation} \label{eq:energy_dep_D}
    D(E) = \frac{v(E)}{n_{\text{H}} \sigma_{\text{H}}(E) + n_{\text{H}_2} \sigma_{\text{H}_2}(E) + n_{\text{He}} \sigma_{\text{He}}(E)},
\end{equation}
where we adopt formulas for $\sigma_{\text{H}}$, $\sigma_{\text{H}_2}$, and $\sigma_{\text{He}}$ from \citet{gryzinski_collisions_1965}.
As positrons propagate through the ISM, they lose energy via Coulomb interactions, bremsstrahlung, and ionization with free electrons, ISM atoms, and molecules.
Positrons with particularly high energies can also lose energy through synchrotron emission and inverse Compton scattering on the ISM radiation field.
These energy losses are encoded in the energy loss rate $b(E) = dE/dt < 0$,
\begin{multline} \label{eq:energy_losses}
    b(E) = b_{\text{SY}}(E) + b_{\text{IC}}(E) + b_{\text{BR}}(E) \\ + b_{\text{COU}}(E) + b_{\text{ION}}(E)
\end{multline}
For synchrotron and inverse Compton losses, we refer to the formulas outlined in \citet{blumenthal_theory_1970}.
Meanwhile, we use the formulas given in \citet{ginzburg_theory_1979} for bremsstrahlung, Coulomb, and ionization losses.

It is important to note that in the Ginzburg-Syrovatskii formalism of particle transfer, the energy loss terms, mean $e^+$ lifetimes, and diffusion coefficients are independent of position.
This constraint means that Equation (\ref{eq:one_transfer_eq}) works for short propagation distances and breaks down otherwise.
For this reason, we cannot reliably estimate the $e^+$ annihilation distribution of pulsars.
However, in making this sacrifice, we can make simple estimates of the flux morphology from X-ray binaries with black holes and weakly-magnetized neutron stars using Green's functions.
We describe this procedure in more detail later.
However, we still have to encode the conditions of the ISM around our population of X-ray binaries.
To this end, we calculate $b,T,D$ at the location of each X-ray binary, then approximate the ISM conditions at small distances around each X-ray binary with these values.
We thus define ``local'' energy loss rates $b_i$, local mean lifetimes $T_i$, and local diffusion coefficients $D_i$ as,
\begin{multline}
    b_i (E) = b(E, \mathbf{r}_i) =
    b_{\text{SY}}(E, \mathbf{r}_i) + b_{\text{IC}}(E, \mathbf{r}_i) \\ + b_{\text{BR}}(E, \mathbf{r}_i) + b_{\text{COU}}(E, \mathbf{r}_i) + b_{\text{ION}}(E, \mathbf{r}_i)
\end{multline}
\begin{equation} \label{eq:local_pos_lifetime}
    T_i (E) = T(E, \mathbf{r}_i) =
    \frac{1}{n(\mathbf{r}_i) \sigma_\text{ann} (E) v(E)}
\end{equation}
\begin{multline}
    D_i (E) = D(E, \mathbf{r}_i) =\\
    \frac{v(E)}{n_{\text{H}} \sigma_{\text{H}}(E) + n_{\text{H}_2} \sigma_{\text{H}_2}(E) + n_{\text{He}} \sigma_{\text{He}}(E)},
\end{multline}

\begin{figure*}
\plottwo{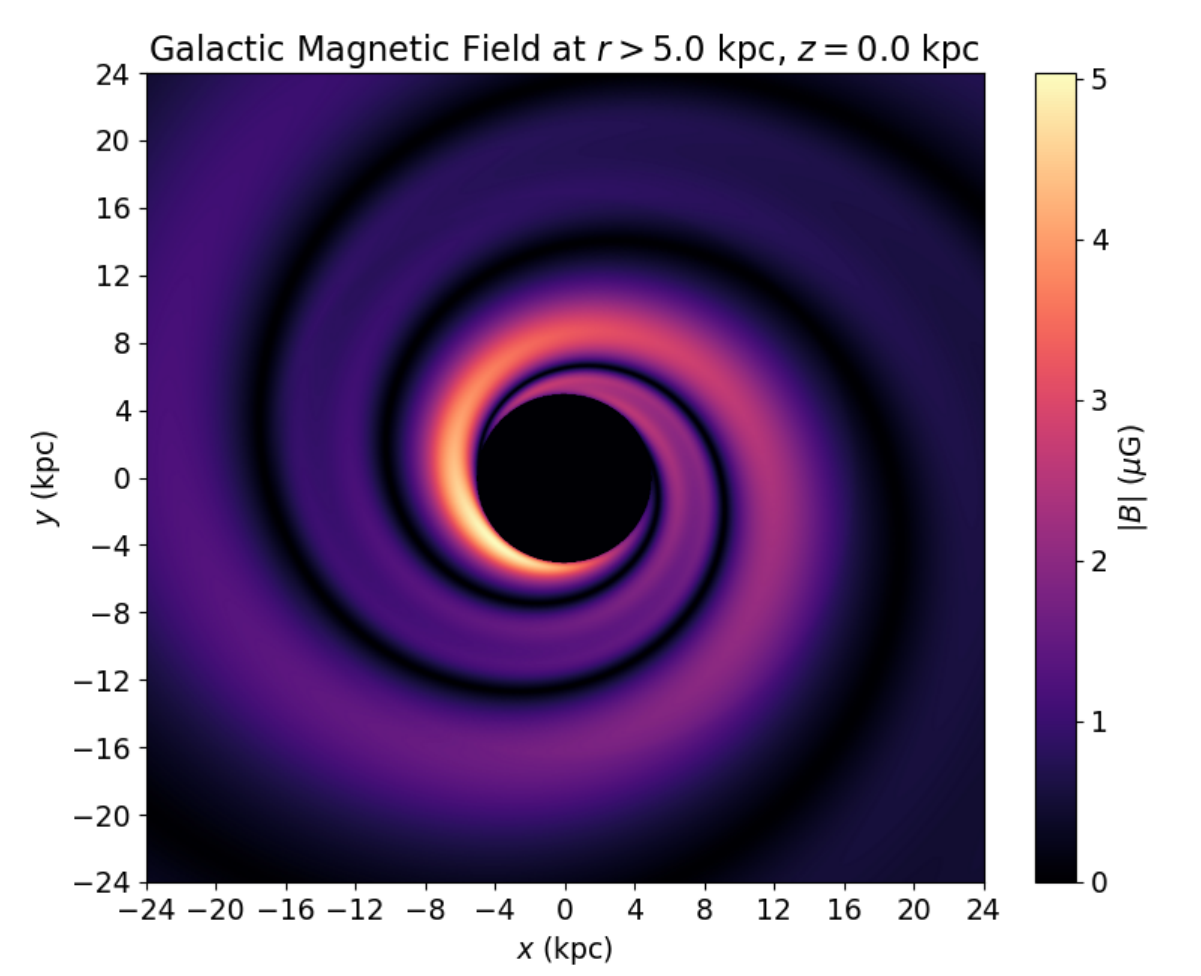}{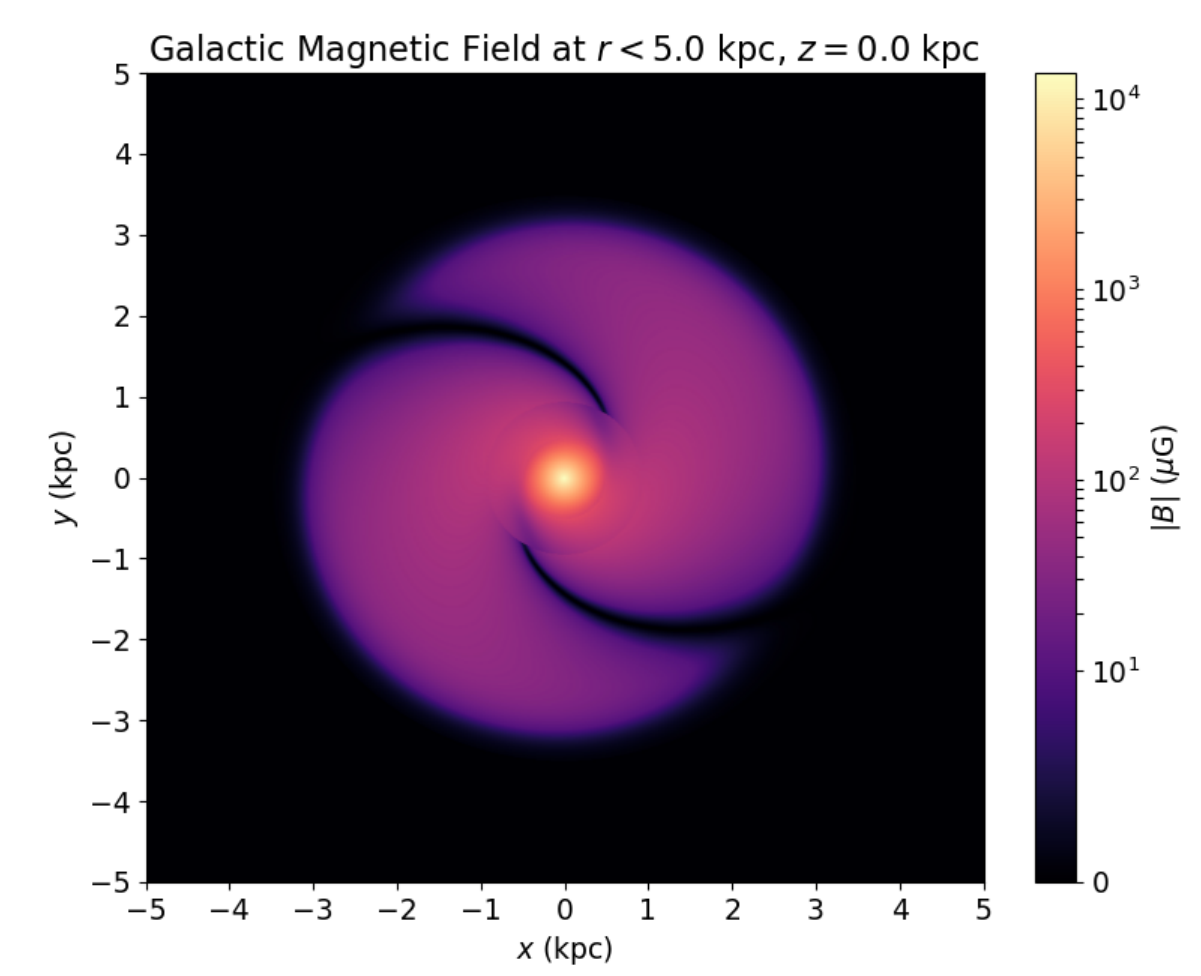}
\caption{
\textbf{Left:} The Galactic Magnetic Field (GMF) in the outer $5\,\text{kpc}$.
We smoothly transition between the spiral regions of \citet{jansson_GMF_2012} by summing over each spiral region, with weights which are dependent on a $\text{sech}^2$ factor.
\textbf{Right:} The GMF in the inner $5\,\text{kpc}$ from \citet{guenduez_GMF_2020}.
Contributions for the inner GMF originate from the diffuse intercloud medium of the Central Molecular Zone, nonthermal radio filaments in the Galactic Center, molecular clouds, and the supermassive black hole Sgr A*.
}
\label{fig:GMF}
\end{figure*}

To model the local conditions of propagating positrons, we incorporate position-dependent models of the Galactic Magnetic Field (GMF), radiation energy density, and ISM density.
The GMF enters our calculations through the synchrotron loss component $b_{\text{SY}}(E,\mathbf{r})$.
For the GMF, we model the outer Galactic regions ($\rho>5\,\text{kpc}$ in Galactocentric coordinates $\tilde{\mathbf{r}}$) with a composite halo and disk field model.
Here, we use a smoothed version of the model from \citet{jansson_GMF_2012} for the disk field and Model C of \citet{ferriere_GMF_2014} for the halo field.
For the inner Galactic regions ($\rho<5\,\text{kpc}$), we model the GMF using the model from \citet{guenduez_GMF_2020}.
We smoothly interpolate the GMF between these two regions.
The GMF magnitude for $z=0$ is shown in Figure~\ref{fig:GMF}.
Meanwhile, the radiation energy density contributes to positron energy losses through the inverse Compton term, $b_{\text{IC}}(E,\mathbf{r})$.
For the total radiation density, we sum contributions from the cosmic microwave background (CMB), infrared (IR), optical, and ultraviolet (UV) spectral bands.
We treat the CMB density as position-independent, $u_{\text{CMB}}=0.26\,\text{eV}\,\text{cm}^{-3}$, while we estimate the contributions of the other bands by adopting the 2D Galactocentric ISRF model outlined in \citet{porter_CRs_2008}.
The ISM density contributes to the remaining energy loss terms.
We assume that positrons propagate through an average of all ISM phases using the model given in \citet{ferriere_global_1998}.

For transient jets, we assume that the positrons are injected as point sources centered on each X-ray binary with an energy spectrum characteristic of first-order Fermi acceleration at the jet termination shock.
The outgoing spectrum is modeled as a power law with an exponential cutoff \citep{bykov_1999_shocks}.
We model the outgoing spectrum for an X-ray binary (arbitrarily indexed as $i$) launching a transient jet by,
\begin{multline} \label{eq:source_func_persistent}
    \frac{d\mathcal{Q}_{\text{t},i}}{dE} (E,\mathbf{r};\mathbf{r}_i) = \\
    \begin{cases}
    \mathcal{C}_{\text{t},i}
    \left( \frac{E}{m_e c^2} \right)^{-p_{\text{t}}}
    \exp \left( -\frac{E}{E_{\text{t,max}}} \right) \\ \times
    \delta (\mathbf{r} - \mathbf{r}_i), & E \geq E_{\text{t,min}} \\
    0, & E < E_{\text{t,min}}
    \end{cases},
\end{multline}
where $\mathcal{C}_{\text{t},i}$ is a constant that normalizes $d\mathcal{Q}_{\text{t},i}/dE$ to $\dot{N}_{+,\text{t},i}$.
For our calculations, we choose $p_{\text{t}}=2.2$, $E_{\text{t,min}}/m_ec^2=\Gamma_{\text{jet}}=2$, and $E_{\text{t,max}}=10\,\text{TeV}$.
This choice of power law index is consistent with a relativistic ideal gas with an adiabatic index of $5/3$ \citep{sironi_2015_shocks}.
The cutoff energy $E_{\text{t,max}}$ corresponds to the energy at which acceleration becomes less efficient.
This turnover occurs because these positrons lose their energies more readily to synchrotron emission, the positrons are capable of escaping the finite-sized acceleration region before reflecting once again, or the jet dissipates after some time.
While this cutoff energy can vary on an event-by-event and source-by-source basis, some X-ray binaries have been observed with transient jets accelerating particles up to TeV energies \citep{sguera_TeV_2009, massi_TeV_2009, aleksic_TeV_2011, sudoh_TeV_2020}.
We assume for our calculations that the 
whole population sample can accelerate particles up to a cutoff energy of $E_{\text{t,max}}=10\,\text{TeV}$.

Meanwhile, persistent jets likely cannot be modeled with first-order Fermi acceleration because observations at gamma-ray energies do not exhibit the sharp spectral cutoffs consistent with the first-order process.
Instead, we model persistent jets with second-order Fermi acceleration, where particles primarily gain energy from resonant scattering off of magnetohydrodynamic (MHD) turbulence in the jet.
The outgoing spectrum for second-order Fermi acceleration is similar to its first-order counterpart, except that the exponential cutoff is softer because acceleration is not as efficient \citep{becker_stochastic_2006, stawarz_stochastic_2008, sudoh_TeV_2020}.
We model the source function for a persistent jet launched by X-ray binary $i$ by,
\begin{multline} \label{eq:source_func_transient}
    \frac{d\mathcal{Q}_{\text{p},i}}{dE} (E,\mathbf{r};\mathbf{r}_i) =\\
    \begin{cases}
    \mathcal{C}_{\text{p},i}
    \left( \frac{E}{m_e c^2} \right)^{-p_{\text{p}}}
    \exp \left[ -\frac{1}{a} \left( \frac{E}{E_{\text{p,max}}} \right)^a \right] \\ \times
    \delta (\mathbf{r} - \mathbf{r}_i), & E \geq E_{\text{p,min}} \\
    0, & E < E_{\text{p,min}}
    \end{cases},
\end{multline}
where the parameter $a$ encodes information about the jet MHD turbulence.
We assume Bohm limit stochastic acceleration, which corresponds to $a = 2$.
For the remaining parameters, we once again normalize $d\mathcal{Q}_{\text{p},i}/dE$ to $\dot{N}_{+,\text{p},i}$ through the constant $\mathcal{C}_{\text{p},i}$.
We choose $p_{\text{p}}=2$, $E_{\text{p,min}}/m_ec^2=\Gamma_{\text{jet}}=1.4$, and $E_{\text{p,max}}=10\,\text{GeV}$ as our parameters to model persistent jets.
For ideal electron gases with an adiabatic index of $5/3$, the power law index deviates from $2.2$ and approaches $2$ for lower bulk Lorentz factors.
We use a lower cutoff energy of $E_{\text{p,max}}=10\,\text{GeV}$, motivated by high-energy gamma-ray observations of steady jets \citep{albert_GeV_2007, zdziarski_GeV_2017}.
Our choice of jet models and their parameters ensures that a significant fraction of positrons annihilate close to their sources, such that the Ginzburg-Syrovatskii formalism holds while still accurately modeling the outgoing positron spectra.

The total source function for an X-ray binary of index $i$ is obtained by adding up the source functions for both jet types.
\begin{equation} \label{eq:source_func_total}
    \frac{d\mathcal{Q}_{i}}{dE} (E,\mathbf{r};\mathbf{r}_i) =
    \varepsilon_{\text{p},i} \frac{d\mathcal{Q}_{\text{p},i}}{dE} (E,\mathbf{r};\mathbf{r}_i) +
    \varepsilon_{\text{t},i} \frac{d\mathcal{Q}_{\text{t},i}}{dE} (E,\mathbf{r};\mathbf{r}_i)
\end{equation}
Here, $\varepsilon_{\text{p},i}$ and $\varepsilon_{\text{t},i}$ are the source-specific duty cycles for persistent and transient jets respectively.
We use duty cycles as a proxy to represent the fraction of time X-ray binaries actively inject positrons into the ISM.
This is a valid approach to make because X-ray binaries cycle through their spectral states on timescales much smaller than the $e^+$ annihilation time in the ISM.
To model each source as accurately as possible, we use measurements of source-specific duty cycles from the literature, then adopt class-dependent fiducial values as default duty cycles.
We use the duty cycles in \citet{sidoli_DCs_2018} to estimate fiducial duty cycles for classes of high-mass X-ray binaries.
For the default duty cycles of high-mass X-ray binaries (for classes which don't have class-dependent default values), we assume fiducial values of $\varepsilon_{\text{t},i} = 0$ (effectively no transient jets) and $\varepsilon_{\text{p},i} = 0.5$.
It is important to note that the duty cycles of low-mass X-ray binaries in outburst can vary from $\sim0.01$ to $0.1$ \citep{carbone_DC_2019}.
As an estimate, we assume a default duty cycle of $\varepsilon_{\text{p},i} = 0.1$ for persistent jets.
This value corresponds to the average duty cycle of low-mass X-ray binaries undergoing outburst \citep{tetarenko_watchdog_2016}.
For transient jets, we assume $\varepsilon_{\text{t},i}=0.01\varepsilon_{\text{p},i}$ as the default duty cycle.
The scaling factor is an estimate for the fraction of outburst time also spent in transition from the HIMS to the SIMS.
We note that while it is known that the outburst time spent within the HIMS-to-SIMS transition is a small fraction of the time spent cycling through spectral states, the transition time is not well-constrained, ranging from several hours to several days.
We obtain a scaling factor of $0.01$ by assuming an average transition time of $1\,\text{d}$ \citep{belloni_states_2010}, estimating the average time spent in outburst from the outburst times calculated in \citet{tetarenko_watchdog_2016}, then taking the ratio of the transition time and mean outburst time.

%
%
We solve the positron transfer equation (Equation (\ref{eq:one_transfer_eq})) for 
an X-ray binary located at $\mathbf{r}_i$ with the associated Green's function $\mathcal{G}_i$.
\begin{multline} \label{eq:XRB_green_func}
    \mathcal{G}_i(E,\mathbf{r},t;E_0,\mathbf{r}_0,t_0) =
    \frac{P_i(E, E_0)}{ \left| b_i(E) \right| \left[ 4\pi \lambda_i (E, E_0) \right]^{3/2}} \\ \times
    \exp \left[ -\frac{(\mathbf{r}-\mathbf{r}_0)^2}{4\lambda_i (E, E_0)} \right]
    \delta \left( t-t_0 - \int_E^{E_0}\frac{dE}{\left|b_i(E)\right|} \right)
\end{multline}
In Equation (\ref{eq:XRB_green_func}), $P_i(E,E_0)$ represents the probability for a positron with initial energy $E_0$ to survive until energy $E$.
The quantity $\lambda_i = \lambda_i(E,E_0)$ represents the squared effective $e^+$ propagation distance for a positron injected at X-ray binary $i$ with energy $E_0$ and annihilating at $E$.
\begin{equation} \label{eq:survival_prob}
    P_i(E,E_0) = \exp \left[ -\int_{E}^{E_0} \frac{dE'}{\left|b_i(E')\right|T_i(E')} \right]
\end{equation}
\begin{equation} \label{eq:green_func_lambda}
    \lambda_i (E, E_0) =
    \int_{E_0}^E dE' \frac{D_i(E')}{\left| b_i(E') \right|}
\end{equation}
where $T_i$ is the local mean $e^+$ lifetime from Equation (\ref{eq:local_pos_lifetime}).
The $\delta$ function in Equation (\ref{eq:XRB_green_func}) corresponds to the instantaneous time at which a positron injected with initial energy $E_0$ at $t_0$ annihilates at final energy $E$ after a finite elapsed time.
Convolving the source term with the Green's function gives the differential number density of positrons at $0.511\,\text{MeV}$.
\begin{multline} \label{eq:XRB_positron_concentration}
    \frac{d{N}_{\text{511},i}}{d\mathbf{r}} (\mathbf{r};\mathbf{r}_i) =
    \int_{E_{\text{th}}}^\infty dE_0 \int_{-\infty}^t dt_0 \frac{d\mathcal{Q}_{i}}{dE} (E_0,\mathbf{r};\mathbf{r}_i) \\ \times \mathcal{G}_i(E_{\text{th}},\mathbf{r},t;E_0,\mathbf{r}_0,t_0) 
\end{multline}
The differential $e^+$ annihilation rate density at $0.511\,\text{MeV}$ (in units of $e^+\,\text{s}^{-1}\,\text{cm}^{-3}\,\text{MeV}^{-1}$) for the same X-ray binary can then be found by multiplying the positron differential number density with the annihilation rate.
\begin{equation} \label{eq:XRB_sol_511}
    \frac{d\dot{N}_{\text{ann},i}}{d\mathbf{r}} (\mathbf{r};\mathbf{r}_i) =
    \frac{1}{T_i(E_{\text{th}})}
    \frac{d{N}_{\text{511},i}}{d\mathbf{r}} (\mathbf{r};\mathbf{r}_i)
\end{equation}
We define $E_{\text{th}}$ as the energy at which positrons have thermalized to the ambient medium.
For flux calculations, we integrate Equation (\ref{eq:XRB_positron_concentration}) down to $E_{\text{th}}$ and evaluate the annihilation rate at $E_{\text{th}}$.
Positrons that do not annihilate in flight will propagate until they cool to energies characterized by the ISM temperature $T_{\text{ISM}}$.
We define $E_{\text{th}} \approx m_e c^2 + \frac{3}{2}k_B T_{\text{ISM}}$ and take $T_{\text{ISM}} \approx 8000\,\text{K}$, a typical temperature for the warm ISM phase as an estimate because measurements performed by SPI-INTEGRAL showed that the majority of positrons annihilate in warm media (see \citet{churazov_2005_emission} and \citet{jean_spectral_2006}).
We choose a prefactor of $3/2$ for $E_{\text{th}}-m_ec^2$ because $\frac{3}{2}k_B T$ corresponds to the lowest kinetic energy where positrons can charge exchange and form positronium in the WNM or thermalize in the WIM \citep{jean_spectral_2006}.
This effective annihilation energy is also an appropriate mathematical tool due to the singular nature of $\mathcal{G}_i$ at $m_e c^2$.
The 0.511 MeV flux is then obtained by,
\begin{multline} \label{eq:F_XRBs_511}
    E_\gamma \frac{dF_{511,\text{X}}}{dE_\gamma} =
    (0.511\,\text{MeV})
    \int \frac{d\mathbf{r}}{4\pi\mathbf{r}^2} \\ \times
    \left[2 \left( 1 - \frac{3}{4} f_{\text{Ps}} (\mathbf{r}) \right) \right]
    \sum_i \frac{d\dot{N}_{\text{ann},i}}{d\mathbf{r}} (\mathbf{r};\mathbf{r}_i)
\end{multline}
We adopt the same $f_{\text{Ps}}$ mentioned in Section~\ref{sec:SFs} for the same reasons as before.

\subsection{Continuum Flux} \label{ssec:XRB_continuum_methods}

While the positrons injected by X-ray binaries will annihilate and contribute to the 0.511 MeV line, these positrons will also produce a gamma-ray continuum which can be decomposed into several components.
Relativistic positrons can decelerate due to Coulomb interactions with the ISM gas, emitting bremsstrahlung before they annihilate.
The continuum also originates from the annihilation of o-Ps due to thermalized positrons and positrons annihilating in-flight (e.g., they annihilate without thermalizing to the ambient medium).
To estimate the in-flight annihilation flux, we generalize Equation (\ref{eq:XRB_sol_511}) from annihilation at 0.511 MeV to arbitrary annihilation energies.
The generalized differential number density of positrons is given by,
\begin{multline} \label{eq:XRB_sol_general}
    \frac{dN_{+,i}}{d\mathbf{r}} (E,\mathbf{r};\mathbf{r}_i) =
    \int_{E_{\text{th}}}^\infty dE_0 \int_{-\infty}^t dt_0 \frac{d\mathcal{Q}_{i}}{dE} (E_0,\mathbf{r};\mathbf{r}_i) \\ \times \mathcal{G}_i(E,\mathbf{r},t;E_0,\mathbf{r}_0,t_0)
\end{multline}
where we recover Equation (\ref{eq:XRB_sol_511}) from the generalized number density distribution in the limit where $E\rightarrow E_{\text{th}}$.
For the in-flight continuum, we calculate the photon spectral distribution of annihilating positrons by integrating the energy-differential reaction rate over time to obtain the in-flight annihilation emissivity.
\begin{multline} \label{eq:XRB_IA_kernel}
    H_{\text{IA},i} (E_\gamma, \mathbf{r}) =
    \int dE \frac{dN_{+,i}}{d\mathbf{r}} (E,\mathbf{r};\mathbf{r}_i) \\ \times
    n_e (\mathbf{r})
    \frac{d\sigma_{\text{IA}}}{dE_{\gamma}} (E_\gamma, E)
    v(E)
\end{multline}
We use the $e^+$ in-flight annihilation cross section defined in \citet{aharonian_cosmic_1981}.
Then the gamma-ray spectrum from in-flight annihilation can be calculated by integrating the in-flight emissivity over all space.
\begin{equation} \label{eq:F_XRBs_IA}
    E_\gamma \frac{dF_{\text{IA},\text{X}}}{dE_\gamma} =
    E_\gamma
    \int \frac{d\mathbf{r}}{4\pi\mathbf{r}^2}
    \sum_i H_{\text{IA},i} (E_\gamma, \mathbf{r})
\end{equation}
The physical interpretation of Equation (\ref{eq:F_XRBs_IA}) is that all positrons that annihilate in-flight at energy $E$ will emit photons at energy $E_\gamma$ according to a probability distribution, given by the energy-differential reaction rate at constant $E$.
We also calculate the bremsstrahlung radiation from positrons injected by X-ray binaries due to interactions with ISM hydrogen and helium.

The procedure for obtaining the bremsstrahlung continuum is essentially the same as in-flight annihilation.
For the bremsstrahlung component, we convolve the generalized differential positron number density from Equation (\ref{eq:XRB_sol_general}) with the bremsstrahlung reaction rate using the bremsstrahlung differential cross section from \citet{bethe_heitler_1934}.
The Bethe-Heitler cross section $d\tilde{\sigma}_{\text{BR}}/dE_\gamma$ is also corrected by scaling with the Elwert-Sommerfeld (ES) factor $\mathcal{F}_{\pm}$ \citep{elwert_bremsstrahlung_1939, gould_bremsstrahlung_1990}.
The correction is given by,
\begin{equation} \label{eq:bethe_heitler_correction}
    \frac{d\sigma_{\text{BR},{\pm}}}{dE_{\gamma}}(E_\gamma, E, Z) =
    \mathcal{F}_{\pm}(E_\gamma, E, Z)
    \frac{d\tilde{\sigma}_{\text{BR}}}{dE_{\gamma}}(E_\gamma, E, Z)
\end{equation}
\begin{equation} \label{eq:elwert_factor}
    \mathcal{F}_{\pm}(E_\gamma, E, Z) =
    \frac{v_i}{v_f}
    \frac{1 - \exp \left( \mp 2\pi Z\alpha / \beta_i \right)}{1 - \exp \left( \mp 2\pi Z\alpha / \beta_f \right)}
\end{equation}
where $d\sigma_{\text{BR}}/dE_\gamma$ denotes the ES-corrected Bethe-Heitler cross section.
Here, $Z$ is the atomic number, $\alpha$ is the fine structure constant, and $v_i=v(E), v_f=v(E-E_\gamma)$ respectively are the initial and final velocities for electrons ($-$) and positrons ($+$).
The bremsstrahlung component from positrons is then given by,
\begin{multline} \label{eq:XRB_BR_kernel}
    H_{\text{BR},+,i} (E_\gamma, \mathbf{r}) =
    \int dE
    \frac{dN_{+,i}}{d\mathbf{r}} (E,\mathbf{r};\mathbf{r}_i) \\ \times
    \left[ n_{\text{H}} (\mathbf{r}) \frac{d\sigma_{\text{BR},{e^+\text{-}\text{H}}}}{dE_{\gamma}}(E_\gamma, E) + n_{\text{He}} (\mathbf{r}) \frac{d\sigma_{\text{BR},e^+\text{-}\text{He}}}{dE_{\gamma}}(E_\gamma, E) \right] \\ \times
    v(E)
\end{multline}
\begin{equation} \label{eq:F_XRBs_BR}
    E_\gamma \frac{dF_{\text{BR},+,\text{X}}}{dE_\gamma} =
    E_\gamma
    \int \frac{d\mathbf{r}}{4\pi\mathbf{r}^2}
    \sum_i
    H_{\text{BR},+,i} (E_\gamma, \mathbf{r})
\end{equation}
Along with bremsstrahlung from the positrons, electrons injected into the ISM by jets from X-ray binaries will also produce a bremsstrahlung component.
The electronic bremsstrahlung cross section differs from positronic bremsstrahlung due to the Coulomb attraction/repulsion of electrons/positrons with ISM nuclei, encapsulated in the ES factor given in Equation (\ref{eq:elwert_factor}).
The electronic bremsstrahlung component is then obtained through Equations (\ref{eq:XRB_BR_kernel}) and (\ref{eq:F_XRBs_BR}).

We estimate the continuum due to the annihilation of o-Ps by using the differential o-Ps spectrum $df_{3\gamma}/dE_\gamma$ from \citet{ore_three-photon_1949} and the differential $e^+$ annihilation rate density at $0.511\,\text{MeV}$ from Equation (\ref{eq:XRB_sol_511}).
\begin{multline} \label{eq:F_XRBs_oPs}
    E_\gamma \frac{dF_{\text{oPs},\text{X}}}{dE_\gamma} =
    E_\gamma^2 \frac{df_{3\gamma}}{dE_\gamma} (E_\gamma) \\ \times
    \int \frac{d\mathbf{r}}{4\pi\mathbf{r}^2}
    \left[ 3 \frac{3}{4} f_{\text{Ps}} (\mathbf{r}) \right]
    \sum_i \frac{d\dot{N}_{\text{ann},i}}{d\mathbf{r}} (\mathbf{r};\mathbf{r}_i)
\end{multline}

\section{Flux Calculations} \label{sec:calculations}

To obtain flux maps, we divide the sky into $1^\circ \times 1^\circ$ solid angles (lines of sight).
Then we integrate Equation (\ref{eq:F_SFs}) and Equation (\ref{eq:F_XRBs_511}) for all solid angles.
We integrate Equation (\ref{eq:F_SFs}) for each line of sight using numerical methods.
Equation (\ref{eq:F_XRBs_511}) requires integration over all energy and all distance within each solid angle.
We calculate the stellar flare flux numerically while we compute the X-ray binary flux using Monte Carlo methods.
We use Monte Carlo integration because Equation (\ref{eq:F_XRBs_511}) requires integration over all space and positron energies to infinity.
To ensure the results are accurate, we train our Monte Carlo integrator with $10^4$ samples and 10 iterations to allow our integrator to adapt to the integrand for each line of sight.
After training, we calculate the flux with $10^5$ samples and 10 iterations.
We finally obtain the total $e^+$ annihilation flux by adding the flux from all lines of sight.
To calculate flux uncertainties, we propagate errors from the parameters used in our models.
The number of samples we use for Monte Carlo integration produces numerical integration errors comparably lower in orders of magnitude to our model parameters, so we neglect the contribution of numerical errors during error propagation.

For the gamma-ray continuum from X-ray binaries, we calculate $E F_E$ for the o-Ps continuum, in-flight annihilation, and $e^+$ bremsstrahlung at 4000 different energies.
We define 2000 energies in a geometric scale from $m_e c^2$ to 10 keV, and 2000 energies in a geometric scale from $m_e c^2$ to $10^8 m_e c^2$.
We then calculate the in-flight annihilation continuum with Equation (\ref{eq:F_XRBs_IA}), the bremsstrahlung continuum with Equation (\ref{eq:F_XRBs_BR}), and the o-Ps continuum with Equation (\ref{eq:F_XRBs_oPs}).
The bremsstrahlung integral is simple enough to calculate without Monte Carlo methods.
We use Monte Carlo methods to obtain the in-flight annihilation component due to its extra dependence on the annihilation energy.
While calculating fluxes, we neglect attenuation by the ISM.
The ISM is expected to be optically thin to electron scattering because measurements of the Galactic hydrogen column densities indicate maximum column densities of $\mathcal{O}(10^{21} \, \text{cm}^{-2})$ \citep{hi4pi_collaboration_hi4pi_2016} whereas the Klein Nishina cross section is $\sigma_{\text{KN}} \sim \mathcal{O}(10^{-25}\,\text{cm}^2)$ at 0.511 MeV.

\section{Results and Discussion} \label{sec:results}

\begin{figure*}
\plottwo{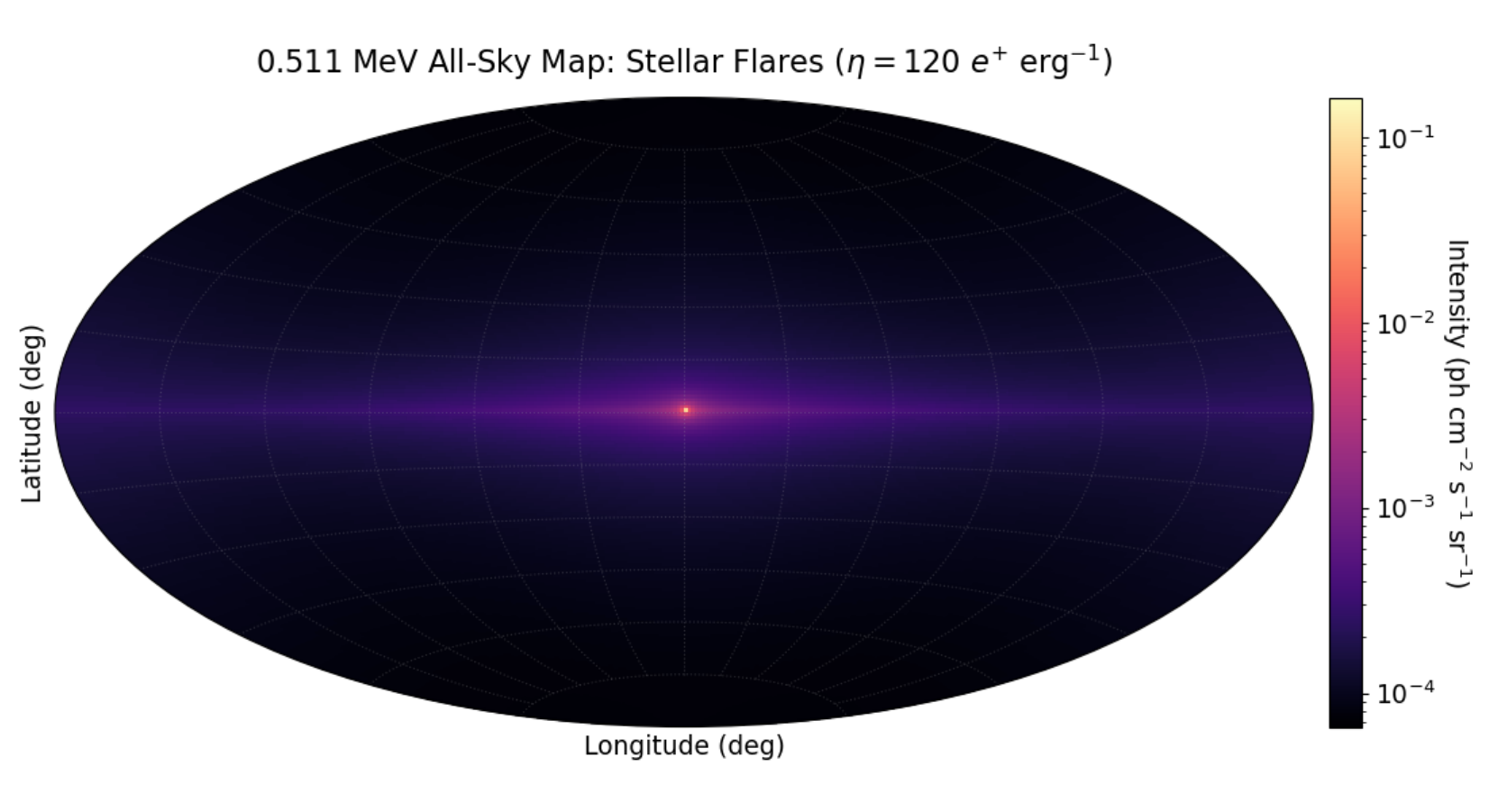}{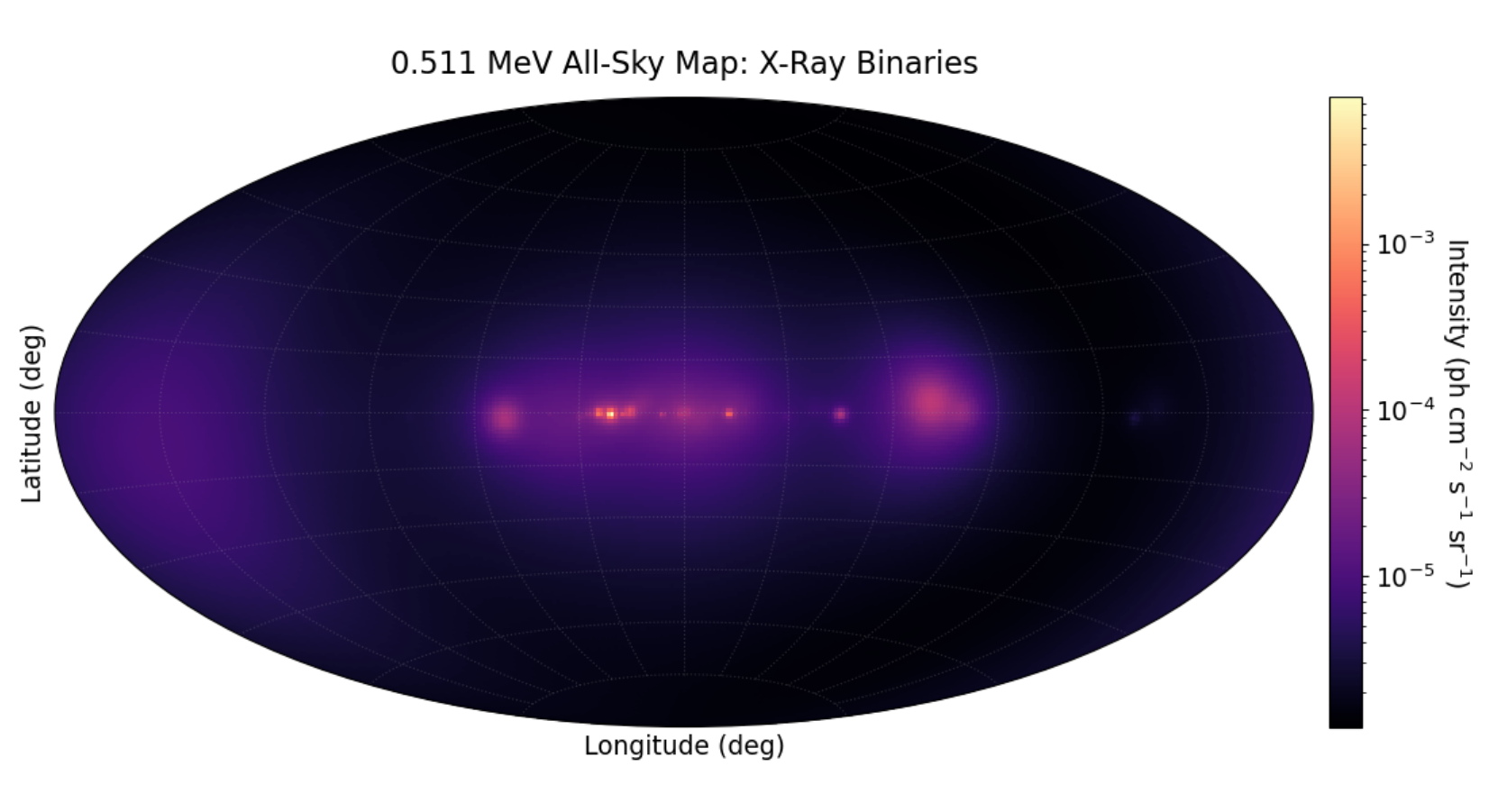}
\caption{
All-sky maps for our 0.511 MeV flux calculations separated by positron source.
\textbf{Left:} The $e^+$ annihilation flux for stellar flares.
Stellar flares trace out the distribution of stars in the Galaxy, so the morphology is heavily disk-dominated.
\textbf{Right:} The $e^+$ annihilation flux for X-ray binaries.
The morphology for X-ray binaries creates a faint bulge component along with point emission along the Galactic plane.
}
\label{fig:separate_maps}
\end{figure*}

%
%
We obtain a total 0.511 MeV flux of $(2.423 \pm 0.260)\times10^{-3}\,\text{ph}\,\text{cm}^{-2}\,\text{s}^{-1}$.
Stellar flares contribute a flux of $(2.354 \pm 0.260)\times10^{-3}\,\text{ph}\,\text{cm}^{-2}\,\text{s}^{-1}$ while X-ray binaries contribute a flux of $(6.940\pm 0.283) \times10^{-5}\,\text{ph}\,\text{cm}^{-2}\,\text{s}^{-1}$, which is about 34 times fainter.
The morphologies of both sources differ significantly.
Whereas the flux from stellar flares is disk-dominated, the flux from X-ray binaries without pulsars produces a faint bulge component and point emission across the disk (see Figure~\ref{fig:separate_maps}).
We reiterate that \citet{bisnovatyi-kogan_SFs_2017} argue that a positron yield per unit energy of $\eta=340\,e^+\,\text{erg}^{-1}$ is required to match $e^+$ annihilation line measurements.
When including the stellar population in the Galactic disk and accounting for other stellar classes capable of flaring activity, a value of $\eta=120\,e^+\,\text{erg}^{-1}$ reproduces our best measurements of the 0.511 MeV line flux quite well.

Based on the total flux of X-ray binaries and number of X-ray binaries considered, this corresponds to a total injection rate of $\sim1.025\times10^{42} \, e^+ \, \text{s}^{-1}$.
Such an injection rate is on the same order of magnitude as the expected injection rate from the $\beta^+$ decay of ${}^{26}\text{Al}$ and ${}^{44}\text{Ti}$ \citep{prantzos_511_2011}.
Measurements from SPI-INTEGRAL indicate that ${}^{26}\text{Al}$ is expected to contribute $\sim3.5\times10^{42}\,e^+\,\text{s}^{-1}$ to the global positron rate \citep{wang_26Al_2009}.
Meanwhile, direct measurements from supernova remnant Cassiopeia A imply that the decay of $^{44}\text{Ti}$ results in a positron rate contribution on the same order of magnitude: $\sim1.6\times10^{42}\,e^+\,\text{s}^{-1}$, assuming a core-collapse supernova rate of $2\times10^{-2} \,\text{yr}^{-1}$ \citep{diehl_26Al_2006, siegert_positron_2023}.
Our maps indicate that X-ray binaries may create some 0.511 MeV hotspots (see Section~\ref{ssec:pt_srcs}).
However, when also factoring in potential contributions from flaring stars, these measurements have consequences for the detectability of these hotspots.

It has been hypothesized that low-mass X-ray binaries could produce the bulge component of the $e^+$ annihilation emission because a large number of low-mass X-ray binaries populate the bulge region \citep{grimm_XRBs_2002, prantzos_XRBs_2004}.
However, our maps indicate that their contribution is negligible when compared to stellar flares.
We obtain a flux of $2.090\times10^{-4}\,\text{ph}\,\text{cm}^{-2}\,\text{s}^{-1}$ from the inner $16^\circ$ of our combined all-sky map, underestimating the lower bound bulge emission detected by previous missions.
As a result, the combined morphology of the two positron sources is dominated by disk emission with a faint bulge component (see Figure~\ref{fig:XRB_pt_src_regions}).
The morphology is therefore more consistent with the model in \citet{skinner_galactic_2015} and the $\text{G}_{12}$ + thin disk model in \citet{siegert_cosi_2020}.
In contrast, the model in \citet{siegert_flux_2016} and the $\text{G}_{12}$ + thick disk model in \citet{siegert_cosi_2020} do not adequately describe our maps because the vertical extents of the disk component in these models are too large.
The prominent disk emission in our calculations warrants comparisons to other disk-tracing sources like $^{26}\text{Al}$.
Based on the intensity maps we obtained, stellar flares and X-ray binaries within the $|l| \leq 30^\circ,|b|\leq10^\circ$ region of the galaxy would contribute a flux of $2.454\times10^{-4}\,\text{ph}\,\text{cm}^{-2}\,\text{s}^{-1}$.
This flux is comparable to, but slightly less than, the estimated 0.511 MeV flux of $4.3\times10^{-4}\,\text{ph}\,\text{cm}^{-2}\,\text{s}^{-1}$ due to $^{26}\text{Al}$ in the same region as measured by SPI-INTEGRAL \citep{bouchet_26Al_2015}.
Factoring in the contributions of $^{26}\text{Al}$ and other $\beta^+$--unstable astrophysical isotopes would mainly strengthen the disk component.
Regardless, our model does not capture the strong bulge component found in 0.511 MeV all-sky observations.
The bulge component either (1) does not originate solely from low-mass X-ray binaries and stellar flares, or (2) hints at $e^+$ transport from the disk into the bulge region.

\begin{figure*}
\plottwo{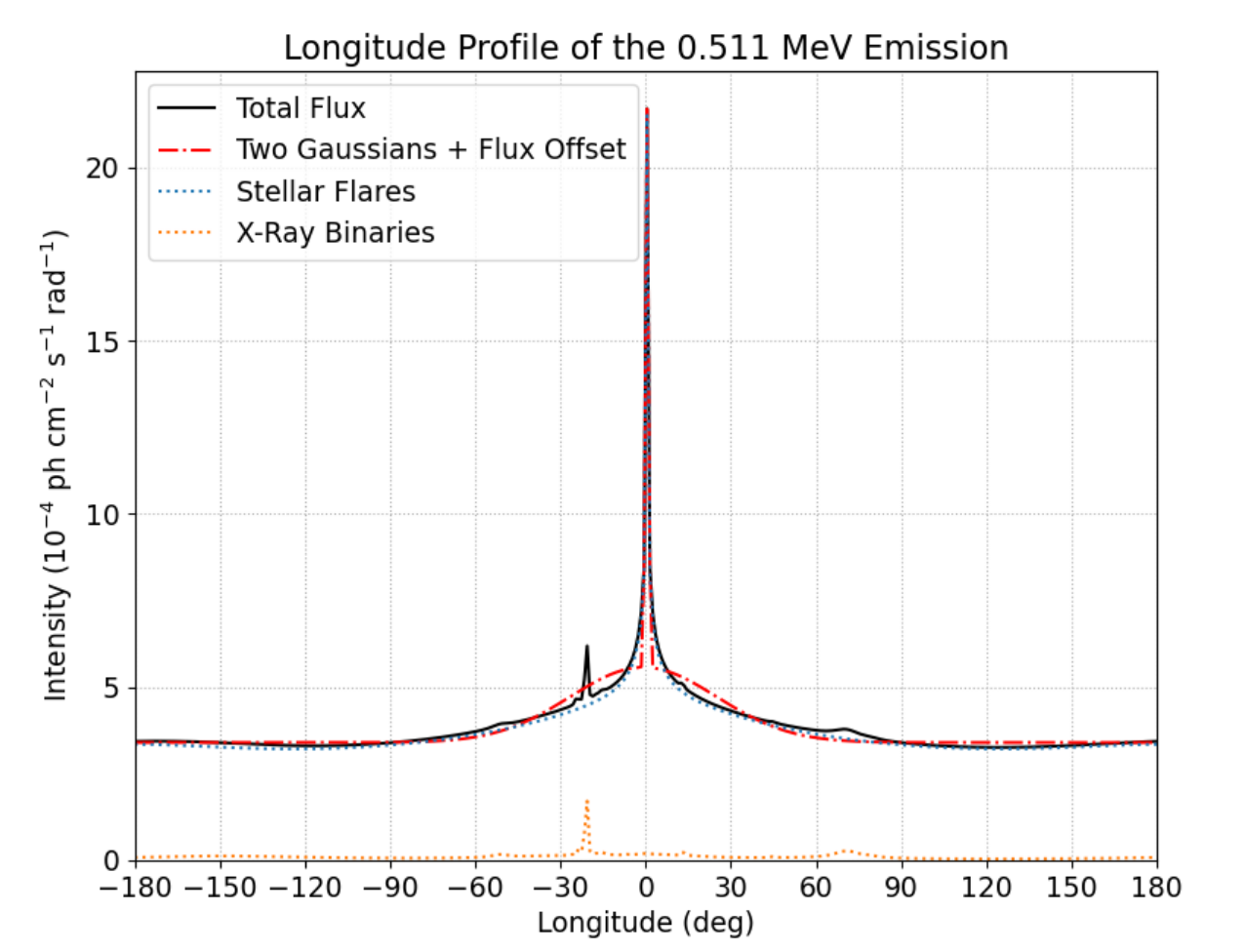}{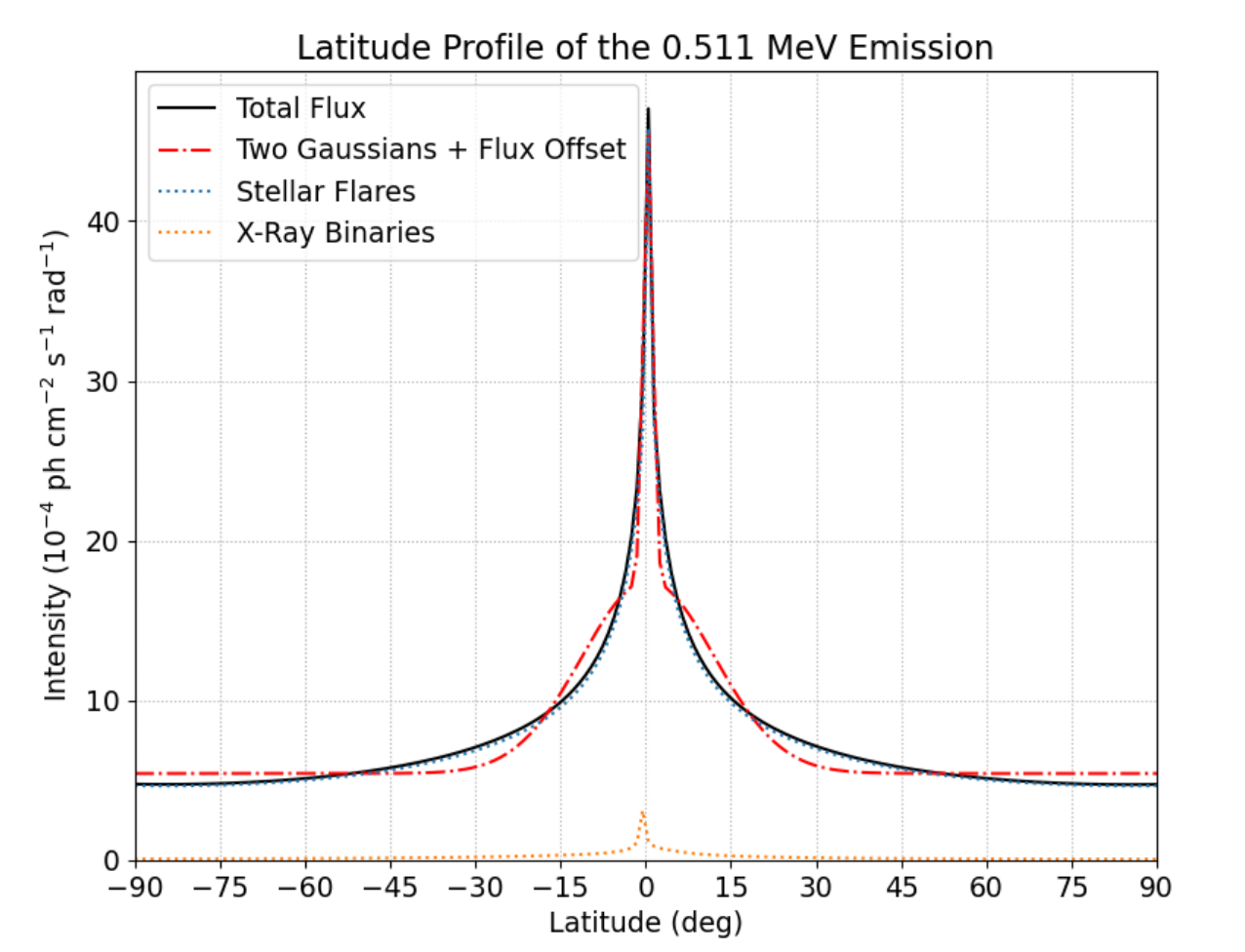}
\caption{
\textbf{Left:} The narrow Gaussian, wide Gaussian, and offset fit for the latitude-integrated total flux (shown by the red dash-dotted line).
The total flux (black solid line) and fluxes from stellar flares (blue dotted line) and X-ray binaries (orange dotted line) are also shown here.
When considered alongside stellar flares, the negative longitudinal asymmetry is relatively small -- only by about $\sim0.832^\circ$.
\textbf{Right:} The narrow Gaussian, wide Gaussian, and offset fit for the longitude-integrated total flux.
The overall flux exhibits a positive latitudinal asymmetry of $\sim0.465^\circ$.
}
\label{fig:flux_profiles}
\end{figure*}

X-ray binaries also introduce a small degree of asymmetry in Galactic longitude and latitude.
To estimate the level of longitudinal asymmetry in our flux maps, we integrate the flux for all latitudes along each longitude.
Then we fit a narrow Gaussian, wide Gaussian, and a flux offset to the integrated flux.
We reverse the roles of Galactic longitude and latitude to estimate latitudinal asymmetry.
We show our fits in Figure~\ref{fig:flux_profiles}.
We find that the 0.511 MeV flux is asymmetric towards negative longitudes by $\sim0.832^\circ$.
We believe that our flux appears asymmetrical in longitude because of the distribution of X-ray binaries in the bulge.
Despite observing one extra hotspot at $b>0^\circ$ over $b<0^\circ$ (see Section~\ref{ssec:pt_srcs}), the strength of the negative-longitude hotspots may also shift our two-Gaussian fit towards negative longitudes.
We also find a positive latitude asymmetry of $\sim0.465^\circ$.
The small asymmetry in Galactic latitude contrasts the results found in \citet{yoneda_2025_SPI}, where twenty years of SPI-INTEGRAL data shows asymmetry towards positive latitudes. 
\citet{yoneda_2025_SPI} proposes that the latitudinal asymmetry originates either from massive stars in OB associations, foreground contamination by cosmic ray interactions with asteroids \citep{siegert_2024_foreground}, or image deconvolution artifacts from SPI-INTEGRAL's non-uniform exposure at high latitudes.
Neither of these sources are captured in our flux calculations, which explains the lack of a positive latitude enhancement in our results.

\subsection{Gamma-Ray Continuum from X-Ray Binaries}

\begin{figure*}
\plotone{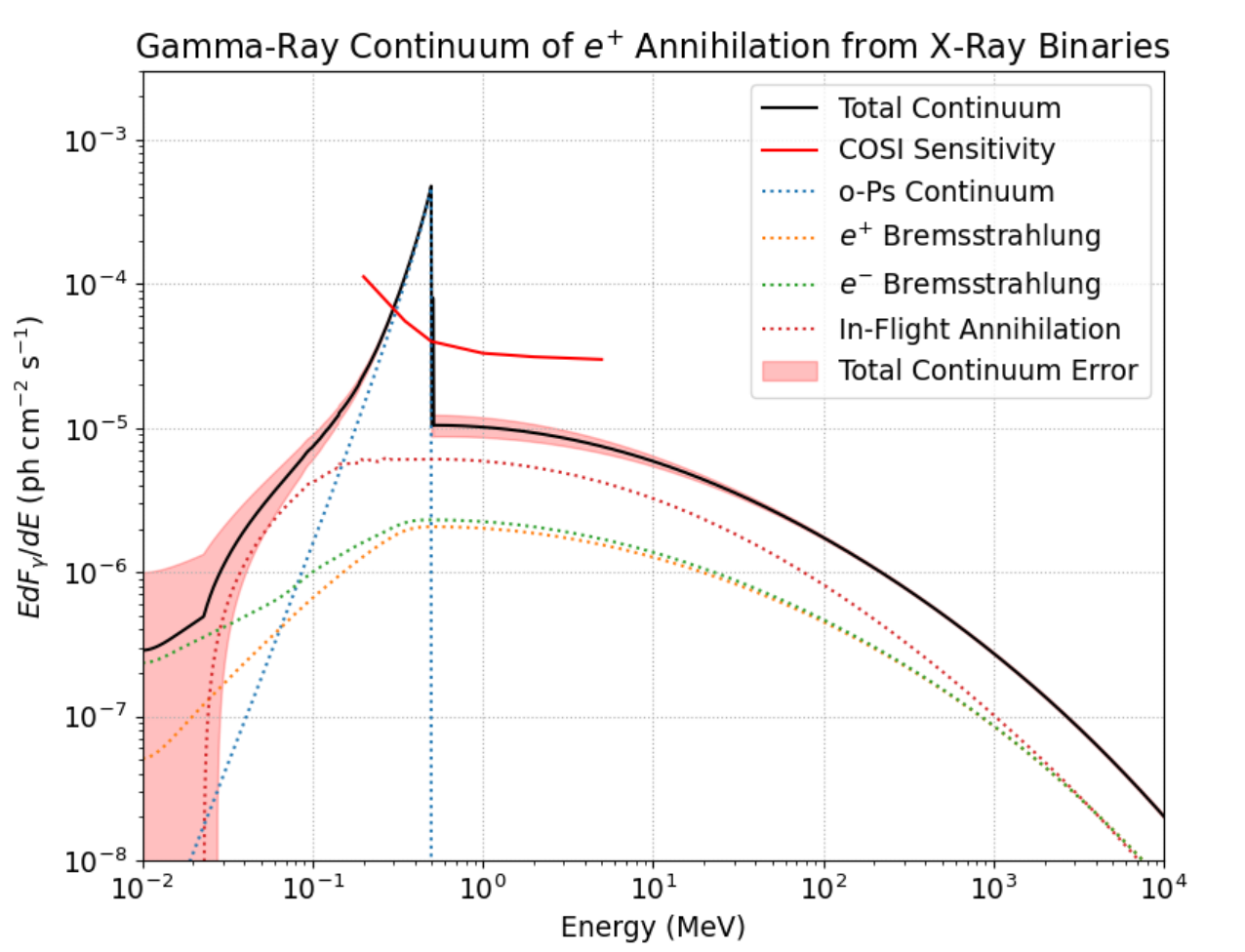}
\caption{
The gamma-ray continuum from positrons and electrons injected by X-ray binaries with the parameters described in Section~\ref{ssec:XRB_inj_rates} and Section~\ref{ssec:XRB_511_methods} (plotted in black).
The total uncertainty in the continuum is given in the red shaded region.
We also plot the individual components with dotted lines.
The red curve shows the continuum sensitivity of the upcoming Compton Spectrometer and Imager (COSI) mission within its bandpass.
The sensitivity curve assumes a point source with average exposure for 2 years at the $3\sigma$ level \citep{tomsick_cosi_2024}.
}
\label{fig:XRB_continuum}
\end{figure*}

Figure~\ref{fig:XRB_continuum} shows the gamma-ray continuum due to the injection of $e^\pm$ into the ISM by X-ray binaries, along with the associated integration error.
Bremsstrahlung from jet electrons dominates up to 10 keV.
Afterwards, the annihilation of o-Ps dominates the continuum from $\sim200\,\text{keV}$ up to 0.511 MeV due to preferential emission of photons just below the electron rest energy.
The total continuum drops just above 0.511 MeV, corresponding with the upper end of the o-Ps continuum.
From here, the only contributions to the continuum are from in-flight annihilation and bremsstrahlung from $e^+$ and $e^-$.
These three components extend into higher energies.
The gamma-ray continuum from positrons injected by X-ray binaries would supposedly contribute to the total diffuse gamma-ray continuum.
Diffuse gamma-rays originate from Galactic point sources and cosmic ray interactions with the ISM.
The Galactic diffuse component dominates over the extragalactic background, whose origin is still undetermined \citep{moskalenko_2004_diffuse}.
To pinpoint sources of the 0.511 MeV excess from the Galactic Center, the positron continuum from these sources must be disentangled from the components of the diffuse gamma-ray emission while looking at data taken by current and future gamma-ray telescopes.
%
%
It is important to mention that the upcoming Compton Spectrometer and Imager (COSI) mission is expected to launch in 2027.
Alongside the gamma-ray continuum, Figure~\ref{fig:XRB_continuum} plots the continuum sensitivity of COSI across its bandpass ($0.2$--$5\,\text{MeV}$).
The sensitivity curve assumes observation of isolated point sources over 2 years of average exposure time at the $3\sigma$ level.
COSI should be able to detect the o-Ps continuum down to $\sim200\,\text{keV}$ up to the 0.511 MeV line of the predicted gamma-ray continuum.

\subsection{0.511 MeV Hotspots from X-Ray Binaries} \label{ssec:pt_srcs}
\begin{figure*}
\plotone{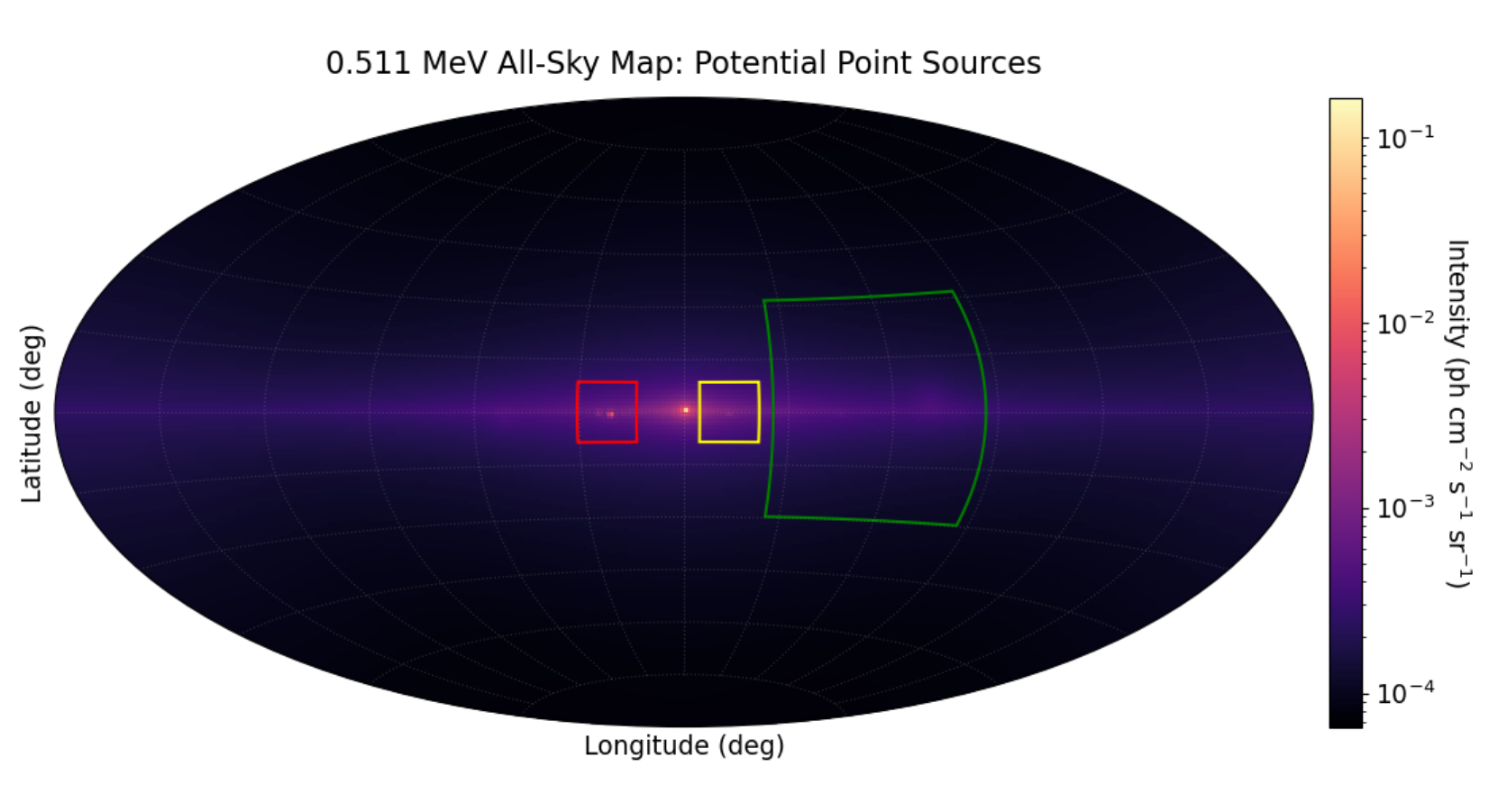}
\caption{
The combined all-sky map for $e^+$ annihilation for stellar flares and X-ray binaries.
The overall 0.511 MeV morphology looks very similar to the morphology from stellar flares, but with a slightly stronger bulge component due to the higher density of X-ray binaries within the inner $15^\circ$ of the sky.
Even if the combined X-ray binary flux is weaker than stellar flares, some X-ray binaries may be distinguishable given the duty cycles from Section~\ref{ssec:XRB_511_methods}.
Regions of interest are marked here in red, yellow, and green.
The hotspots identified within each region are shown in Figure~\ref{fig:XRB_pt_srcs}.
}
\label{fig:XRB_pt_src_regions}
\end{figure*}

\begin{figure*}
\plotone{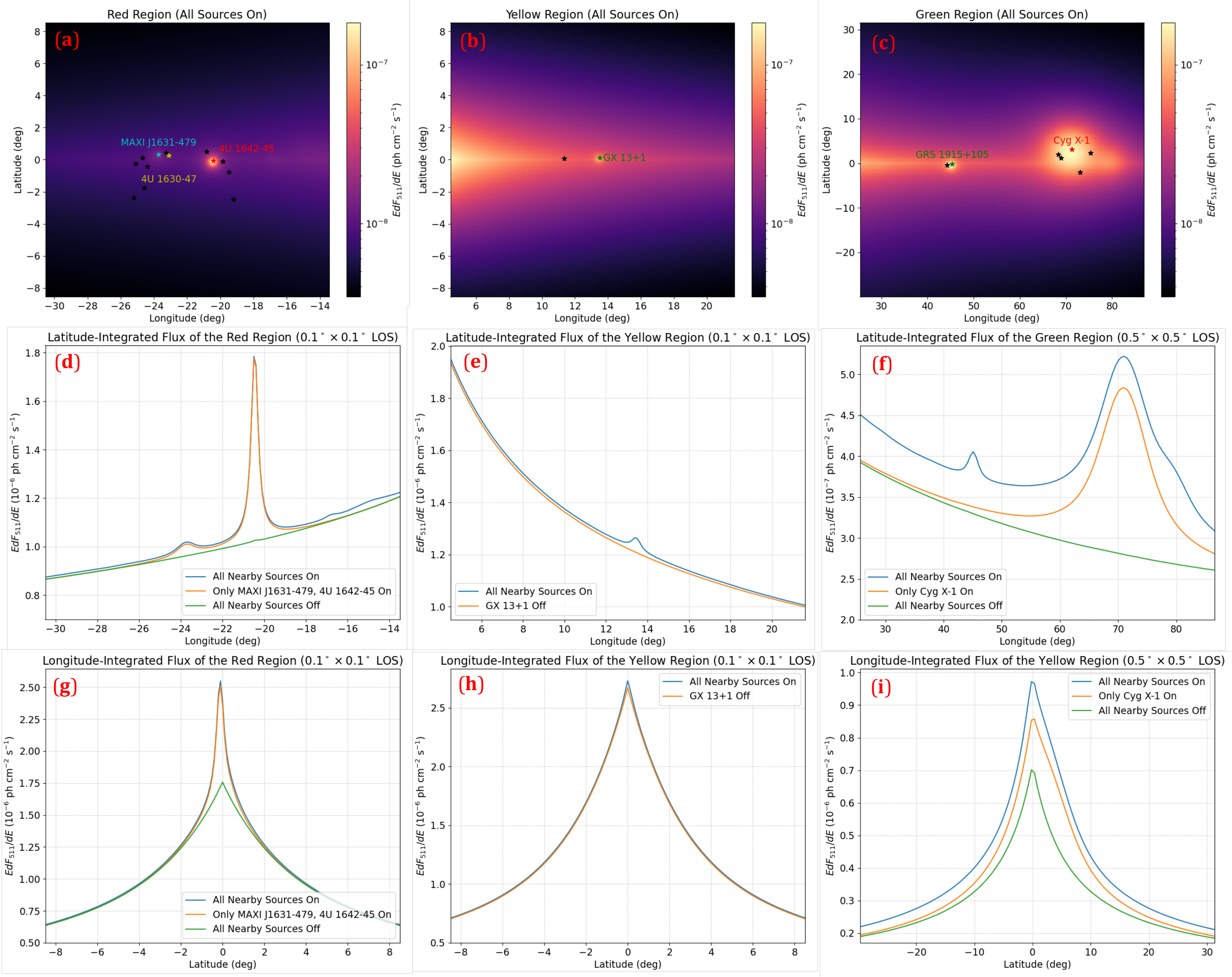}
\caption{
All 0.511 MeV hotspots in ascending order of Galactic longitude.
Plots (a), (d), and (g) show the regional flux map of 0.511 MeV with $0.1^\circ \times0.1^\circ$ lines of sight, latitude-integrated flux, and longitude-integrated flux for the red region in Figure~\ref{fig:XRB_pt_src_regions} respectively.
Plots (b), (e), and (h) are the same respective plots, but for the yellow region.
Plots (c), (f), and (i) show the same plots but with $0.5^\circ \times0.5^\circ$ lines of sight for the green region.
For plots (a--c), we also plot sources within $3^\circ$ (``nearby'' sources), except for the hotspot at $l=70^\circ$, where we show sources within $5^\circ$.
These sources are plotted as stars.
The colored stars indicate the brightest sources associated with hotspots.
For the remaining plots, we plot the total flux for all nearby sources, then gradually disable nearby sources to demonstrate their contributions to the regional flux.
The X-ray binaries associated with each hotspot are given in Table~\ref{tab:XRB_pt_srcs}.
}
\label{fig:XRB_pt_srcs}
\end{figure*}

In the following, we discuss the possibility of detecting hotspots in 0.511 MeV flux maps and determine the dominant contributors to these hotspots.
The three regions of interest in our all-sky maps are shown in Figure~\ref{fig:XRB_pt_src_regions}.
We first compute flux maps for these regions and increase the angular resolution to investigate the morphology of these hotspots.
The red and yellow regions are calculated with $0.1^\circ\times0.1^\circ$ lines of sight while the green region has $0.5^\circ\times0.5^\circ$ lines of sight.
After doing so, we identify X-ray binaries within a $3^\circ$ radius of the hotspot peak.
Our choice of radius is motivated by the fact that the red region contains two hotspots in close proximity to each other.
For the green region, we relax our search condition to $5^\circ$ due to the scarcity of sources and the relatively lower density of the ISM.
To study the contribution of individual X-ray binaries to their respective hotspots, we plot regional longitude and latitude profiles, then re-run calculations with different sources turned off.
We also empirically estimate the angular diameter of these hotspots by fitting a Gaussian + constant model onto the total flux data, then extracting the Full-Width-at-Half-Maximum (FWHM).
We plot the identified hotspots in Figure~\ref{fig:XRB_pt_srcs}, then report hotspot data and information on their nearby X-ray binaries in Table~\ref{tab:XRB_pt_srcs}.

It should not be surprising that the hotspots which appear in our maps lie within the Galactic disk.
In the ISM model of \citet{ferriere_global_1998}, each ISM component drops exponentially with scale height with respect to the disk.
As a result, X-ray binaries residing outside of the disk have their $e^+$ annihilation distributions smeared out to larger distances than X-ray binaries in the disk.
Observations, whether pointed towards specific X-ray binaries or all-sky surveys, can determine whether X-ray binaries may contribute a notable fraction of the Galactic Center $e^+$ annihilation flux.
Here we describe the dominant X-ray binaries contributing to the 0.511 MeV hotspots we have identified in our all-sky maps in descending order of Galactic longitude (from right-to-left in Figure~\ref{fig:XRB_pt_srcs}).

\startlongtable
\begin{deluxetable*}{c|c|c|c|c}
\tablecaption{The sky locations $(l, b)$ for all identified hotspots in degrees.
For all hotspots, we also show their angular diameter (FWHM) in arcminutes, nearby X-ray binaries, and their 0.511 MeV flux contributions.
The X-ray binary with the strongest contribution for each hotspot is shown in bold text.
\label{tab:XRB_pt_srcs}}
\tablehead{
\colhead{Longitude, $l$\tablenotemark{a}} & \colhead{Latitude, $b$\tablenotemark{a}} & \colhead{Angular Diameter\tablenotemark{b} } & \colhead{Nearby Sources} & \colhead{Line Flux, $E_\gamma (dF_{511,\text{X}}/dE_{\gamma})$} \\
\colhead{($\deg$)} & \colhead{($\deg$)} & \colhead{($^{\prime}$)} & \colhead{} & \colhead{($\text{ph}\,\text{cm}^{-2}\,\text{s}^{-1}$)}
}
\startdata
$-23.7$ & $+0.2$ & $0.448$ & \textbf{MAXI J1631-479} & $\mathbf{6.617\times10^{-7}}$ \\
{ } & { } & { } & 4U 1630-47 & $2.206\times10^{-7}$ \\
{ } & { } & { } & 4U 1624-49 & $5.136\times10^{-9}$ \\
{ } & { } & { } & XTE J1637-498 & $2.599\times10^{-9}$ \\
{ } & { } & { } & IGR J16374-5043 & $1.427\times10^{-9}$ \\
{ } & { } & { } & IGR J16318-4848 & $6.115\times10^{-10}$ \\
{ } & { } & { } & IGR J16374-5043 & $3.520\times10^{-10}$ \\
{ } & { } & { } & IGR J16283-4838 & $4.427\times10^{-11}$ \\
\hline
$-20.5$ & $-1.0$ & $31.409$ & \textbf{4U 1642-45} & $\mathbf{1.811\times10^{-6}}$ \\
{ } & { } & { } & XTE J1652-453 & $8.668\times10^{-8}$ \\
{ } & { } & { } & IGR J16418-4532 & $4.061\times10^{-9}$ \\
{ } & { } & { } & XTE J1701-462 & $1.831\times10^{-9}$ \\
{ } & { } & { } & IGR J16479-4514 & $1.459\times10^{-9}$ \\
\hline
$+13.5$ & $+0.1$ & $0.457$ & \textbf{GX 13+1} & $\mathbf{2.826\times10^{-7}}$ \\
{ } & { } & { } & XTE J1810-189 & $3.348\times10^{-9}$ \\
\hline
$+45.1$ & $-0.7$ & $13.069$ & \textbf{GRS 1915+105} & $\mathbf{2.206\times10^{-7}}$ \\
{ } & { } & { } & IGR J19140+951 & $1.696\times10^{-9}$ \\
\hline
$+71.1$ & $+2.8$ & $472.285$ & \textbf{Cygnus X-1} & $\mathbf{5.208\times10^{-6}}$ \\
{ } & { } & { } & XTE J2012+381 & $5.889\times10^{-8}$ \\
{ } & { } & { } & 4U 1954+31 & $8.283\times10^{-9}$ \\
{ } & { } & { } & 2MASS J20002185+3211232 & $3.389\times10^{-9}$ \\
{ } & { } & { } & GS 2023+338 & $2.295\times10^{-10}$ \\
\enddata
\tablenotetext{a}{Sky locations are reported based on the regional flux maps with the angular resolutions given in Figure~\ref{fig:XRB_pt_srcs}.}
\tablenotetext{b}{The angular diameter is represented by the hotspot's FWHM. We obtain the FWHM by fitting a Gaussian and offset on the total flux.}
\end{deluxetable*}

\textbf{Cygnus X-1:} Cygnus X-1 (Cyg X-1) is a high-mass X-ray binary with a black hole which was discovered in 1965 \citep{bowyer_Cyg_X-1_1965}.
The binary system consists of a black hole with a mass of $(21.2\pm 2.2)\,M_{\odot}$ and a type-O supergiant stellar companion of mass $44.6^{+7.7}_{-7.1}\,M_{\odot}$ \citep{miller-jones_Cyg_X-1_2021}.
Cyg X-1 is one of the brightest persistent sources of X-rays in our galaxy and contributes prominently to the 0.511 MeV hotspot. Assuming a future extremely sensitive mission, the Cyg X-1 hotspot can be distinguished from the stellar flare emission due to its spatial extent and its location outside the inner $1^\circ$ of Galactic latitudes.
Figure~\ref{fig:XRB_pt_src_regions} shows that the hotspot near Cyg X-1 is relatively faint, making its flux difficult to detect. 
Indeed, \citet{jourdain_cyg_X-1_2012} report hard X-ray and MeV-band observations of Cyg X-1 with SPI-INTEGRAL that do not reveal evidence for a hotspot at the location of Cyg X-1. The negative result can be attributed 
to SPI-INTEGRAL's angular resolution of $2.5^\circ$ \citep{winkler_INTEGRAL_2003} and the presence of the disk component that drowns out the hotspot flux.
The 0.511 MeV emission from positrons injected by jets from Cyg X-1 could also be drowned out by annihilation emission from other sources, like positrons from the $\beta^+$ decay of $^{26}\text{Al}$ in the Cygnus region.
For example, SPI-INTEGRAL measured a 1.806--1.812 MeV flux of $(6.0\pm1.0)\times10^{-5}\,\text{ph}\,\text{cm}^{-2}\,\text{s}^{-1}$ from the Cygnus region \citep{martin_26Al_2009}.
This flux would correspond to a 0.511 MeV line flux of $(7.3\pm1.2)\times10^{-5}\,\text{ph}\,\text{cm}^{-2}\,\text{s}^{-1}$, given a branching ratio of $82\%$.
The $e^+$ annihilation flux from decaying $^{26}\text{Al}$ would then easily mask the 0.511 MeV emission from Cyg X-1.

\textbf{GRS 1915+105:} GRS 1915+105 was the first source to be detected with apparent superluminal motion and thus, the first microquasar identified \citep{mirabel_1994_grs_1915}.
This microquasar contains a $12.4^{+2.0}_{-1.8}\,M_{\odot}$ black hole \citep{reid_GRS_1915+105_2014} and a K III companion star with a mass of $(0.81\pm0.53)\,M_{\odot}$ \citep{harlaftis_GRS_1915+105_2004}.
GRS 1915+105 exhibits variability which is often tied to ejecting at least part of its inner accretion flow, possibly including the corona, through transient jets \citep{pooley_GRS_1915_1997}.
Interestingly, GRS 1915+105 differs from other low-mass X-ray binaries in the sense that it has remained active since its discovery in 1992.
It makes sense for GRS 1915+105 to create a 0.511 MeV hotspot on account of its high activity.
The $e^+$ annihilation morphology of GRS 1915+105 differs from Cyg X-1 in the sense that GRS 1915+105's flux is more concentrated whereas Cyg X-1's flux is spread out across the sky.
GRS 1915+105's flux appears much fainter, but this fact is relative rather than absolute.
The reasoning for this is three-fold.
First, if we assume our jet models are correct, then Cyg X-1 has a higher injection rate per unit energy than GRS 1915+105 ($3.400\times10^{42} \, e^+ \, \text{s}^{-1} \, \text{MeV}^{-1}$ vs. $2.122\times10^{41} \, e^+ \, \text{s}^{-1} \, \text{MeV}^{-1}$).
Second, GRS 1915+105 is situated much closer to the Galactic plane ($b=-0.219^\circ$) than Cyg X-1, so the former's positrons have to propagate through a more dense ISM environment.
This results in a hotspot of small radius centered on GRS 1915+105 in spite of the smaller injection rate with respect to Cyg X-1.
Third, GRS 1915+105's relative proximity to the disk and lower absolute longitude ($l=45.366^\circ$) means that the flux will be drowned out by a relatively stronger flux from stellar flares than Cyg X-1.
We conclude that we should not expect to see hotspots within the disk unless they have a high positron injection rate, they inhabit a location of the galaxy with strong local ISM conditions, or they reside in a sky location with a relatively weaker disk flux component.

\textbf{GX 13+1:}
GX 13+1 (or 4U 1811-17) is a persistently bright low-mass X-ray binary with a neutron star and likely a K5 III star \citep{bandyopadhyay_GX_13+1_2002}.
Although initially classified as an atoll source, observations by the Rossi X-ray Timing Explorer (RXTE) during its lifetime have shown that GX 13+1 also exhibits properties shared by Z-type sources \citep{hasinger_GX_13+1_1989, fridriksson_NSLMXBs_2015}.
The X-ray binary GX 13+1 is located close to us ($5.600\,\text{kpc}$) in a region where the local GMF strength ($8.195\,\mu\text{G}$) and radiation energy density ($3.008\,\text{eV}\,\text{cm}^{-3}$) are relatively high, so the injected positrons cool down and annihilate relatively quicker.
Similar to GRS 1915+105, GX 13+1 is close in sky location to the Galactic Center ($l=13.516^\circ,b=0.108^\circ$), so the flux from stellar flares makes the detection of the source in 0.511~MeV $\gamma$ rays difficult.

\textbf{4U 1642-45:}
The low-mass X-ray binary 4U 1642-45 (also known as GX 340+0 or SWIFT J1645.7-4538) has not been widely studied in the literature, but the XRBcats catalogue cites that it is likely a Z-type source \citep{avakyan_xrbcats_2023}.
As stated before, 4U 1642-45 is the dominant contributor out of five nearby X-ray binaries contributing to the rightmost hotspot in Figure~\ref{fig:XRB_pt_srcs} (left).

\textbf{MAXI J1631-479 and 4U 1630-47:} MAXI J1631-479 is an X-ray transient with a black hole candidate \citep{miyasaka_MAXI_2018, xu_MAXI_2020}.
4U 1630-47 is also an X-ray transient with an accreting black hole, but exhibits a short outburst recurrence rate of about 2--3 years \citep{kuulkers_4U_1630-47_1998}.
Based on our analysis of each source and their contributed fluxes, the hotspot originates primarily from both MAXI J1631-479 and 4U 1630-47.
MAXI J1631-479 and 4U 1630-47 contribute fluxes of similar magnitude to their associated hotspot, although MAXI J1631-479 has a more important role (see Table~\ref{tab:XRB_pt_srcs}).
The orange curve in Figure~\ref{fig:XRB_pt_srcs} (d) disables all nearby sources, including 4U 1630-47, except for MAXI J1631-479 and 4U 1642-45.
While the hotspot still remains, its peak is slightly smaller when removing the contribution from 4U 1630-47.
MAXI J1631-479 injects $1.423\times10^{41}\,e^{+}\,\text{s}^{-1}\,\text{MeV}^{-1}$ from $4.5\,\text{kpc}$ away.
While 4U 1630-47 is located further away at $8.1\,\text{kpc}$, the local GMF strength is relatively strong, at $79.426\,\mu\text{G}$, so that the high-energy positrons drop to lower energies very fast.

\subsection{Effects of Varying Model Parameters on the Flux and Continuum}

%
%
Finally, we discuss the effect of altering model parameters on the predicted 0.511 MeV flux, its spatial morphology, and the gamma-ray continuum.
To start, our default duty cycle values apply to all X-ray binaries which do not have measured duty cycles in the literature, meaning any large change in the assumed default values may change our results.
For example, if the default persistent jet duty cycle $\varepsilon$ is lower; e.g., on the order of $1\%$ instead of $10\%$; then these X-ray binaries may not generate discernable 0.511 MeV hotspots.
In fact, low-mass X-ray binaries could exhibit persistent duty cycles as low as $\mathcal{O}(0.1\%)$ \citep{tetarenko_watchdog_2016}.
The $e^+$ yield of outbursting X-ray binaries can be extracted from measurements of jet luminosity, and from the accretion luminosity by association, then assuming the empirical scaling relation as outlined in \citet{fender_jet_model_2004}.
While we use values of $A_{\text{BH}}=0.1$ and $A_{\text{NS}}=0.3A_{\text{BH}}$ (see Equation (\ref{eq:XRB_empirical})), these proportionality factors could assume values from about $8\times10^{-3}$ to 0.3 \citep{fender_2003_jets}.

Another source of uncertainty comes from the observationally and theoretically poorly constrained positron injection spectrum.
While a power law and exponential cutoff describes particles accelerated through Fermi acceleration well, this prescription simplifies the rich microphysics of the jet acceleration region.
As stated in Section~\ref{ssec:XRB_511_methods}, the spectra of particles energized by stochastic acceleration differs depending on the type of turbulence.
This information is carried by the parameter $a$, and is related to the spectral index of turbulence by the relation $a=3-q$.
A Kolmogorov-type turbulent spectrum (spectral index $q=5/3$) corresponds to $a = 4/3$ in Equation (\ref{eq:source_func_persistent}).
Kolmogorov turbulence results in a smoother cutoff for persistent jets.
In the context of the composite source function, a change in the turbulent spectral index would only have a major effect on the energy-differential positron rate around $E_{\text{p,max}}$.
We therefore do not expect a significant change in the total flux and morphology by tweaking the type of jet turbulence, especially since our focus is on obtaining results at MeV energies rather than GeV and TeV energies.
In addition, transient jets have been observed expelling material with a range of bulk Lorentz factors up to $\Gamma_{\text{jet}}\sim10$.
Calculations using the Maxwell-J\"uttner distribution function show that jets lead to $p=2$ and $p\rightarrow2.2$ in the 
non-relativistic and ultra-relativistic limits, respectively \citep{keshet_shocks_2005}.
This implies that $p$ and $\Gamma_{\text{jet}}$ are coupled until the bulk Lorentz factor reaches the ultra-relativistic limit.
We expect steeper $e^+$/$e^-$ bremsstrahlung and in-flight annihilation components for highly relativistic jets since these components trace the $e^+$ injection spectrum, but it is unclear how the 0.511 MeV flux and its morphology would differ.
For transient jets, we could increase $E_{\text{t,min}}$ and explore these questions ourselves, but the Green's function approximation would begin to break down if a non-negligible fraction of positrons begin to propagate far from their sources.
Therefore it is unclear to us if the hotspots described earlier would still be 0.511 MeV hotspots under these new parameters.

The predicted stellar flare flux depends on the stellar mass distribution in our Galaxy, class-dependent FFD parameters, and $e^+$ yield per bolometric energy.
For the latter, a parameter like $\eta$ approximates the physics of $e^+$ production, $e^+$ interactions with particles in the stellar corona, and $e^+$ annihilation well, but sweeps this physics under the rug.
The stellar flare flux strongly impacts the detectability of X-ray binary hotspots.

\section{Conclusions} \label{sec:conclusion}

We have calculated the 0.511 MeV flux and gamma-ray continuum from positrons injected by X-ray binaries in our Galaxy, where our sample of X-ray binaries comes from the XRBcats catalogue \citep{avakyan_xrbcats_2023, neumann_xrbcats_2023}.
%
%
%
%
We summarize the key assumptions entering our positron production, propagation, and annihilation calculations in Table~\ref{tab:assumptions_SFs} (stellar flare properties), Table~\ref{tab:assumptions_XRBs} (X-ray binary properties) and Table~\ref{tab:assumptions_ISM} (the Milky way properties).
Our results indicate that X-ray binaries may be an important source of the Galactic $e^+$ annihilation emission.
Stellar flares generally outshine the X-ray binary sources. The brightest X-ray binaries may be detectable with future instruments
(see Table~\ref{tab:XRB_pt_srcs}).
Finding signatures of electron-positron annihilation in the direction of these hotspots could be useful in constraining the contribution of X-ray binaries to the Galactic Center 0.511 MeV excess.
A deep observation of the hotspot at $(71.1^\circ,2.8^\circ)$, near Cyg X-1, can test the hypothesis that X-ray binaries are significant Galactic positron sources.
Yet, we caution that this hotspot could be drowned out by $\beta^+$ decay positrons from $^{26}\text{Al}$ due to OB associations in the Cygnus region.
The upcoming COSI mission with a projected angular resolution of $4.1^\circ$ at $0.511\,\text{MeV}$ \citep{tomsick_cosi_2024} might detect the $\sim7.8^\circ$--diameter hotspot.
We also analyzed hotspots with diameters between $0.5^{\prime}$ and $31.4^{\prime}$ diameters.
It would thus be best to combine wide-field-of-view instruments like the upcoming COSI mission with narrow-field-of-view instruments like the proposed 511-CAM mission \citep{2023JATIS...9b4006S} to search for the hotspots.
In terms of the $e^+$ annihilation emission, we expect COSI to be sensitive to the continuum from $\sim0.3\,\text{MeV}$ up to $0.511\,\text{MeV}$, given the assumptions of the reported COSI continuum sensitivity curves.

%
%
General Relativistic MHD (GRMHD), resistive GRMHD, and General Relativistic Particle-in-Cell (GRPIC) simulations are now starting to give us new insights into the production of pairs close to compact objects. 
GRMHD simulations of jets 
\citep{dieckmann_jet_2019, pohl_PIC_2020, nishikawa_PIC_2021}
are still missing pair creation physics.
Whereas GRPIC simulations capture kinetic effects that GRMHD simulations miss, they are limited to small scales due to their computational demand. Furthermore, GRMHD and GRPIC simulations suffer from a large separation of their scales that is hard to reconcile.  Pair processes have historically been a difficult issue to tackle in PIC simulations.
For many GRPIC simulations of  magnetized plasmas, electron-positron pairs are artificially introduced as a floor for the calculations.  GRPIC simulations also often mimic electron-positron pair cascades with a magnetization threshold.
At the frontier of astrophysical plasma simulation is the use of GRPIC in the regime of quantum electrodynamics (QED).
In relation to the Galactic 0.511 MeV excess, GRPIC simulations have begun to incorporate $\gamma+\gamma\rightarrow e^+ + e^-$ \citep{hakobyan_PIC_2019}, $\gamma+B\rightarrow e^+ + e^-$ \citep{schoeffler_PIC_2019}, and $e^+ + e^-\rightarrow\gamma+\gamma$ \citep{nattila_turbulence_2024} processes, among other QED processes, in a self-consistent manner.
As soon as GRMHD and GRPIC methods develop enough, it can be useful to simulate the production of positrons in the inner accretion flow and their ejection through jets to make use of the common origin of the two components. Such studies could furthermore constrain the pair-to-proton ratio of the jet particle content, which is one of the leading uncertainties when considering X-ray binaries as Galactic positron sources.

%
%
\begin{figure*}
\plotone{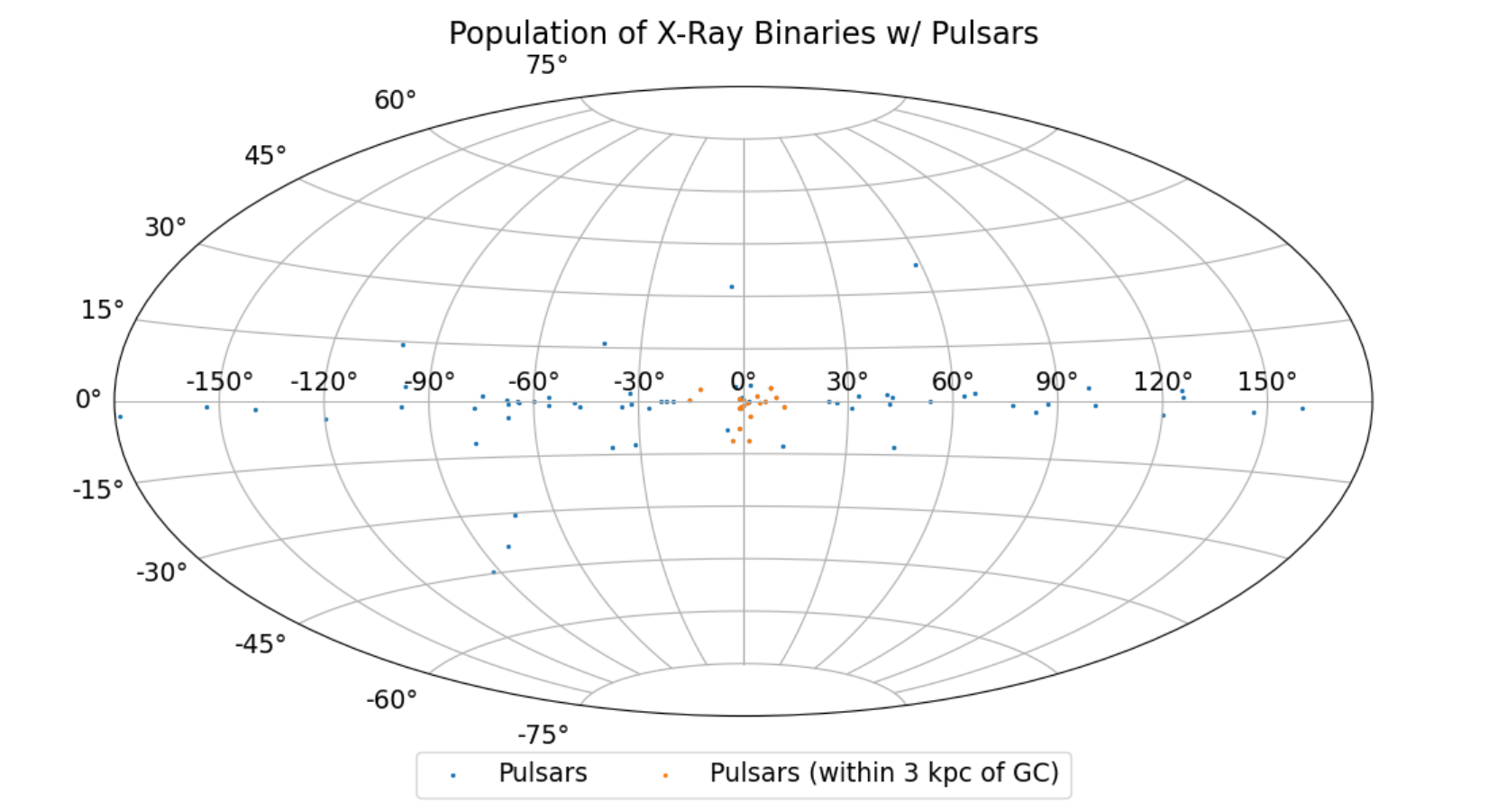}
\caption{
The population of X-ray binaries with pulsars reported with well-constrained sky locations $(l,b)$ and distances in the XRBcats catalogue \citep{avakyan_xrbcats_2023, neumann_xrbcats_2023}.
The pulsars located within the inner $3\,\text{kpc}$ of the galaxy are shown in orange.
}
\label{fig:pulsar_locations}
\end{figure*}
As stated before, using Green's functions are a good approximation for estimating the flux from X-ray binaries as long as the positrons annihilate close to their injection sites.
We are confident that the hotspots which appear in our calculations should still be hotspots if $e^+$ propagation was simulated for X-ray binaries.
Low energy positrons should still annihilate close to their injection sites.
Instead of isotropic annihilation, we expect the $e^+$ annihilation sites to lie preferentially within the disk because the GMF within the disk has been found to align itself with the Galactic plane \citep{jansson_GMF_2012, han_GMF_2017, korochkin_GMF_2025}.
Since the number of annihilation sites should be conserved between the Green's function and transport methods, we predict that the hotspots identified in this work will be easier to identify.
Unfortunately, the Green's function approximation is incompatible with predicting the 0.511 MeV flux from pulsars because their positrons are injected at high energies.
It is possible that including these pulsars might bump up the total contribution of X-ray binaries to the 0.511 MeV flux a notable amount.

%
%
To break free of this restriction would mean carrying out transport simulations of positrons injected by X-ray binaries.
Such simulations would center on the particle transfer equation (Equation (\ref{eq:one_transfer_eq})) with position-dependent energy losses, diffusion, and mean positron lifetimes using Monte Carlo methods and weighted test particles (see \citet{jean_positron_2009, alexis_MC_2014}).
Future work modeling the injection and propagation of positrons from X-ray binaries (including those with pulsars), would be very useful in favoring or ruling out candidate Galactic positron sources.
However, it is important to mention that determining the GMF magnitude and direction in our galaxy is still an open research question.
Even then, the propagation of positrons through the ISM is not well understood.
A propagation study for all X-ray binaries could test our hypothesis about our aforementioned hotspots being easier to identify when accounting for the GMF directional information.
A numerical code like GALPROP would be well suited to estimate the propagation of high energy positrons from X-ray binaries, including pulsars \citep{strong_GALPROP_1998}.
Indeed, GALPROP has been used in the past to study the propagation of positrons produced by $\beta^+$ decay of radioactive ${}^{26}\text{Al}$, ${}^{56}\text{Ni}$, and ${}^{44}\text{Ti}$. \citep{martin_nucleosynthesis_2012}.
Regardless, this direction of work would also accurately identify the contribution of pulsars to the Galactic 0.511 MeV excess and its morphology.
There are $\sim 20$ X-ray binaries with pulsars definitively located within the inner $3\,\text{kpc}$ of the galaxy (see Figure~\ref{fig:pulsar_locations}) and more elsewhere, although there may be even more pulsars we have not detected yet.
While we believe the GMF near the Galactic Center is strong (see Figure~\ref{fig:GMF}, right), pulsars may inject positrons with high enough energies so that they disperse throughout the bulge to produce a strong bulge-centered flux component.
These positrons may also escape confinement in the Galactic Center and potentially escape the galaxy.

\section*{Acknowledgements}

This work was supported by funding from NASA grants GR0044377-80NSSC26K0791 and GR0026987-80NSSC22K1883, the McDonnell Center for Space Sciences, and the National Science Foundation grant 2152221.
The authors would like to thank the anonymous referee for constructive, insightful feedback on the paper during the review process.
We are also grateful to Jim Buckley and Kun Hu for useful discussions, and John Tomsick for providing the COSI continuum sensitivity data.

\startlongtable
\begin{deluxetable*}{l|c|c|c}
\tablecaption{Assumptions made for our models of the stellar flare flux.
\label{tab:assumptions_SFs}}
\tablehead{
\colhead{Parameters (Stellar Flares)} & \colhead{Units} & \colhead{Values} & \colhead{References}
}
\startdata
$\eta$ & $e^+\ \text{erg}^{-1}$ & $120$ & \citet{bisnovatyi-kogan_SFs_2017} \\
\hline
$f_{\text{C}}\dot{E}_{\text{C}}$ & $\text{erg}\ \text{s}^{-1}$ & $\cdots$ & \citet{chabrier_galactic_2003} \\
{     } & {     } & {     } & \citet{yang_flare_2019} \\
{     } & {     } & {     } & See Table~\ref{tab:SF_FFD_table} \\
\hline
$n_{\text{0,B}}$ & $M_{\odot}\ \text{kpc}^{-3}$ & $(9.84 \pm 0.10)\times10^{10}$ & \citet{mcmillan_mass_2017} \\
\hline
$\rho_{0}$ & $\text{kpc}$ & $0.075$ & \vdots \\
\hline
$\beta$ & $\text{no units}$ & $1.8$ & \vdots \\
\hline
$\rho_{\text{cut}}$ & $\text{kpc}$ & $2.1$ & \vdots \\
\hline
$q$ & $\text{no units}$ & $0.5$ & \vdots \\
\hline
$\Sigma_0$ & $M_{\odot}\ \text{kpc}^{-2}$ & $8.95\times10^8$ & \vdots \\
\hline
$z_{\text{D,thin}}$ & $\text{kpc}$ & $0.3$ & \vdots \\
\hline
$\rho_{\text{D,thin}}$ & $\text{kpc}$ & $2.6\pm 0.5$ & \vdots \\
\hline
$z_{\text{D,thick}}$ & $\text{kpc}$ & $0.9$ & \vdots \\
\hline
$\rho_{\text{D,thick}}$ & $\text{kpc}$ & $3.6\pm 0.7$ & \vdots \\
\enddata
\end{deluxetable*}

\startlongtable
\begin{deluxetable*}{l|c|c|c}
\tablecaption{Assumptions made for our models of the X-ray binary flux.
\label{tab:assumptions_XRBs}}
\tablehead{
\colhead{Parameters (X-Ray Binaries)} & \colhead{Units} & \colhead{Values} & \colhead{References}
}
\startdata
Number of X-ray binaries & $\text{no units}$ & $362$ & \citet{avakyan_xrbcats_2023} \\
{     } & {     } & {     } & \citet{neumann_xrbcats_2023} \\
\hline
$D, l, b$ for all X-ray binaries & $\text{kpc},\text{deg},\text{deg}$ & $\cdots$ & \vdots \\
\hline
$A_{\text{BH}}$ (persistent) & $\text{no units}$ & $0.1$ & \citet{bartels_galactic_2018} \\
\hline
$A_{\text{BH}}$ (transient) & $\text{no units}$ & $0.7$ & \citet{fender_2003_jets} \\
\hline
$A_{\text{NS}}$ & $\text{no units}$ & $0.3A_{\text{BH}}$ & \citet{bartels_galactic_2018} \\
\hline
$L_{\text{jet}}$\tablenotemark{a} & $\text{erg}\ \text{s}^{-1}$ & $\cdots$ & \citet{avakyan_xrbcats_2023} \\
{     } & {     } & {     } & \citet{neumann_xrbcats_2023} \\
\hline
$\langle \gamma \rangle$ & $\text{no units}$ & $1$ & \citet{reynolds_jet_1996} \\
\hline
$\Gamma_{\text{jet}}$ (persistent) & $\text{no units}$ & $1.4$ & \citet{fender_2003_jets} \\
\hline
$\Gamma_{\text{jet}}$ (transient) & $\text{no units}$ & $2$ & \citet{fender_2003_jets} \\
\hline
$p_{\text{t}}$ & $\text{no units}$ & $2.2$ & \citet{sironi_2015_shocks} \\
\hline
$E_{\text{t,max}}$ & $\text{MeV}$ & $10^7$ & \citet{aleksic_TeV_2011} \\
{     } & {     } & {     } & \citet{massi_TeV_2009} \\
{     } & {     } & {     } & \citet{sguera_TeV_2009} \\
{     } & {     } & {     } & \citet{sudoh_TeV_2020} \\
\hline
$p_{\text{p}}$ & $\text{no units}$ & $2$ & \citet{sironi_2015_shocks} \\
\hline
$E_{\text{p,max}}$ & $\text{MeV}$ & $10^4$ & \citet{albert_GeV_2007} \\
{     } & {     } & {     } & \citet{zdziarski_GeV_2017} \\
\hline
$a$ & $\text{no units}$ & $2$ & \citet{becker_stochastic_2006} \\
{     } & {     } & {     } & \citet{stawarz_stochastic_2008} \\
{     } & {     } & {     } & \citet{sudoh_TeV_2020} \\
\hline
$\varepsilon_{\text{t},i}, \varepsilon_{\text{p},i}$ (LMXRBs) & $\text{no units}$ & $0.01\varepsilon_{\text{p},i}, 0.1$ & \citet{belloni_states_2010} \\
{     } & {     } & {     } & \citet{tetarenko_watchdog_2016} \\
\hline
$\varepsilon_{\text{t},i}, \varepsilon_{\text{p},i}$ (HMXRBs)\tablenotemark{b} & $\text{no units}$ & $0, 0.5$ & \citet{sidoli_DCs_2018} \\
\hline
$\varepsilon_{\text{t},i}, \varepsilon_{\text{p},i}$ (Z-type sources)\tablenotemark{b} & $\text{no units}$ & $0.01, 0.5$ & \citet{fender_persistent_jets_2000} \\
{     } & {     } & {     } & \citet{migliari_jets_2006} \\
{     } & {     } & {     } & \citet{van_den_eijnden_NSXRB_2021} \\
\hline
$\varepsilon_{\text{t},i}, \varepsilon_{\text{p},i}$ (atoll sources)\tablenotemark{b} & $\text{no units}$ & $0.001, 0.05$ & \vdots \\
\hline
$\varepsilon_{\text{p},i}$ (sgHMXRBs)\tablenotemark{b} & $\text{no units}$ & $0.3$ & \citet{sidoli_DCs_2018} \\
\hline
$\varepsilon_{\text{t},i}$ (SFXTs)\tablenotemark{b} & $\text{no units}$ & $0.01$ & \vdots \\
\hline
$\varepsilon_{\text{t},i}$ (BeXRBs)\tablenotemark{b} & $\text{no units}$ & $0.1$ & \vdots \\
\enddata
\tablenotetext{a}{Calculated from the soft X-ray flux and $D$ in \citet{avakyan_xrbcats_2023, neumann_xrbcats_2023}.}
\tablenotetext{b}{Estimated by averaging over duty cycle measurements.}
\end{deluxetable*}

\startlongtable
\begin{deluxetable*}{l|c|c|c}
\tablecaption{Assumptions made for the ISM properties; including models of the ISM densities, GMF, and ISM radiation fields (CMB plus IR, optical, and UV bands).
\label{tab:assumptions_ISM}}
\tablehead{
\colhead{Parameters (Interstellar Medium)} & \colhead{Units} & \colhead{Values} & \colhead{References}
}
\startdata
Positronium Fraction & $\text{no units}$ & $0.97\pm0.02$ & \citet{jean_spectral_2006} \\
\hline
ISM H densities & $\text{cm}^{-3}$ & $\cdots$ & \citet{ferriere_global_1998} \\
\hline
He-to-H ratio & $\text{no units}$ & $0.09$ & \citet{ferriere_global_1998} \\
\hline
GMF (excluding Galactic Center) & $\mu\text{G}$ & $\cdots$ &  \citet{ferriere_GMF_2014} \\
{     } & {     } & {     } & \citet{jansson_GMF_2012} \\
\hline
GMF (Galactic Center region) & $\mu\text{G}$ & $\cdots$ & \citet{guenduez_GMF_2020} \\
\hline
ISM Radiation Energy Density & $\text{eV}\ \text{cm}^{-3}$ & $\cdots$ & \citet{porter_CRs_2008} \\
(CMB, IR, optical, UV) & {     } & {     } & {     } \\
\hline
Optical Depth to 0.511 MeV + Continuum & $\text{no units}$ & $\approx0$ & \citet{hi4pi_collaboration_hi4pi_2016} \\
\enddata
\end{deluxetable*}

\bibliography{511_paper}

@ARTICLE{2023JATIS...9b4006S,
       author = {{Shirazi}, Farzane and {Gau}, Ephraim and {Hossen}, Md. Arman and {Becker}, Daniel and {Schmidt}, Daniel and {Swetz}, Daniel and {Bennett}, Douglas and {Braun}, Dana and {Kislat}, Fabian and {Gard}, Johnathon and {Mates}, John and {Weber}, Joel and {Rodriguez Cavero}, Nicole and {Chun}, Sohee and {Lisalda}, Lindsey and {West}, Andrew and {Dev}, Bhupal and {Ferrer}, Francesc and {Bose}, Richard and {Ullom}, Joel and {Krawczynski}, Henric},
        title = "{511-CAM mission: a pointed 511 keV gamma-ray telescope with a focal plane detector made of stacked transition edge sensor microcalorimeter arrays}",
      journal = {Journal of Astronomical Telescopes, Instruments, and Systems},
         year = 2023,
        month = apr,
       volume = {9},
          eid = {024006},
        pages = {024006},
          doi = {10.1117/1.JATIS.9.2.024006},
archivePrefix = {arXiv},
       eprint = {2206.14652},
 primaryClass = {astro-ph.IM},
       adsurl = {https://ui.adsabs.harvard.edu/abs/2023JATIS...9b4006S}
}

@article{siegert_positron_2023,
	title = {The {Positron} {Puzzle}},
	volume = {368},
	issn = {1572-946X},
	url = {https://doi.org/10.1007/s10509-023-04184-4},
	doi = {10.1007/s10509-023-04184-4},
	language = {en},
	number = {4},
	urldate = {2025-04-17},
	journal = {Astrophys Space Sci},
	author = {Siegert, Thomas},
	month = apr,
	year = {2023},
	pages = {27},
}

@article{prantzos_511_2011,
	title = {The 511 {keV} emission from positron annihilation in the {Galaxy}},
	volume = {83},
	url = {https://link.aps.org/doi/10.1103/RevModPhys.83.1001},
	doi = {10.1103/RevModPhys.83.1001},
	number = {3},
	urldate = {2025-04-17},
	journal = {Rev. Mod. Phys.},
	author = {Prantzos, N. and Boehm, C. and Bykov, A. M. and Diehl, R. and Ferrière, K. and Guessoum, N. and Jean, P. and Knoedlseder, J. and Marcowith, A. and Moskalenko, I. V. and Strong, A. and Weidenspointner, G.},
	month = sep,
	year = {2011},
	note = {Publisher: American Physical Society},
	pages = {1001--1056},
}

@article{murphy_high-energy_1987,
	title = {High-{Energy} {Processes} in {Solar} {Flares}},
	volume = {63},
	issn = {0067-0049},
	url = {https://ui.adsabs.harvard.edu/abs/1987ApJS...63..721M},
	doi = {10.1086/191180},
	urldate = {2025-05-14},
	journal = {The Astrophysical Journal Supplement Series},
	author = {Murphy, R. J. and Dermer, C. D. and Ramaty, R.},
	month = mar,
	year = {1987},
	note = {Publisher: IOP
ADS Bibcode: 1987ApJS...63..721M},
	pages = {721},
}

@article{kowalski_stellar_2024,
	title = {Stellar flares},
	volume = {21},
	url = {https://ui.adsabs.harvard.edu/abs/2024LRSP...21....1K},
	doi = {10.1007/s41116-024-00039-4},
	urldate = {2025-05-14},
	journal = {Living Reviews in Solar Physics},
	author = {Kowalski, Adam F.},
	month = dec,
	year = {2024},
	note = {ADS Bibcode: 2024LRSP...21....1K},
	pages = {1},
}

@ARTICLE{bisnovatyi-kogan_SFs_2017,
    author = {{Bisnovatyi-Kogan}, G.~S. and {Pozanenko}, A.~S.},
    title = "{Can Flare Stars Explain the Annihilation Line from the Galactic Bulge?}",
    journal = {Astrophysics},
    year = 2017,
    month = jun,
    volume = {60},
    number = {2},
    pages = {223-227},
    doi = {10.1007/s10511-017-9477-6},
    adsurl = {https://ui.adsabs.harvard.edu/abs/2017Ap.....60..223B}
}

@article{mcmillan_mass_2017,
	title = {The mass distribution and gravitational potential of the {Milky} {Way}},
	volume = {465},
	issn = {0035-8711},
	url = {https://ui.adsabs.harvard.edu/abs/2017MNRAS.465...76M},
	doi = {10.1093/mnras/stw2759},
	urldate = {2025-05-15},
	journal = {Monthly Notices of the Royal Astronomical Society},
	author = {McMillan, Paul J.},
	month = feb,
	year = {2017},
	note = {Publisher: OUP
ADS Bibcode: 2017MNRAS.465...76M},
	pages = {76--94},
}

@article{weidenspointner_asymmetric_2008,
	title = {An asymmetric distribution of positrons in the {Galactic} disk revealed by γ-rays},
	volume = {451},
	copyright = {2008 Springer Nature Limited},
	issn = {1476-4687},
	url = {https://www.nature.com/articles/nature06490},
	doi = {10.1038/nature06490},
	language = {en},
	number = {7175},
	urldate = {2025-06-15},
	journal = {Nature},
	author = {Weidenspointner, Georg and Skinner, Gerry and Jean, Pierre and Knödlseder, Jürgen and von Ballmoos, Peter and Bignami, Giovanni and Diehl, Roland and Strong, Andrew W. and Cordier, Bertrand and Schanne, Stéphane and Winkler, Christoph},
	month = jan,
	year = {2008},
	note = {Publisher: Nature Publishing Group},
	pages = {159--162},
}

@article{avakyan_xrbcats_2023,
	title = {{XRBcats}: {Galactic} low-mass {X}-ray binary catalogue},
	volume = {675},
	issn = {0004-6361},
	shorttitle = {{XRBcats}},
	url = {https://ui.adsabs.harvard.edu/abs/2023A&A...675A.199A},
	doi = {10.1051/0004-6361/202346522},
	urldate = {2025-06-18},
	journal = {Astronomy and Astrophysics},
	author = {Avakyan, A. and Neumann, M. and Zainab, A. and Doroshenko, V. and Wilms, J. and Santangelo, A.},
	month = jul,
	year = {2023},
	note = {Publisher: EDP
ADS Bibcode: 2023A\&A...675A.199A},
	pages = {A199},
}

@article{neumann_xrbcats_2023,
	title = {{XRBcats}: {Galactic} {High} {Mass} {X}-ray {Binary} {Catalogue}★},
	volume = {677},
	issn = {0004-6361},
	shorttitle = {{XRBcats}},
	url = {https://ui.adsabs.harvard.edu/abs/2023A&A...677A.134N},
	doi = {10.1051/0004-6361/202245728},
	urldate = {2025-06-18},
	journal = {Astronomy and Astrophysics},
	author = {Neumann, M. and Avakyan, A. and Doroshenko, V. and Santangelo, A.},
	month = sep,
	year = {2023},
	note = {Publisher: EDP
ADS Bibcode: 2023A\&A...677A.134N},
	pages = {A134},
}

@article{jean_positron_2009,
	title = {Positron transport in the interstellar medium},
	volume = {508},
	issn = {0004-6361},
	url = {https://ui.adsabs.harvard.edu/abs/2009A&A...508.1099J},
	doi = {10.1051/0004-6361/200809830},
	urldate = {2025-06-19},
	journal = {Astronomy and Astrophysics},
	author = {Jean, P. and Gillard, W. and Marcowith, A. and Ferrière, K.},
	month = dec,
	year = {2009},
	note = {ADS Bibcode: 2009A\&A...508.1099J},
	pages = {1099--1116},
}

@article{bartels_galactic_2018,
	title = {Galactic binaries can explain the {Fermi} {Galactic} centre excess and 511 {keV} emission},
	volume = {480},
	issn = {0035-8711},
	url = {https://ui.adsabs.harvard.edu/abs/2018MNRAS.480.3826B},
	doi = {10.1093/mnras/sty2135},
	urldate = {2025-06-19},
	journal = {Monthly Notices of the Royal Astronomical Society},
	author = {Bartels, R. and Calore, F. and Storm, E. and Weniger, C.},
	month = nov,
	year = {2018},
	note = {Publisher: OUP
ADS Bibcode: 2018MNRAS.480.3826B},
	pages = {3826--3841},
}

@article{jean_spectral_2006,
	title = {Spectral analysis of the {Galactic} e+e- annihilation emission},
	volume = {445},
	copyright = {© ESO, 2005},
	issn = {0004-6361, 1432-0746},
	url = {https://www.aanda.org/articles/aa/abs/2006/02/aa3765-05/aa3765-05.html},
	doi = {10.1051/0004-6361:20053765},
	language = {en},
	number = {2},
	urldate = {2025-06-20},
	journal = {A\&A},
	author = {Jean, P. and Knödlseder, J. and Gillard, W. and Guessoum, N. and Ferrière, K. and Marcowith, A. and Lonjou, V. and Roques, J. P.},
	month = jan,
	year = {2006},
	note = {Number: 2
Publisher: EDP Sciences},
	pages = {579--589},
}

@article{ferriere_global_1998,
	title = {Global {Model} of the {Interstellar} {Medium} in {Our} {Galaxy} with {New} {Constraints} on the {Hot} {Gas} {Component}},
	volume = {497},
	issn = {0004-637X},
	url = {https://ui.adsabs.harvard.edu/abs/1998ApJ...497..759F},
	doi = {10.1086/305469},
	urldate = {2025-07-13},
	journal = {The Astrophysical Journal},
	author = {Ferrière, Katia},
	month = apr,
	year = {1998},
	note = {Publisher: IOP
ADS Bibcode: 1998ApJ...497..759F},
	pages = {759--776},
}

@article{johnson_spectrum_1972,
	title = {The {Spectrum} of {Low}-{Energy} {Gamma} {Radiation} from the {Galactic}-{Center} {Region}.},
	volume = {172},
	issn = {0004-637X},
	url = {https://ui.adsabs.harvard.edu/abs/1972ApJ...172L...1J},
	doi = {10.1086/180878},
	urldate = {2025-10-20},
	journal = {The Astrophysical Journal},
	author = {Johnson, III, W. N. and Harnden, Jr., F. R. and Haymes, R. C.},
	month = feb,
	year = {1972},
	note = {Publisher: IOP
ADS Bibcode: 1972ApJ...172L...1J},
	pages = {L1},
}

@article{leventhal_detection_1978,
	title = {Detection of 511 {keV} positron annihilation radiation from the galactic center direction.},
	volume = {225},
	issn = {0004-637X},
	url = {https://ui.adsabs.harvard.edu/abs/1978ApJ...225L..11L},
	doi = {10.1086/182782},
	urldate = {2025-10-20},
	journal = {The Astrophysical Journal},
	author = {Leventhal, M. and MacCallum, C. J. and Stang, P. D.},
	month = oct,
	year = {1978},
	note = {Publisher: IOP
ADS Bibcode: 1978ApJ...225L..11L},
	pages = {L11--L14},
}

@article{haymes_observation_1969,
	title = {Observation of {Hard} {Radiation} from the {Region} of the {Galactic} {Center}},
	volume = {157},
	issn = {0004-637X},
	url = {https://ui.adsabs.harvard.edu/abs/1969ApJ...157.1455H},
	doi = {10.1086/150164},
	urldate = {2025-10-31},
	journal = {The Astrophysical Journal},
	author = {Haymes, R. C. and Ellis, D. V. and Fishman, G. J. and Glenn, S. W. and Kurfess, J. D.},
	month = sep,
	year = {1969},
	note = {Publisher: IOP
ADS Bibcode: 1969ApJ...157.1455H},
	pages = {1455},
}

@article{yang_flare_2019,
	title = {The {Flare} {Catalog} and the {Flare} {Activity} in the {Kepler} {Mission}},
	volume = {241},
	issn = {0067-0049},
	url = {https://ui.adsabs.harvard.edu/abs/2019ApJS..241...29Y},
	doi = {10.3847/1538-4365/ab0d28},
	urldate = {2025-11-06},
	journal = {The Astrophysical Journal Supplement Series},
	author = {Yang, Huiqin and Liu, Jifeng},
	month = apr,
	year = {2019},
	note = {Publisher: IOP
ADS Bibcode: 2019ApJS..241...29Y},
	pages = {29},
}

@article{chabrier_galactic_2003,
	title = {Galactic {Stellar} and {Substellar} {Initial} {Mass} {Function}},
	volume = {115},
	issn = {0004-6280},
	url = {https://ui.adsabs.harvard.edu/abs/2003PASP..115..763C},
	doi = {10.1086/376392},
	urldate = {2025-11-14},
	journal = {Publications of the Astronomical Society of the Pacific},
	author = {Chabrier, Gilles},
	month = jul,
	year = {2003},
	note = {Publisher: IOP
ADS Bibcode: 2003PASP..115..763C},
	pages = {763--795},
}

@article{murphy_radioactive_2014,
	title = {{RADIOACTIVE} {POSITRON} {EMITTER} {PRODUCTION} {BY} {ENERGETIC} {ALPHA} {PARTICLES} {IN} {SOLAR} {FLARES}},
	volume = {215},
	issn = {0067-0049},
	url = {https://doi.org/10.1088/0067-0049/215/2/18},
	doi = {10.1088/0067-0049/215/2/18},
	language = {en},
	number = {2},
	urldate = {2025-11-14},
	journal = {ApJS},
	author = {Murphy, R. J. and Kozlovsky, B. and Share, G. H.},
	month = dec,
	year = {2014},
	note = {Publisher: The American Astronomical Society},
	pages = {18},
}

@article{guessoum_lives_2005,
	title = {The lives and deaths of positrons in the interstellar medium},
	volume = {436},
	copyright = {© ESO, 2005},
	issn = {0004-6361, 1432-0746},
	url = {https://www.aanda.org/articles/aa/abs/2005/22/aa2454-04/aa2454-04.html},
	doi = {10.1051/0004-6361:20042454},
	language = {en},
	number = {1},
	urldate = {2025-04-19},
	journal = {A\&A},
	author = {Guessoum, N. and Jean, P. and Gillard, W.},
	month = jun,
	year = {2005},
	note = {Number: 1
Publisher: EDP Sciences},
	pages = {171--185},
}

@article{hi4pi_collaboration_hi4pi_2016,
	title = {{HI4PI}: {A} full-sky {H} {I} survey based on {EBHIS} and {GASS}},
	volume = {594},
	issn = {0004-6361},
	shorttitle = {{HI4PI}},
	url = {https://ui.adsabs.harvard.edu/abs/2016A&A...594A.116H},
	doi = {10.1051/0004-6361/201629178},
	urldate = {2025-11-15},
	journal = {Astronomy and Astrophysics},
	author = {{HI4PI Collaboration} and Ben Bekhti, N. and Flöer, L. and Keller, R. and Kerp, J. and Lenz, D. and Winkel, B. and Bailin, J. and Calabretta, M. R. and Dedes, L. and Ford, H. A. and Gibson, B. K. and Haud, U. and Janowiecki, S. and Kalberla, P. M. W. and Lockman, F. J. and McClure-Griffiths, N. M. and Murphy, T. and Nakanishi, H. and Pisano, D. J. and Staveley-Smith, L.},
	month = oct,
	year = {2016},
	note = {Publisher: EDP
ADS Bibcode: 2016A\&A...594A.116H},
	pages = {A116},
}

@article{lacy_uv_1976,
	title = {{UV} {Ceti} stars: statistical analysis of observational data.},
	volume = {30},
	issn = {0067-0049},
	shorttitle = {{UV} {Ceti} stars},
	url = {https://ui.adsabs.harvard.edu/abs/1976ApJS...30...85L},
	doi = {10.1086/190358},
	urldate = {2025-11-16},
	journal = {The Astrophysical Journal Supplement Series},
	author = {Lacy, C. H. and Moffett, T. J. and Evans, D. S.},
	month = jan,
	year = {1976},
	note = {Publisher: IOP
ADS Bibcode: 1976ApJS...30...85L},
	pages = {85--96},
}

@article{gershberg_results_1972,
	title = {Some results of the cooperative photometric observations of the {UV} {Cet}-type flare stars in the years 1967–71},
	volume = {19},
	issn = {1572-946X},
	url = {https://doi.org/10.1007/BF00643168},
	doi = {10.1007/BF00643168},
	language = {en},
	number = {1},
	urldate = {2025-11-16},
	journal = {Astrophys Space Sci},
	author = {Gershberg, R. E.},
	month = nov,
	year = {1972},
	pages = {75--92},
}

@article{blandford_electromagnetic_1977,
	title = {Electromagnetic extraction of energy from {Kerr} black holes.},
	volume = {179},
	issn = {0035-8711},
	url = {https://ui.adsabs.harvard.edu/abs/1977MNRAS.179..433B},
	doi = {10.1093/mnras/179.3.433},
	urldate = {2025-11-16},
	journal = {Monthly Notices of the Royal Astronomical Society},
	author = {Blandford, R. D. and Znajek, R. L.},
	month = may,
	year = {1977},
	note = {Publisher: OUP
ADS Bibcode: 1977MNRAS.179..433B},
	pages = {433--456},
}

@article{shibata_hot-plasma_1995,
	title = {Hot-{Plasma} {Ejections} {Associated} with {Compact}-{Loop} {Solar} {Flares}},
	volume = {451},
	issn = {0004-637X},
	url = {https://iopscience.iop.org/article/10.1086/309688/meta},
	doi = {10.1086/309688},
	language = {en},
	number = {2},
	urldate = {2025-11-16},
	journal = {ApJ},
	author = {Shibata, K. and Masuda, S. and Shimojo, M. and Hara, H. and Yokoyama, T. and Tsuneta, S. and Kosugi, T. and Ogawara, Y.},
	month = oct,
	year = {1995},
	note = {Publisher: IOP Publishing},
	pages = {L83},
}

@article{chintzoglou_origin_2019,
	title = {The {Origin} of {Major} {Solar} {Activity}: {Collisional} {Shearing} between {Nonconjugated} {Polarities} of {Multiple} {Bipoles} {Emerging} within {Active} {Regions}},
	volume = {871},
	issn = {0004-637X},
	shorttitle = {The {Origin} of {Major} {Solar} {Activity}},
	url = {https://doi.org/10.3847/1538-4357/aaef30},
	doi = {10.3847/1538-4357/aaef30},
	language = {en},
	number = {1},
	urldate = {2025-11-17},
	journal = {ApJ},
	author = {Chintzoglou, Georgios and Zhang, Jie and Cheung, Mark C. M. and Kazachenko, Maria},
	month = jan,
	year = {2019},
	note = {Publisher: The American Astronomical Society},
	pages = {67},
}

@article{audard_extreme-ultraviolet_2000,
	title = {Extreme-{Ultraviolet} {Flare} {Activity} in {Late}-{TypeStars}},
	volume = {541},
	issn = {0004-637X},
	url = {https://iopscience.iop.org/article/10.1086/309426/meta},
	doi = {10.1086/309426},
	language = {en},
	number = {1},
	urldate = {2025-11-17},
	journal = {ApJ},
	author = {Audard, Marc and Güdel, Manuel and Drake, Jeremy J. and Kashyap, Vinay L.},
	month = sep,
	year = {2000},
	note = {Publisher: IOP Publishing},
	pages = {396},
}

@article{barnes_rotational_2003,
	title = {On the {Rotational} {Evolution} of {Solar}- and {Late}-{Type} {Stars}, {Its} {Magnetic} {Origins}, and the {Possibility} of {Stellar} {Gyrochronology}*},
	volume = {586},
	issn = {0004-637X},
	url = {https://iopscience.iop.org/article/10.1086/367639/meta},
	doi = {10.1086/367639},
	language = {en},
	number = {1},
	urldate = {2025-11-17},
	journal = {ApJ},
	author = {Barnes, Sydney A.},
	month = mar,
	year = {2003},
	note = {Publisher: IOP Publishing},
	pages = {464},
}

@article{share_high-resolution_2003,
	title = {High-{Resolution} {Observation} of the {Solar} {Positron}-{Electron} {Annihilation} {Line}},
	volume = {595},
	issn = {0004-637X},
	url = {https://ui.adsabs.harvard.edu/abs/2003ApJ...595L..85S},
	doi = {10.1086/378174},
	urldate = {2025-11-17},
	journal = {The Astrophysical Journal},
	author = {Share, Gerald H. and Murphy, Ronald J. and Skibo, Jeffrey G. and Smith, David M. and Hudson, Hugh S. and Lin, Robert P. and Shih, Albert Y. and Dennis, Brian R. and Schwartz, Richard A. and Kozlovsky, Benzion},
	month = oct,
	year = {2003},
	note = {Publisher: IOP
ADS Bibcode: 2003ApJ...595L..85S},
	pages = {L85--L88},
}

@INPROCEEDINGS{mittal_flares_2024,
       author = {{Mittal}, Saurabh and {Siegert}, Thomas},
        title = "{Stellar Flares to Explain the Galactic 511 keV Emission}",
    booktitle = {AAS High Energy Astrophysics Division Meeting \#21},
         year = 2024,
       series = {AAS/High Energy Astrophysics Division},
       volume = {21},
        month = may,
          eid = {104.01},
        pages = {104.01},
       adsurl = {https://ui.adsabs.harvard.edu/abs/2024HEAD...2110401M}
}

@BOOK{ginzburg_cosmic_rays_1964,
    author = {{Ginzburg}, V.~L. and {Syrovatskii}, S.~I.},
    title = "{The Origin of Cosmic Rays}",
    year={1964},
    publisher={Pergamon Press},
    address={Oxford},
    adsurl = {https://ui.adsabs.harvard.edu/abs/1964ocr..book.....G}
}

@ARTICLE{romero_jet_2017,
    author = {{Romero}, Gustavo E. and {Boettcher}, M. and {Markoff}, S. and {Tavecchio}, F.},
    title = "{Relativistic Jets in Active Galactic Nuclei and Microquasars}",
    journal = {\ssr},
    year = 2017,
    month = jul,
    volume = {207},
    number = {1-4},
    pages = {5-61},
    doi = {10.1007/s11214-016-0328-2},
    archivePrefix = {arXiv},
    eprint = {1611.09507},
    primaryClass = {astro-ph.HE},
    adsurl = {https://ui.adsabs.harvard.edu/abs/2017SSRv..207....5R}
}

@INPROCEEDINGS{coppi_hybrid_plasma_1999,
    author = {{Coppi}, P.~S.},
    title = "{The Physics of Hybrid Thermal/Non-Thermal Plasmas}",
    booktitle = {High Energy Processes in Accreting Black Holes},
    year = 1999,
    editor = {{Poutanen}, Juri and {Svensson}, Roland},
    series = {Astronomical Society of the Pacific Conference Series},
    volume = {161},
    month = jan,
    pages = {375},
    doi = {10.48550/arXiv.astro-ph/9903158},
    archivePrefix = {arXiv},
    eprint = {astro-ph/9903158},
    primaryClass = {astro-ph},
    adsurl = {https://ui.adsabs.harvard.edu/abs/1999ASPC..161..375C}
}

@INPROCEEDINGS{tomsick_cosi_2024,
    author = {{Tomsick}, J. and {Boggs}, S. and {Zoglauer}, A. and {Hartmann}, D.~H. and {Ajello}, M. and {Burns}, E. and {Fryer}, C. and {Karwin}, C. and {Kierans}, C. and {Lowell}, A. and {Malzac}, J. and {Roberts}, J. and {Saint-Hilaire}, P. and {Shih}, A. and {Siegert}, T. and {Sleator}, C. and {Takahashi}, T. and {Tavecchio}, F. and {Wulf}, E. and {Beechert}, J. and {Gulick}, H. and {Joens}, A. and {Lazar}, H. and {Neights}, E. and {Martinez Oliveros}, J.~C. and {Matsumoto}, S. and {Melia}, T. and {Yoneda}, H. and {Amman}, M. and {Bal}, D. and {von Ballmoos}, P. and {Bates}, H. and {B{\"o}ttcher}, M. and {Bulgarelli}, A. and {Cavazzuti}, E. and {Chang}, H.~K. and {Chen}, C. and {Chu}, C.~Y. and {Ciabattoni}, A. and {Costamante}, L. and {Dreyer}, L. and {Fioretti}, V. and {Fenu}, F. and {Gallego}, S. and {Ghirlanda}, G. and {Grove}, E. and {Huang}, C.~Y. and {Jean}, P. and {Khatiya}, N. and {Kn{\"o}dlseder}, J. and {Kraus}, M. and {Leising}, M. and {Lewis}, T. and {Lommler}, J. and {Marcotulli}, L. and {Martinez Castellanos}, I. and {Mittal}, S. and {Negro}, M. and {Al Nussirat}, S. and {Nakazawa}, K. and {Oberlack}, U. and {Palmore}, D. and {Panebianco}, G. and {Parmiggiani}, N. and {Pike}, S. and {Rogers}, F. and {Schutte}, H. and {Sheng}, Y. and {Smale}, A. and {Smith}, J.~R. and {Trigg}, A. and {Venters}, T. and {Watanabe}, Y. and {Zhang}, H.},
    title = "{The Compton Spectrometer and Imager}",
    booktitle = {38th International Cosmic Ray Conference},
    year = 2024,
    month = sep,
    eid = {745},
    pages = {745},
    doi = {10.22323/1.444.0745},
    archivePrefix = {arXiv},
    eprint = {2308.12362},
    primaryClass = {astro-ph.HE},
    adsurl = {https://ui.adsabs.harvard.edu/abs/2024icrc.confE.745T}
}

@article{aharonian_cosmic_1981,
    title = {Cosmic gamma-rays associated with annihilation of relativistic e+ −e− pairs},
    volume = {99},
    issn = {0370-2693},
    url = {https://www.sciencedirect.com/science/article/pii/0370269381911308},
    doi = {10.1016/0370-2693(81)91130-8},
    number = {3},
    urldate = {2025-12-22},
    journal = {Physics Letters B},
    author = {Aharonian, F. A. and Atoyan, A. M.},
    month = feb,
    year = {1981},
    pages = {301--304},
}

@ARTICLE{bethe_heitler_1934,
    author = {{Bethe}, H. and {Heitler}, W.},
    title = "{On the Stopping of Fast Particles and on the Creation of Positive Electrons}",
    journal = {Proceedings of the Royal Society of London Series A},
    year = 1934,
    month = aug,
    volume = {146},
    number = {856},
    pages = {83-112},
    doi = {10.1098/rspa.1934.0140},
    adsurl = {https://ui.adsabs.harvard.edu/abs/1934RSPSA.146...83B}
}

@article{ore_three-photon_1949,
	title = {Three-{Photon} {Annihilation} of an {Electron}-{Positron} {Pair}},
	volume = {75},
	url = {https://link.aps.org/doi/10.1103/PhysRev.75.1696},
	doi = {10.1103/PhysRev.75.1696},
	number = {11},
	urldate = {2025-12-21},
	journal = {Phys. Rev.},
	publisher = {American Physical Society},
	author = {Ore, A. and Powell, J. L.},
	month = jun,
	year = {1949},
	pages = {1696--1699},
}

@ARTICLE{bykov_1999_shocks,
    author = {{Bykov}, A.~M. and {Uvarov}, Yu. A.},
    title = "{Electron kinetics in collisionless shock waves}",
    journal = {Soviet Journal of Experimental and Theoretical Physics},
    year = 1999,
    month = mar,
    volume = {88},
    number = {3},
    pages = {465-475},
    doi = {10.1134/1.558817},
    adsurl = {https://ui.adsabs.harvard.edu/abs/1999JETP...88..465B}
}

@ARTICLE{sironi_2015_shocks,
    author = {{Sironi}, L. and {Keshet}, U. and {Lemoine}, M.},
    title = "{Relativistic Shocks: Particle Acceleration and Magnetization}",
    journal = {\ssr},
    year = 2015,
    month = oct,
    volume = {191},
    number = {1-4},
    pages = {519-544},
    doi = {10.1007/s11214-015-0181-8},
    archivePrefix = {arXiv},
    eprint = {1506.02034},
    primaryClass = {astro-ph.HE},
    adsurl = {https://ui.adsabs.harvard.edu/abs/2015SSRv..191..519S}
}

@article{jansson_GMF_2012,
    title = {The {Galactic} {Magnetic} {Field}},
    volume = {761},
    issn = {0004-637X},
    url = {https://ui.adsabs.harvard.edu/abs/2012ApJ...761L..11J},
    doi = {10.1088/2041-8205/761/1/L11},
    urldate = {2025-10-09},
    journal = {The Astrophysical Journal},
    publisher = {IOP},
    author = {Jansson, Ronnie and Farrar, Glennys R.},
    month = dec,
    year = {2012},
    note = {ADS Bibcode: 2012ApJ...761L..11J},
    pages = {L11},
}

@article{han_GMF_2017,
	title = {Observing {Interstellar} and {Intergalactic} {Magnetic} {Fields}},
	volume = {55},
	issn = {0066-4146},
	url = {https://ui.adsabs.harvard.edu/abs/2017ARA&A..55..111H},
	doi = {10.1146/annurev-astro-091916-055221},
	urldate = {2025-06-20},
	journal = {Annual Review of Astronomy and Astrophysics},
	author = {Han, J. L.},
	month = aug,
	year = {2017},
	note = {ADS Bibcode: 2017ARA\&A..55..111H},
	pages = {111--157},
}

@article{korochkin_GMF_2025,
	title = {The coherent magnetic field of the {Milky} {Way} halo, the {Local} {Bubble}, and the {Fan} region},
	volume = {693},
	issn = {0004-6361},
	url = {https://ui.adsabs.harvard.edu/abs/2025A&A...693A.284K},
	doi = {10.1051/0004-6361/202451440},
	urldate = {2025-10-09},
	journal = {Astronomy and Astrophysics},
	publisher = {EDP},
	author = {Korochkin, Alexander and Semikoz, Dmitri and Tinyakov, Peter},
	month = jan,
	year = {2025},
	note = {ADS Bibcode: 2025A\&A...693A.284K},
	pages = {A284},
}

@article{keshet_shocks_2005,
    title = {Energy {Spectrum} of {Particles} {Accelerated} in {Relativistic} {Collisionless} {Shocks}},
    volume = {94},
    url = {https://link.aps.org/doi/10.1103/PhysRevLett.94.111102},
    doi = {10.1103/PhysRevLett.94.111102},
    number = {11},
    urldate = {2026-01-17},
    journal = {Phys. Rev. Lett.},
    publisher = {American Physical Society},
    author = {Keshet, Uri and Waxman, Eli},
    month = mar,
    year = {2005},
    pages = {111102},
}

@ARTICLE{fender_2003_jets,
    author = {{Fender}, R.~P. and {Gallo}, E. and {Jonker}, P.~G.},
    title = "{Jet-dominated states: an alternative to advection across black hole event horizons in `quiescent' X-ray binaries}",
    journal = {\mnras},
    year = 2003,
    month = aug,
    volume = {343},
    number = {4},
    pages = {L99-L103},
    doi = {10.1046/j.1365-8711.2003.06950.x},
    archivePrefix = {arXiv},
    eprint = {astro-ph/0306614},
    primaryClass = {astro-ph},
    adsurl = {https://ui.adsabs.harvard.edu/abs/2003MNRAS.343L..99F}
}

@ARTICLE{dubus_2001_disk_instability,
    author = {{Dubus}, G. and {Hameury}, J.-M. and {Lasota}, J.-P.},
    title = "{The disc instability model for X-ray transients: Evidence for truncation and irradiation}",
    journal = {\aap},
    year = 2001,
    month = jul,
    volume = {373},
    pages = {251-271},
    doi = {10.1051/0004-6361:20010632},
    archivePrefix = {arXiv},
    eprint = {astro-ph/0102237},
    primaryClass = {astro-ph},
    adsurl = {https://ui.adsabs.harvard.edu/abs/2001A&A...373..251D}
}

@ARTICLE{yoneda_2025_SPI,
    author = {{Yoneda}, Hiroki and {Siegert}, Thomas and {Mittal}, Saurabh},
    title = "{Imaging the positron annihilation line with 20 years of INTEGRAL/SPI observations}",
    journal = {\aap},
    year = 2025,
    month = oct,
    volume = {702},
    eid = {A220},
    pages = {A220},
    doi = {10.1051/0004-6361/202555895},
    archivePrefix = {arXiv},
    eprint = {2509.01066},
    primaryClass = {astro-ph.HE},
    adsurl = {https://ui.adsabs.harvard.edu/abs/2025A&A...702A.220Y}
}

@ARTICLE{mirabel_1994_grs_1915,
    author = {{Mirabel}, I.~F. and {Rodr{\'\i}guez}, L.~F.},
    title = "{A superluminal source in the Galaxy}",
    journal = {\nat},
    year = 1994,
    month = sep,
    volume = {371},
    number = {6492},
    pages = {46-48},
    doi = {10.1038/371046a0},
    adsurl = {https://ui.adsabs.harvard.edu/abs/1994Natur.371...46M}
}

@INPROCEEDINGS{moskalenko_2004_diffuse,
    author = {{Moskalenko}, Igor V. and {Strong}, Andrew W. and {Reimer}, Olaf},
    title = "{Diffuse Gamma Rays}",
    booktitle = {Cosmic Gamma-Ray Sources},
    year = 2004,
    editor = {{Cheng}, K.~S. and {Romero}, Gustavo E.},
    series = {Astrophysics and Space Science Library},
    volume = {304},
    month = jan,
    pages = {279},
    doi = {10.1007/978-1-4020-2256-2_12},
    archivePrefix = {arXiv},
    eprint = {astro-ph/0402243},
    primaryClass = {astro-ph},
    adsurl = {https://ui.adsabs.harvard.edu/abs/2004ASSL..304..279M}
}

@INPROCEEDINGS{siegert_2024_foreground,
    author = {{Siegert}, Thomas},
    title = "{Time-variable diffuse gamma-ray foreground}",
    booktitle = {AAS High Energy Astrophysics Division Meeting \#21},
    year = 2024,
    series = {AAS/High Energy Astrophysics Division},
    volume = {21},
    month = may,
    eid = {104.07},
    pages = {104.07},
    adsurl = {https://ui.adsabs.harvard.edu/abs/2024HEAD...2110407S}
}

@ARTICLE{churazov_2011_emission,
    author = {{Churazov}, E. and {Sazonov}, S. and {Tsygankov}, S. and {Sunyaev}, R. and {Varshalovich}, D.},
    title = "{Positron annihilation spectrum from the Galactic Centre region observed by SPI/INTEGRAL revisited: annihilation in a cooling ISM?}",
    journal = {\mnras},
    year = 2011,
    month = mar,
    volume = {411},
    number = {3},
    pages = {1727-1743},
    doi = {10.1111/j.1365-2966.2010.17804.x},
    archivePrefix = {arXiv},
    eprint = {1010.0864},
    primaryClass = {astro-ph.HE},
    adsurl = {https://ui.adsabs.harvard.edu/abs/2011MNRAS.411.1727C}
}

@article{siegert_cosi_2020,
    title = {Imaging the 511 {keV} {Positron} {Annihilation} {Sky} with {COSI}},
    volume = {897},
    issn = {0004-637X},
    url = {https://dx.doi.org/10.3847/1538-4357/ab9607},
    doi = {10.3847/1538-4357/ab9607},
    language = {en},
    number = {1},
    urldate = {2025-09-08},
    journal = {ApJ},
    publisher = {The American Astronomical Society},
    author = {Siegert, Thomas and Boggs, Steven E. and Tomsick, John A. and Zoglauer, Andreas C. and Kierans, Carolyn A. and Sleator, Clio C. and Beechert, Jacqueline and Brandt, Theresa J. and Jean, Pierre and Lazar, Hadar and Lowell, Alex W. and Roberts, Jarred M. and Ballmoos, Peter von},
    month = jul,
    year = {2020},
    pages = {45},
}

@article{kierans_cosi_2020,
    title = {Detection of the 511 {keV} {Galactic} {Positron} {Annihilation} {Line} with {COSI}},
    volume = {895},
    issn = {0004-637X},
    url = {https://dx.doi.org/10.3847/1538-4357/ab89a9},
    doi = {10.3847/1538-4357/ab89a9},
    language = {en},
    number = {1},
    urldate = {2025-09-08},
    journal = {ApJ},
    publisher = {The American Astronomical Society},
    author = {Kierans, C. A. and Boggs, S. E. and Zoglauer, A. and Lowell, A. W. and Sleator, C. and Beechert, J. and Brandt, T. J. and Jean, P. and Lazar, H. and Roberts, J. and Siegert, T. and Tomsick, J. A. and Ballmoos, P. von},
    month = may,
    year = {2020},
    pages = {44},
}

@ARTICLE{churazov_2005_emission,
    author = {{Churazov}, E. and {Sunyaev}, R. and {Sazonov}, S. and {Revnivtsev}, M. and {Varshalovich}, D.},
    title = "{Positron annihilation spectrum from the Galactic Centre region observed by SPI/INTEGRAL}",
    journal = {\mnras},
    year = 2005,
    month = mar,
    volume = {357},
    number = {4},
    pages = {1377-1386},
    doi = {10.1111/j.1365-2966.2005.08757.x},
    archivePrefix = {arXiv},
    eprint = {astro-ph/0411351},
    primaryClass = {astro-ph},
    adsurl = {https://ui.adsabs.harvard.edu/abs/2005MNRAS.357.1377C}
}

@ARTICLE{purcell_1997_imaging,
    author = {{Purcell}, W.~R. and {Cheng}, L.-X. and {Dixon}, D.~D. and {Kinzer}, R.~L. and {Kurfess}, J.~D. and {Leventhal}, M. and {Saunders}, M.~A. and {Skibo}, J.~G. and {Smith}, D.~M. and {Tueller}, J.},
    title = "{OSSE Mapping of Galactic 511 keV Positron Annihilation Line Emission}",
    journal = {\apj},
    year = 1997,
    month = dec,
    volume = {491},
    number = {2},
    pages = {725-748},
    doi = {10.1086/304994},
    adsurl = {https://ui.adsabs.harvard.edu/abs/1997ApJ...491..725P}
}

@ARTICLE{knodlseder_2005_imaging,
    author = {{Kn{\"o}dlseder}, J. and {Jean}, P. and {Lonjou}, V. and {Weidenspointner}, G. and {Guessoum}, N. and {Gillard}, W. and {Skinner}, G. and {von Ballmoos}, P. and {Vedrenne}, G. and {Roques}, J.-P. and {Schanne}, S. and {Teegarden}, B. and {Sch{\"o}nfelder}, V. and {Winkler}, C.},
    title = "{The all-sky distribution of 511 keV electron-positron annihilation emission}",
    journal = {\aap},
    year = 2005,
    month = oct,
    volume = {441},
    number = {2},
    pages = {513-532},
    doi = {10.1051/0004-6361:20042063},
    archivePrefix = {arXiv},
    eprint = {astro-ph/0506026},
    primaryClass = {astro-ph},
    adsurl = {https://ui.adsabs.harvard.edu/abs/2005A&A...441..513K}
}

@article{siegert_flux_2016,
    title = {Gamma-ray spectroscopy of positron annihilation in the {Milky} {Way}},
    volume = {586},
    issn = {0004-6361},
    url = {https://ui.adsabs.harvard.edu/abs/2016A&A...586A..84S},
    doi = {10.1051/0004-6361/201527510},
    urldate = {2025-06-19},
    journal = {Astronomy and Astrophysics},
    author = {Siegert, Thomas and Diehl, Roland and Khachatryan, Gerasim and Krause, Martin G. H. and Guglielmetti, Fabrizia and Greiner, Jochen and Strong, Andrew W. and Zhang, Xiaoling},
    month = feb,
    year = {2016},
    note = {ADS Bibcode: 2016A\&A...586A..84S},
    pages = {A84},
}

@inproceedings{skinner_galactic_2015,
	address = {Annapolis, MD, USA},
	title = {The {Galactic} distribution of the 511 {keV} e+/e- annihilation radiation},
	url = {https://pos.sissa.it/228/054},
	doi = {10.22323/1.228.0054},
	language = {en},
	urldate = {2025-11-16},
	booktitle = {Proceedings of 10th {INTEGRAL} {Workshop}: {A} {Synergistic} {View} of the {High}-{Energy} {Sky} — {PoS}({Integral2014})},
	publisher = {Sissa Medialab},
	author = {Skinner, Gerald and Diehl, Roland and Zhang, Xiao-Ling and Bouchet, Laurent and Jean, Pierre},
	month = mar,
	year = {2015},
	pages = {054},
}

@article{guessoum_microquasars_2006,
    title = {Microquasars as sources of positron annihilation radiation},
    volume = {457},
    issn = {0004-6361},
    url = {https://ui.adsabs.harvard.edu/abs/2006A&A...457..753G},
    doi = {10.1051/0004-6361:20065240},
    urldate = {2025-04-17},
    journal = {Astronomy and Astrophysics},
    author = {Guessoum, N. and Jean, P. and Prantzos, N.},
    month = oct,
    year = {2006},
    note = {ADS Bibcode: 2006A\&A...457..753G},
    pages = {753--762},
}

@ARTICLE{bouchet_2010_imaging,
    author = {{Bouchet}, L. and {Roques}, J.~P. and {Jourdain}, E.},
    title = "{On the Morphology of the Electron-Positron Annihilation Emission as Seen by Spi/integral}",
    journal = {\apj},
    year = 2010,
    month = sep,
    volume = {720},
    number = {2},
    pages = {1772-1780},
    doi = {10.1088/0004-637X/720/2/1772},
    archivePrefix = {arXiv},
    eprint = {1007.4753},
    primaryClass = {astro-ph.HE},
    adsurl = {https://ui.adsabs.harvard.edu/abs/2010ApJ...720.1772B}
}

@article{martin_nucleosynthesis_2012,
	title = {Galactic annihilation emission from nucleosynthesis positrons},
	volume = {543},
	issn = {0004-6361},
	url = {https://ui.adsabs.harvard.edu/abs/2012A&A...543A...3M},
	doi = {10.1051/0004-6361/201118721},
	urldate = {2025-04-19},
	journal = {Astronomy and Astrophysics},
	author = {Martin, P. and Strong, A. W. and Jean, P. and Alexis, A. and Diehl, R.},
	month = jul,
	year = {2012},
	note = {ADS Bibcode: 2012A\&A...543A...3M},
	pages = {A3},
}

@article{panther_propagation_2018,
	title = {Positron {Transport} and {Annihilation} in the {Galactic} {Bulge}},
	volume = {6},
	url = {https://ui.adsabs.harvard.edu/abs/2018Galax...6...39P},
	doi = {10.3390/galaxies6020039},
	urldate = {2025-05-16},
	journal = {Galaxies},
	author = {Panther, Fiona Helen},
	month = mar,
	year = {2018},
	note = {ADS Bibcode: 2018Galax...6...39P},
	pages = {39},
}

@article{alexis_MC_2014,
    title = {Monte {Carlo} modelling of the propagation and annihilation of nucleosynthesis positrons in the {Galaxy}},
    volume = {564},
    copyright = {© ESO, 2014},
    issn = {0004-6361, 1432-0746},
    url = {https://www.aanda.org/articles/aa/abs/2014/04/aa22393-13/aa22393-13.html},
    doi = {10.1051/0004-6361/201322393},
    language = {en},
    urldate = {2025-06-20},
    journal = {A\&A},
    publisher = {EDP Sciences},
    author = {Alexis, A. and Jean, P. and Martin, P. and Ferrière, K.},
    month = apr,
    year = {2014},
    pages = {A108},
}

@ARTICLE{siegert_2022_propagation,
    author = {{Siegert}, Thomas and {Crocker}, Roland M. and {Macias}, Oscar and {Panther}, Fiona H. and {Calore}, Francesca and {Song}, Deheng and {Horiuchi}, Shunsaku},
    title = "{Measuring the smearing of the Galactic 511-keV signal: positron propagation or supernova kicks?}",
    journal = {\mnras},
    year = 2022,
    month = jan,
    volume = {509},
    number = {1},
    pages = {L11-L16},
    doi = {10.1093/mnrasl/slab113},
    archivePrefix = {arXiv},
    eprint = {2109.03691},
    primaryClass = {astro-ph.HE},
    adsurl = {https://ui.adsabs.harvard.edu/abs/2022MNRAS.509L..11S}
}

@article{fender_jet_model_2004,
	title = {Towards a unified model for black hole {X}-ray binary jets},
	volume = {355},
	issn = {0035-8711},
	url = {https://doi.org/10.1111/j.1365-2966.2004.08384.x},
	doi = {10.1111/j.1365-2966.2004.08384.x},
	number = {4},
	urldate = {2025-06-24},
	journal = {Monthly Notices of the Royal Astronomical Society},
	author = {Fender, R. P. and Belloni, T. M. and Gallo, E.},
	month = dec,
	year = {2004},
	pages = {1105--1118},
}

@article{fender_persistent_jets_2000,
	title = {The radio luminosity of persistent {X}-ray binaries},
	volume = {317},
	issn = {0035-8711},
	url = {https://ui.adsabs.harvard.edu/abs/2000MNRAS.317....1F},
	doi = {10.1046/j.1365-8711.2000.03443.x},
	urldate = {2026-01-24},
	journal = {Monthly Notices of the Royal Astronomical Society},
	publisher = {OUP},
	author = {Fender, R. P. and Hendry, M. A.},
	month = sep,
	year = {2000},
	note = {ADS Bibcode: 2000MNRAS.317....1F},
	pages = {1--8},
}

@article{stawarz_stochastic_2008,
	title = {On the {Momentum} {Diffusion} of {Radiating} {Ultrarelativistic} {Electrons} in a {Turbulent} {Magnetic} {Field}},
	volume = {681},
	issn = {0004-637X},
	url = {https://ui.adsabs.harvard.edu/abs/2008ApJ...681.1725S},
	doi = {10.1086/588813},
	urldate = {2026-01-24},
	journal = {The Astrophysical Journal},
	publisher = {IOP},
	author = {Stawarz, \L{}ukasz and Petrosian, Vahe},
	month = jul,
	year = {2008},
	note = {ADS Bibcode: 2008ApJ...681.1725S},
	pages = {1725--1744},
}

@article{tetarenko_watchdog_2016,
	title = {{WATCHDOG}: {A} {Comprehensive} {All}-sky {Database} of {Galactic} {Black} {Hole} {X}-ray {Binaries}},
	volume = {222},
	issn = {0067-0049},
	shorttitle = {{WATCHDOG}},
	url = {https://ui.adsabs.harvard.edu/abs/2016ApJS..222...15T},
	doi = {10.3847/0067-0049/222/2/15},
	urldate = {2025-11-19},
	journal = {The Astrophysical Journal Supplement Series},
	publisher = {IOP},
	author = {Tetarenko, B. E. and Sivakoff, G. R. and Heinke, C. O. and Gladstone, J. C.},
	month = feb,
	year = {2016},
	note = {ADS Bibcode: 2016ApJS..222...15T},
	pages = {15},
}

@article{sidoli_DCs_2018,
	title = {An {INTEGRAL} overview of {High}-{Mass} {X}–ray {Binaries}: classes or transitions?},
	volume = {481},
	issn = {0035-8711},
	shorttitle = {An {INTEGRAL} overview of {High}-{Mass} {X}–ray {Binaries}},
	url = {https://doi.org/10.1093/mnras/sty2428},
	doi = {10.1093/mnras/sty2428},
	number = {2},
	urldate = {2026-01-24},
	journal = {Mon Not R Astron Soc},
	author = {Sidoli, L and Paizis, A},
	month = dec,
	year = {2018},
	pages = {2779--2803},
}

@article{van_den_eijnden_NSXRB_2021,
	title = {A new radio census of neutron star {X}-ray binaries},
	volume = {507},
	issn = {0035-8711},
	url = {https://ui.adsabs.harvard.edu/abs/2021MNRAS.507.3899V},
	doi = {10.1093/mnras/stab1995},
	urldate = {2026-01-25},
	journal = {Monthly Notices of the Royal Astronomical Society},
	publisher = {OUP},
	author = {van den Eijnden, J. and Degenaar, N. and Russell, T. D. and Wijnands, R. and Bahramian, A. and Miller-Jones, J. C. A. and Hernández Santisteban, J. V. and Gallo, E. and Atri, P. and Plotkin, R. M. and Maccarone, T. J. and Sivakoff, G. and Miller, J. M. and Reynolds, M. and Russell, D. M. and Maitra, D. and Heinke, C. O. and Armas Padilla, M. and Shaw, A. W.},
	month = nov,
	year = {2021},
	note = {ADS Bibcode: 2021MNRAS.507.3899V},
	pages = {3899--3922},
}

@article{migliari_jets_2006,
	title = {Jets in neutron star {X}-ray binaries: a comparison with black holes},
	volume = {366},
	issn = {0035-8711},
	shorttitle = {Jets in neutron star {X}-ray binaries},
	url = {https://doi.org/10.1111/j.1365-2966.2005.09777.x},
	doi = {10.1111/j.1365-2966.2005.09777.x},
	number = {1},
	urldate = {2026-01-24},
	journal = {Mon Not R Astron Soc},
	author = {Migliari, S. and Fender, R. P.},
	month = feb,
	year = {2006},
	pages = {79--91},
}

@article{bandyopadhyay_emission_2009,
	title = {On the origin of the 511-{keV} emission in the {Galactic} {Centre}},
	volume = {392},
	issn = {0035-8711},
	url = {https://ui.adsabs.harvard.edu/abs/2009MNRAS.392.1115B},
	doi = {10.1111/j.1365-2966.2008.14113.x},
	urldate = {2025-06-19},
	journal = {Monthly Notices of the Royal Astronomical Society},
	publisher = {OUP},
	author = {Bandyopadhyay, Reba M. and Silk, Joseph and Taylor, James E. and Maccarone, Thomas J.},
	month = jan,
	year = {2009},
	note = {ADS Bibcode: 2009MNRAS.392.1115B},
	pages = {1115--1123},
}

@ARTICLE{goodenough_2009_GeV_excess,
    author = {{Goodenough}, Lisa and {Hooper}, Dan},
    title = "{Possible Evidence For Dark Matter Annihilation In The Inner Milky Way From The Fermi Gamma Ray Space Telescope}",
    journal = {arXiv e-prints},
    year = 2009,
    month = oct,
    eid = {arXiv:0910.2998},
    pages = {arXiv:0910.2998},
    doi = {10.48550/arXiv.0910.2998},
    archivePrefix = {arXiv},
    eprint = {0910.2998},
    primaryClass = {hep-ph},
    adsurl = {https://ui.adsabs.harvard.edu/abs/2009arXiv0910.2998G}
}

@article{picciotto_DM_2005,
	title = {Unstable relics as a source of galactic positrons},
	volume = {605},
	issn = {0370-2693},
	url = {https://www.sciencedirect.com/science/article/pii/S0370269304015813},
	doi = {10.1016/j.physletb.2004.11.025},
	number = {1},
	urldate = {2026-01-27},
	journal = {Physics Letters B},
	author = {Picciotto, Charles and Pospelov, Maxim},
	month = jan,
	year = {2005},
	pages = {15--25},
}

@article{hooper_DM_2004,
	title = {Possible evidence for axino dark matter in the galactic bulge},
	volume = {70},
	issn = {1550-79980556-2821},
	url = {https://ui.adsabs.harvard.edu/abs/2004PhRvD..70f3506H},
	doi = {10.1103/PhysRevD.70.063506},
	urldate = {2026-01-27},
	journal = {Physical Review D},
	publisher = {APS},
	author = {Hooper, Dan and Wang, Lian-Tao},
	month = sep,
	year = {2004},
	note = {ADS Bibcode: 2004PhRvD..70f3506H},
	pages = {063506},
}

@article{boehm_DM_2004,
	title = {{MeV} {Dark} {Matter}: {Has} {It} {Been} {Detected}?},
	volume = {92},
	issn = {0031-9007},
	shorttitle = {{MeV} {Dark} {Matter}},
	url = {https://ui.adsabs.harvard.edu/abs/2004PhRvL..92j1301B},
	doi = {10.1103/PhysRevLett.92.101301},
	urldate = {2026-01-27},
	journal = {Physical Review Letters},
	publisher = {APS},
	author = {Boehm, Céline and Hooper, Dan and Silk, Joseph and Casse, Michel and Paul, Jacques},
	month = mar,
	year = {2004},
	note = {ADS Bibcode: 2004PhRvL..92j1301B},
	pages = {101301},
}

@article{gunion_DM_2006,
	title = {Light neutralino dark matter in the next-to-minimal supersymmetric standard model},
	volume = {73},
	issn = {1550-79980556-2821},
	url = {https://ui.adsabs.harvard.edu/abs/2006PhRvD..73a5011G},
	doi = {10.1103/PhysRevD.73.015011},
	urldate = {2026-01-27},
	journal = {Physical Review D},
	publisher = {APS},
	author = {Gunion, John F. and Hooper, Dan and McElrath, Bob},
	month = jan,
	year = {2006},
	note = {ADS Bibcode: 2006PhRvD..73a5011G},
	pages = {015011},
}

@article{finkbeiner_DM_2007,
	title = {Exciting dark matter and the {INTEGRAL}/{SPI} {511keV} signal},
	volume = {76},
	issn = {1550-79980556-2821},
	url = {https://ui.adsabs.harvard.edu/abs/2007PhRvD..76h3519F},
	doi = {10.1103/PhysRevD.76.083519},
	urldate = {2026-01-27},
	journal = {Physical Review D},
	publisher = {APS},
	author = {Finkbeiner, Douglas P. and Weiner, Neal},
	month = oct,
	year = {2007},
	note = {ADS Bibcode: 2007PhRvD..76h3519F},
	pages = {083519},
}

@article{sizun_DM_2006,
	title = {Continuum γ-ray emission from light dark matter positrons and electrons},
	volume = {74},
	issn = {1550-79980556-2821},
	url = {https://ui.adsabs.harvard.edu/abs/2006PhRvD..74f3514S},
	doi = {10.1103/PhysRevD.74.063514},
	urldate = {2025-09-09},
	journal = {Physical Review D},
	author = {Sizun, P. and Cassé, M. and Schanne, S.},
	month = sep,
	year = {2006},
	note = {ADS Bibcode: 2006PhRvD..74f3514S},
	pages = {063514},
}

@article{clayton_SNe_1973,
	title = {Positronium {Origin} of 476 {keV} {Galactic} {Feature}},
	volume = {244},
	copyright = {1973 Springer Nature Limited},
	issn = {2058-1106},
	url = {https://www.nature.com/articles/physci244137b0},
	doi = {10.1038/physci244137b0},
	language = {en},
	number = {139},
	urldate = {2026-01-27},
	journal = {Nature Physical Science},
	publisher = {Nature Publishing Group},
	author = {Clayton, Donald D.},
	month = aug,
	year = {1973},
	pages = {137--138},
}

@ARTICLE{diehl_massive_stars_1995,
    author = {{Diehl}, R. and {Dupraz}, C. and {Bennett}, K. and {Bloemen}, H. and {Hermsen}, W. and {Knoedlseder}, J. and {Lichti}, G. and {Morris}, D. and {Ryan}, J. and {Schoenfelder}, V. and {Steinle}, H. and {Strong}, A. and {Swanenburg}, B. and {Varendorff}, M. and {Winkler}, C.},
    title = "{COMPTEL observations of Galactic \^26\^Al emission.}",
    journal = {\aap},
    year = 1995,
    month = jun,
    volume = {298},
    pages = {445},
    adsurl = {https://ui.adsabs.harvard.edu/abs/1995A&A...298..445D}
}

@ARTICLE{prantzos_massive_stars_1996,
           author = {{Prantzos}, N. and {Diehl}, R.},
            title = "{Radioactive 26Al in the galaxy: observations versus theory}",
          journal = {\physrep},
             year = 1996,
            month = mar,
           volume = {267},
            pages = {1-69},
              doi = {10.1016/0370-1573(95)00055-0},
           adsurl = {https://ui.adsabs.harvard.edu/abs/1996PhR...267....1P}
}

@ARTICLE{casse_hypernovae_2004,
    author = {{Cass{\'e}}, M. and {Cordier}, B. and {Paul}, J. and {Schanne}, S.},
    title = "{Hypernovae/Gamma-Ray Bursts in the Galactic Center as Possible Sources of Galactic Positrons}",
    journal = {\apjl},
    year = 2004,
    month = feb,
    volume = {602},
    number = {1},
    pages = {L17-L20},
    doi = {10.1086/381884},
    archivePrefix = {arXiv},
    eprint = {astro-ph/0309824},
    primaryClass = {astro-ph},
    adsurl = {https://ui.adsabs.harvard.edu/abs/2004ApJ...602L..17C}
}

@ARTICLE{parizot_GRB_2005,
    author = {{Parizot}, E. and {Cass{\'e}}, M. and {Lehoucq}, R. and {Paul}, J.},
    title = "{GRBs and the 511 keV emission of the Galactic bulge}",
    journal = {\aap},
    year = 2005,
    month = mar,
    volume = {432},
    number = {3},
    pages = {889-894},
    doi = {10.1051/0004-6361:20042215},
    archivePrefix = {arXiv},
    eprint = {astro-ph/0411656},
    primaryClass = {astro-ph},
    adsurl = {https://ui.adsabs.harvard.edu/abs/2005A&A...432..889P}
}

@ARTICLE{bertone_GRB_2006,
    author = {{Bertone}, Gianfranco and {Kusenko}, Alexander and {Palomares-Ruiz}, Sergio and {Pascoli}, Silvia and {Semikoz}, Dmitry},
    title = "{Gamma-ray bursts and the origin of galactic positrons}",
    journal = {Physics Letters B},
    year = 2006,
    month = apr,
    volume = {636},
    number = {1},
    pages = {20-24},
    doi = {10.1016/j.physletb.2006.03.022},
    archivePrefix = {arXiv},
    eprint = {astro-ph/0405005},
    primaryClass = {astro-ph},
    adsurl = {https://ui.adsabs.harvard.edu/abs/2006PhLB..636...20B}
}

@ARTICLE{milne_SNe_1999,
    author = {{Milne}, P.~A. and {The}, L.-S. and {Leising}, M.~D.},
    title = "{Positron Escape from Type IA Supernovae}",
    journal = {\apjs},
    year = 1999,
    month = oct,
    volume = {124},
    number = {2},
    pages = {503-526},
    doi = {10.1086/313262},
    archivePrefix = {arXiv},
    eprint = {astro-ph/9901206},
    primaryClass = {astro-ph},
    adsurl = {https://ui.adsabs.harvard.edu/abs/1999ApJS..124..503M}
}

@ARTICLE{mazzali_SNe_2007,
    author = {{Mazzali}, Paolo A. and {R{\"o}pke}, Friedrich K. and {Benetti}, Stefano and {Hillebrandt}, Wolfgang},
    title = "{A Common Explosion Mechanism for Type Ia Supernovae}",
    journal = {Science},
    year = 2007,
    month = feb,
    volume = {315},
    number = {5813},
    pages = {825},
    doi = {10.1126/science.1136259},
    archivePrefix = {arXiv},
    eprint = {astro-ph/0702351},
    primaryClass = {astro-ph},
    adsurl = {https://ui.adsabs.harvard.edu/abs/2007Sci...315..825M}
}

@ARTICLE{seitenzahl_SNe_2009,
    author = {{Seitenzahl}, I.~R. and {Taubenberger}, S. and {Sim}, S.~A.},
    title = "{Late-time supernova light curves: the effect of internal conversion and Auger electrons}",
    journal = {\mnras},
    year = 2009,
    month = nov,
    volume = {400},
    number = {1},
    pages = {531-535},
    doi = {10.1111/j.1365-2966.2009.15478.x},
    archivePrefix = {arXiv},
    eprint = {0908.0247},
    primaryClass = {astro-ph.SR},
    adsurl = {https://ui.adsabs.harvard.edu/abs/2009MNRAS.400..531S}
}

@ARTICLE{moskalenko_CRs_1998,
    author = {{Moskalenko}, I.~V. and {Strong}, A.~W.},
    title = "{Production and Propagation of Cosmic-Ray Positrons and Electrons}",
    journal = {\apj},
    year = 1998,
    month = jan,
    volume = {493},
    number = {2},
    pages = {694-707},
    doi = {10.1086/305152},
    archivePrefix = {arXiv},
    eprint = {astro-ph/9710124},
    primaryClass = {astro-ph},
    adsurl = {https://ui.adsabs.harvard.edu/abs/1998ApJ...493..694M}
}

@ARTICLE{adriani_CRs_2009,
    author = {{Adriani}, O. and {Barbarino}, G.~C. and {Bazilevskaya}, G.~A. and {Bellotti}, R. and {Boezio}, M. and {Bogomolov}, E.~A. and {Bonechi}, L. and {Bongi}, M. and {Bonvicini}, V. and {Bottai}, S. and {Bruno}, A. and {Cafagna}, F. and {Campana}, D. and {Carlson}, P. and {Casolino}, M. and {Castellini}, G. and {de Pascale}, M.~P. and {de Rosa}, G. and {de Simone}, N. and {di Felice}, V. and {Galper}, A.~M. and {Grishantseva}, L. and {Hofverberg}, P. and {Koldashov}, S.~V. and {Krutkov}, S.~Y. and {Kvashnin}, A.~N. and {Leonov}, A. and {Malvezzi}, V. and {Marcelli}, L. and {Menn}, W. and {Mikhailov}, V.~V. and {Mocchiutti}, E. and {Orsi}, S. and {Osteria}, G. and {Papini}, P. and {Pearce}, M. and {Picozza}, P. and {Ricci}, M. and {Ricciarini}, S.~B. and {Simon}, M. and {Sparvoli}, R. and {Spillantini}, P. and {Stozhkov}, Y.~I. and {Vacchi}, A. and {Vannuccini}, E. and {Vasilyev}, G. and {Voronov}, S.~A. and {Yurkin}, Y.~T. and {Zampa}, G. and {Zampa}, N. and {Zverev}, V.~G.},
    title = "{An anomalous positron abundance in cosmic rays with energies 1.5-100GeV}",
    journal = {\nat},
    year = 2009,
    month = apr,
    volume = {458},
    number = {7238},
    pages = {607-609},
    doi = {10.1038/nature07942},
    archivePrefix = {arXiv},
    eprint = {0810.4995},
    primaryClass = {astro-ph},
    adsurl = {https://ui.adsabs.harvard.edu/abs/2009Natur.458..607A}
}

@ARTICLE{porter_CRs_2008,
    author = {{Porter}, Troy A. and {Moskalenko}, Igor V. and {Strong}, Andrew W. and {Orlando}, Elena and {Bouchet}, L.},
    title = "{Inverse Compton Origin of the Hard X-Ray and Soft Gamma-Ray Emission from the Galactic Ridge}",
    journal = {\apj},
    year = 2008,
    month = jul,
    volume = {682},
    number = {1},
    pages = {400-407},
    doi = {10.1086/589615},
    archivePrefix = {arXiv},
    eprint = {0804.1774},
    primaryClass = {astro-ph},
    adsurl = {https://ui.adsabs.harvard.edu/abs/2008ApJ...682..400P}
}

@ARTICLE{harding_pulsars_1981,
    author = {{Harding}, A.~K.},
    title = "{Galactic gamma-ray emission from pulsars}",
    journal = {\apj},
    year = 1981,
    month = jul,
    volume = {247},
    pages = {639-649},
    doi = {10.1086/159075},
    adsurl = {https://ui.adsabs.harvard.edu/abs/1981ApJ...247..639H}
}

@ARTICLE{zhang_pulsars_1997,
    author = {{Zhang}, L. and {Cheng}, K.~S.},
    title = "{High-Energy Radiation from Rapidly Spinning Pulsars with Thick Outer Gaps}",
    journal = {\apj},
    year = 1997,
    month = sep,
    volume = {487},
    number = {1},
    pages = {370-379},
    doi = {10.1086/304589},
    adsurl = {https://ui.adsabs.harvard.edu/abs/1997ApJ...487..370Z}
}

@ARTICLE{riegler_sag_A_1981,
    author = {{Riegler}, G.~R. and {Ling}, J.~C. and {Mahoney}, W.~A. and {Wheaton}, W.~A. and {Willett}, J.~B. and {Jacobson}, A.~S. and {Prince}, T.~A.},
    title = "{Variable positron annihilation radiation from the galactic center region}",
    journal = {\apjl},
    year = 1981,
    month = aug,
    volume = {248},
    pages = {L13-L16},
    doi = {10.1086/183613},
    adsurl = {https://ui.adsabs.harvard.edu/abs/1981ApJ...248L..13R}
}

@ARTICLE{fatuzzo_sag_A_2001,
    author = {{Fatuzzo}, Marco and {Melia}, Fulvio and {Rafelski}, Johann},
    title = "{Electron-Positron Annihilation Radiation from Sagittarius A East at the Galactic Center}",
    journal = {\apj},
    year = 2001,
    month = mar,
    volume = {549},
    number = {1},
    pages = {293-302},
    doi = {10.1086/319069},
    archivePrefix = {arXiv},
    eprint = {astro-ph/0007371},
    primaryClass = {astro-ph},
    adsurl = {https://ui.adsabs.harvard.edu/abs/2001ApJ...549..293F}
}

@ARTICLE{cheng_sag_A_2006,
    author = {{Cheng}, K.~S. and {Chernyshov}, D.~O. and {Dogiel}, V.~A.},
    title = "{Annihilation Emission from the Galactic Black Hole}",
    journal = {\apj},
    year = 2006,
    month = jul,
    volume = {645},
    number = {2},
    pages = {1138-1151},
    doi = {10.1086/504583},
    archivePrefix = {arXiv},
    eprint = {astro-ph/0603659},
    primaryClass = {astro-ph},
    adsurl = {https://ui.adsabs.harvard.edu/abs/2006ApJ...645.1138C}
}

@ARTICLE{totani_sag_A_2006,
    author = {{Totani}, Tomonori},
    title = "{A RIAF Interpretation for the Past Higher Activity of the Galactic Center Black Hole and the 511 keV Annihilation Emission}",
    journal = {\pasj},
    year = 2006,
    month = dec,
    volume = {58},
    pages = {965-977},
    doi = {10.1093/pasj/58.6.965},
    archivePrefix = {arXiv},
    eprint = {astro-ph/0607414},
    primaryClass = {astro-ph},
    adsurl = {https://ui.adsabs.harvard.edu/abs/2006PASJ...58..965T}
}

@ARTICLE{massi_TeV_2009,
    author = {{Massi}, M. and {Kaufman Bernad{\'o}}, M.},
    title = "{Radio Spectral Index Analysis and Classes of Ejection in LS I +61{\textdegree}303}",
    journal = {\apj},
    year = 2009,
    month = sep,
    volume = {702},
    number = {2},
    pages = {1179-1189},
    doi = {10.1088/0004-637X/702/2/1179},
    archivePrefix = {arXiv},
    eprint = {0908.2600},
    primaryClass = {astro-ph.HE},
    adsurl = {https://ui.adsabs.harvard.edu/abs/2009ApJ...702.1179M}
}

@ARTICLE{sguera_TeV_2009,
    author = {{Sguera}, V. and {Romero}, G.~E. and {Bazzano}, A. and {Masetti}, N. and {Bird}, A.~J. and {Bassani}, L.},
    title = "{Dissecting the Region of 3EG J1837-0423 and HESS J1841-055 with INTEGRAL}",
    journal = {\apj},
    year = 2009,
    month = jun,
    volume = {697},
    number = {2},
    pages = {1194-1205},
    doi = {10.1088/0004-637X/697/2/1194},
    archivePrefix = {arXiv},
    eprint = {0903.1763},
    primaryClass = {astro-ph.HE},
    adsurl = {https://ui.adsabs.harvard.edu/abs/2009ApJ...697.1194S}
}

@ARTICLE{sudoh_TeV_2020,
    author = {{Sudoh}, Takahiro and {Inoue}, Yoshiyuki and {Khangulyan}, Dmitry},
    title = "{Multiwavelength Emission from Galactic Jets: The Case of the Microquasar SS433}",
    journal = {\apj},
    year = 2020,
    month = feb,
    volume = {889},
    number = {2},
    eid = {146},
    pages = {146},
    doi = {10.3847/1538-4357/ab6442},
    archivePrefix = {arXiv},
    eprint = {1911.00013},
    primaryClass = {astro-ph.HE},
    adsurl = {https://ui.adsabs.harvard.edu/abs/2020ApJ...889..146S}
}

@ARTICLE{aleksic_TeV_2011,
    author = {{Aleksi{\'c}}, J. and {Alvarez}, E.~A. and {Antonelli}, L.~A. and {Antoranz}, P. and {Asensio}, M. and {Backes}, M. and {Barrio}, J.~A. and {Bastieri}, D. and {Becerra Gonz{\'a}lez}, J. and {Bednarek}, W. and {Berdyugin}, A. and {Berger}, K. and {Bernardini}, E. and {Biland}, A. and {Blanch}, O. and {Bock}, R.~K. and {Boller}, A. and {Bonnoli}, G. and {Bordas}, P. and {Borla Tridon}, D. and {Bosch-Ramon}, V. and {Braun}, I. and {Bretz}, T. and {Ca{\~n}ellas}, A. and {Carmona}, E. and {Carosi}, A. and {Colin}, P. and {Colombo}, E. and {Contreras}, J.~L. and {Cortina}, J. and {Cossio}, L. and {Covino}, S. and {Dazzi}, F. and {De Angelis}, A. and {De Cea del Pozo}, E. and {De Lotto}, B. and {Delgado Mendez}, C. and {Diago Ortega}, A. and {Doert}, M. and {Dom{\'\i}nguez}, A. and {Dominis Prester}, D. and {Dorner}, D. and {Doro}, M. and {Elsaesser}, D. and {Ferenc}, D. and {Fonseca}, M.~V. and {Font}, L. and {Fruck}, C. and {Garc{\'\i}a L{\'o}pez}, R.~J. and {Garczarczyk}, M. and {Garrido}, D. and {Giavitto}, G. and {Godinovi{\'c}}, N. and {Hadasch}, D. and {H{\"a}fner}, D. and {Herrero}, A. and {Hildebrand}, D. and {H{\"o}hne-M{\"o}nch}, D. and {Hose}, J. and {Hrupec}, D. and {Huber}, B. and {Jogler}, T. and {Klepser}, S. and {Kr{\"a}henb{\"u}hl}, T. and {Krause}, J. and {La Barbera}, A. and {Lelas}, D. and {Leonardo}, E. and {Lindfors}, E. and {Lombardi}, S. and {L{\'o}pez}, M. and {Lorenz}, E. and {Makariev}, M. and {Maneva}, G. and {Mankuzhiyil}, N. and {Mannheim}, K. and {Maraschi}, L. and {Mariotti}, M. and {Mart{\'\i}nez}, M. and {Mazin}, D. and {Meucci}, M. and {Miranda}, J.~M. and {Mirzoyan}, R. and {Miyamoto}, H. and {Mold{\'o}n}, J. and {Moralejo}, A. and {Munar-Adrover}, P. and {Nieto}, D. and {Nilsson}, K. and {Orito}, R. and {Oya}, I. and {Paneque}, D. and {Paoletti}, R. and {Pardo}, S. and {Paredes}, J.~M. and {Partini}, S. and {Pasanen}, M. and {Pauss}, F. and {Perez-Torres}, M.~A. and {Persic}, M. and {Peruzzo}, L. and {Pilia}, M. and {Pochon}, J. and {Prada}, F. and {Prada Moroni}, P.~G. and {Prandini}, E. and {Puljak}, I. and {Reichardt}, I. and {Reinthal}, R. and {Rhode}, W. and {Rib{\'o}}, M. and {Rico}, J. and {R{\"u}gamer}, S. and {Saggion}, A. and {Saito}, K. and {Saito}, T.~Y. and {Salvati}, M. and {Satalecka}, K. and {Scalzotto}, V. and {Scapin}, V. and {Schultz}, C. and {Schweizer}, T. and {Shayduk}, M. and {Shore}, S.~N. and {Sillanp{\"a}{\"a}}, A. and {Sitarek}, J. and {Sobczynska}, D. and {Spanier}, F. and {Spiro}, S. and {Stamerra}, A. and {Steinke}, B. and {Storz}, J. and {Strah}, N. and {Suri{\'c}}, T. and {Takalo}, L. and {Takami}, H. and {Tavecchio}, F. and {Temnikov}, P. and {Terzi{\'c}}, T. and {Tescaro}, D. and {Teshima}, M. and {Thom}, M. and {Tibolla}, O. and {Torres}, D.~F. and {Treves}, A. and {Vankov}, H. and {Vogler}, P. and {Wagner}, R.~M. and {Weitzel}, Q. and {Zabalza}, V. and {Zandanel}, F. and {Zanin}, R.},
    title = "{A Search for Very High Energy Gamma-Ray Emission from Scorpius X-1 with the Magic Telescopes}",
    journal = {\apjl},
    year = 2011,
    month = jul,
    volume = {735},
    number = {1},
    eid = {L5},
    pages = {L5},
    doi = {10.1088/2041-8205/735/1/L5},
    archivePrefix = {arXiv},
    eprint = {1103.5677},
    primaryClass = {astro-ph.HE},
    adsurl = {https://ui.adsabs.harvard.edu/abs/2011ApJ...735L...5A}
}

@ARTICLE{zdziarski_GeV_2017,
    author = {{Zdziarski}, Andrzej A. and {Malyshev}, Denys and {Chernyakova}, Maria and {Pooley}, Guy G.},
    title = "{High-energy gamma-rays from Cyg X-1}",
    journal = {\mnras},
    year = 2017,
    month = nov,
    volume = {471},
    number = {3},
    pages = {3657-3667},
    doi = {10.1093/mnras/stx1846},
    archivePrefix = {arXiv},
    eprint = {1607.05059},
    primaryClass = {astro-ph.HE},
    adsurl = {https://ui.adsabs.harvard.edu/abs/2017MNRAS.471.3657Z}
}

@ARTICLE{albert_GeV_2007,
    author = {{Albert}, J. and {Aliu}, E. and {Anderhub}, H. and {Antoranz}, P. and {Armada}, A. and {Baixeras}, C. and {Barrio}, J.~A. and {Bartko}, H. and {Bastieri}, D. and {Becker}, J.~K. and {Bednarek}, W. and {Berger}, K. and {Bigongiari}, C. and {Biland}, A. and {Bock}, R.~K. and {Bordas}, P. and {Bosch-Ramon}, V. and {Bretz}, T. and {Britvitch}, I. and {Camara}, M. and {Carmona}, E. and {Chilingarian}, A. and {Coarasa}, J.~A. and {Commichau}, S. and {Contreras}, J.~L. and {Cortina}, J. and {Costado}, M.~T. and {Curtef}, V. and {Danielyan}, V. and {Dazzi}, F. and {De Angelis}, A. and {Delgado}, C. and {de los Reyes}, R. and {De Lotto}, B. and {Domingo-Santamar{\'\i}a}, E. and {Dorner}, D. and {Doro}, M. and {Errando}, M. and {Fagiolini}, M. and {Ferenc}, D. and {Fern{\'a}ndez}, E. and {Firpo}, R. and {Flix}, J. and {Fonseca}, M.~V. and {Font}, L. and {Fuchs}, M. and {Galante}, N. and {Garc{\'\i}a-L{\'o}pez}, R.~J. and {Garczarczyk}, M. and {Gaug}, M. and {Giller}, M. and {Goebel}, F. and {Hakobyan}, D. and {Hayashida}, M. and {Hengstebeck}, T. and {Herrero}, A. and {H{\"o}hne}, D. and {Hose}, J. and {Hsu}, C.~C. and {Jacon}, P. and {Jogler}, T. and {Kosyra}, R. and {Kranich}, D. and {Kritzer}, R. and {Laille}, A. and {Lindfors}, E. and {Lombardi}, S. and {Longo}, F. and {L{\'o}pez}, J. and {L{\'o}pez}, M. and {Lorenz}, E. and {Majumdar}, P. and {Maneva}, G. and {Mannheim}, K. and {Mansutti}, O. and {Mariotti}, M. and {Mart{\'\i}nez}, M. and {Mazin}, D. and {Merck}, C. and {Meucci}, M. and {Meyer}, M. and {Miranda}, J.~M. and {Mirzoyan}, R. and {Mizobuchi}, S. and {Moralejo}, A. and {Nieto}, D. and {Nilsson}, K. and {Ninkovic}, J. and {O{\~n}a-Wilhelmi}, E. and {Otte}, N. and {Oya}, I. and {Panniello}, M. and {Paoletti}, R. and {Paredes}, J.~M. and {Pasanen}, M. and {Pascoli}, D. and {Pauss}, F. and {Pegna}, R. and {Persic}, M. and {Peruzzo}, L. and {Piccioli}, A. and {Prandini}, E. and {Puchades}, N. and {Raymers}, A. and {Rhode}, W. and {Rib{\'o}}, M. and {Rico}, J. and {Rissi}, M. and {Robert}, A. and {R{\"u}gamer}, S. and {Saggion}, A. and {Saito}, T. and {S{\'a}nchez}, A. and {Sartori}, P. and {Scalzotto}, V. and {Scapin}, V. and {Schmitt}, R. and {Schweizer}, T. and {Shayduk}, M. and {Shinozaki}, K. and {Shore}, S.~N. and {Sidro}, N. and {Sillanp{\"a}{\"a}}, A. and {Sobczynska}, D. and {Stamerra}, A. and {Stark}, L.~S. and {Takalo}, L. and {Temnikov}, P. and {Tescaro}, D. and {Teshima}, M. and {Torres}, D.~F. and {Turini}, N. and {Vankov}, H. and {Vitale}, V. and {Wagner}, R.~M. and {Wibig}, T. and {Wittek}, W. and {Zandanel}, F. and {Zanin}, R. and {Zapatero}, J.},
    title = "{Very High Energy Gamma-Ray Radiation from the Stellar Mass Black Hole Binary Cygnus X-1}",
    journal = {\apjl},
    year = 2007,
    month = aug,
    volume = {665},
    number = {1},
    pages = {L51-L54},
    doi = {10.1086/521145},
    archivePrefix = {arXiv},
    eprint = {0706.1505},
    primaryClass = {astro-ph},
    adsurl = {https://ui.adsabs.harvard.edu/abs/2007ApJ...665L..51A}
}

@ARTICLE{reynolds_jet_1996,
    author = {{Reynolds}, C.~S. and {Fabian}, A.~C. and {Celotti}, A. and {Rees}, M.~J.},
    title = "{The matter content of the jet in M87: evidence for an electron-positron jet}",
    journal = {\mnras},
    year = 1996,
    month = dec,
    volume = {283},
    number = {3},
    pages = {873-880},
    doi = {10.1093/mnras/283.3.873},
    archivePrefix = {arXiv},
    eprint = {astro-ph/9603140},
    primaryClass = {astro-ph},
    adsurl = {https://ui.adsabs.harvard.edu/abs/1996MNRAS.283..873R}
}

@ARTICLE{romero_hadrons_2005,
    author = {{Romero}, Gustavo E. and {Christiansen}, Hugo R. and {Orellana}, Mariana},
    title = "{Hadronic High-Energy Gamma-Ray Emission from the Microquasar LS I +61 303}",
    journal = {\apj},
    year = 2005,
    month = oct,
    volume = {632},
    number = {2},
    pages = {1093-1098},
    doi = {10.1086/444446},
    archivePrefix = {arXiv},
    eprint = {astro-ph/0506735},
    primaryClass = {astro-ph},
    adsurl = {https://ui.adsabs.harvard.edu/abs/2005ApJ...632.1093R}
}

@article{becker_stochastic_2006,
	title = {Time-dependent {Stochastic} {Particle} {Acceleration} in {Astrophysical} {Plasmas}: {Exact} {Solutions} {Including} {Momentum}-dependent {Escape}},
	volume = {647},
	issn = {0004-637X},
	shorttitle = {Time-dependent {Stochastic} {Particle} {Acceleration} in {Astrophysical} {Plasmas}},
	url = {https://ui.adsabs.harvard.edu/abs/2006ApJ...647..539B},
	doi = {10.1086/505319},
	urldate = {2026-01-24},
	journal = {The Astrophysical Journal},
	publisher = {IOP},
	author = {Becker, Peter A. and Le, Truong and Dermer, Charles D.},
	month = aug,
	year = {2006},
	note = {ADS Bibcode: 2006ApJ...647..539B},
	pages = {539--551},
}

@book{ginzburg_theory_1979,
    author = {Ginzburg, V. L.},
    title = {Theoretical Physics and Astrophysics},
    year = {1979},
    publisher = {Pergamon Press},
    location = {Oxford},
    series = {International Series in Natural Philosophy},
    volume = {99},
    isbn = {0080230660}
}

@ARTICLE{blumenthal_theory_1970,
    author = {{Blumenthal}, George R. and {Gould}, Robert J.},
    title = "{Bremsstrahlung, Synchrotron Radiation, and Compton Scattering of High-Energy Electrons Traversing Dilute Gases}",
    journal = {Reviews of Modern Physics},
    year = 1970,
    month = jan,
    volume = {42},
    number = {2},
    pages = {237-271},
    doi = {10.1103/RevModPhys.42.237},
    adsurl = {https://ui.adsabs.harvard.edu/abs/1970RvMP...42..237B}
}

@ARTICLE{shakura_BHs_1973,
    author = {{Shakura}, N.~I. and {Sunyaev}, R.~A.},
    title = "{Black holes in binary systems. Observational appearance.}",
    journal = {\aap},
    year = 1973,
    month = jan,
    volume = {24},
    pages = {337-355},
    adsurl = {https://ui.adsabs.harvard.edu/abs/1973A&A....24..337S}
}

@ARTICLE{romero_transient_jet_2007,
    author = {{Romero}, G.~E. and {Okazaki}, A.~T. and {Orellana}, M. and {Owocki}, S.~P.},
    title = "{Accretion vs. colliding wind models for the gamma-ray binary LS I +61 303: an assessment}",
    journal = {\aap},
    year = 2007,
    month = oct,
    volume = {474},
    number = {1},
    pages = {15-22},
    doi = {10.1051/0004-6361:20078035},
    archivePrefix = {arXiv},
    eprint = {0706.1320},
    primaryClass = {astro-ph},
    adsurl = {https://ui.adsabs.harvard.edu/abs/2007A&A...474...15R}
}

@ARTICLE{garcia_SFXTs_2014,
    author = {{Garc{\'\i}a}, Federico and {Aguilera}, Deborah N. and {Romero}, Gustavo E.},
    title = "{Exploring jet-launching conditions for supergiant fast X-ray transients}",
    journal = {\aap},
    year = 2014,
    month = may,
    volume = {565},
    eid = {A122},
    pages = {A122},
    doi = {10.1051/0004-6361/201323157},
    archivePrefix = {arXiv},
    eprint = {1404.7243},
    primaryClass = {astro-ph.HE},
    adsurl = {https://ui.adsabs.harvard.edu/abs/2014A&A...565A.122G}
}

@ARTICLE{miyasaka_MAXI_2018,
    author = {{Miyasaka}, Hiromasa and {Tomsick}, John A. and {Xu}, Yanjun and {Harrison}, Fiona A.},
    title = "{MAXI J1631-479 is a new X-ray transient}",
    journal = {The Astronomer's Telegram},
    year = 2018,
    month = dec,
    volume = {12340},
    pages = {1},
    adsurl = {https://ui.adsabs.harvard.edu/abs/2018ATel12340....1M}
}

@ARTICLE{xu_MAXI_2020,
    author = {{Xu}, Yanjun and {Harrison}, Fiona A. and {Tomsick}, John A. and {Walton}, Dominic J. and {Barret}, Didier and {Garc{\'\i}a}, Javier A. and {Hare}, Jeremy and {Parker}, Michael L.},
    title = "{Studying the Reflection Spectra of the New Black Hole X-Ray Binary Candidate MAXI J1631-479 Observed by NuSTAR: A Variable Broad Iron Line Profile}",
    journal = {\apj},
    year = 2020,
    month = apr,
    volume = {893},
    number = {1},
    eid = {30},
    pages = {30},
    doi = {10.3847/1538-4357/ab7dc0},
    archivePrefix = {arXiv},
    eprint = {2003.03465},
    primaryClass = {astro-ph.HE},
    adsurl = {https://ui.adsabs.harvard.edu/abs/2020ApJ...893...30X}
}

@ARTICLE{dieckmann_jet_2019,
    author = {{Dieckmann}, M.~E. and {Folini}, D. and {Hotz}, I. and {Nordman}, A. and {Dell'Acqua}, P. and {Ynnerman}, A. and {Walder}, R.},
    title = "{Structure of a collisionless pair jet in a magnetized electron-proton plasma: flow-aligned magnetic field}",
    journal = {\aap},
    year = 2019,
    month = jan,
    volume = {621},
    eid = {A142},
    pages = {A142},
    doi = {10.1051/0004-6361/201834393},
    archivePrefix = {arXiv},
    eprint = {1810.05415},
    primaryClass = {astro-ph.HE},
    adsurl = {https://ui.adsabs.harvard.edu/abs/2019A&A...621A.142D}
}

@ARTICLE{nishikawa_PIC_2021,
    author = {{Nishikawa}, Kenichi and {Du{\c{t}}an}, Ioana and {K{\"o}hn}, Christoph and {Mizuno}, Yosuke},
    title = "{PIC methods in astrophysics: simulations of relativistic jets and kinetic physics in astrophysical systems}",
    journal = {Living Reviews in Computational Astrophysics},
    year = 2021,
    month = dec,
    volume = {7},
    number = {1},
    eid = {1},
    pages = {1},
    doi = {10.1007/s41115-021-00012-0},
    archivePrefix = {arXiv},
    eprint = {2008.02105},
    primaryClass = {astro-ph.HE},
    adsurl = {https://ui.adsabs.harvard.edu/abs/2021LRCA....7....1N}
}

@ARTICLE{pohl_PIC_2020,
    author = {{Pohl}, M. and {Hoshino}, M. and {Niemiec}, J.},
    title = "{PIC simulation methods for cosmic radiation and plasma instabilities}",
    journal = {Progress in Particle and Nuclear Physics},
    year = 2020,
    month = mar,
    volume = {111},
    eid = {103751},
    pages = {103751},
    doi = {10.1016/j.ppnp.2019.103751},
    archivePrefix = {arXiv},
    eprint = {1912.02673},
    primaryClass = {astro-ph.HE},
    adsurl = {https://ui.adsabs.harvard.edu/abs/2020PrPNP.11103751P}
}

@ARTICLE{mckinney_reconnection_2012,
    author = {{McKinney}, Jonathan C. and {Uzdensky}, Dmitri A.},
    title = "{A reconnection switch to trigger gamma-ray burst jet dissipation}",
    journal = {\mnras},
    year = 2012,
    month = jan,
    volume = {419},
    number = {1},
    pages = {573-607},
    doi = {10.1111/j.1365-2966.2011.19721.x},
    archivePrefix = {arXiv},
    eprint = {1011.1904},
    primaryClass = {astro-ph.HE},
    adsurl = {https://ui.adsabs.harvard.edu/abs/2012MNRAS.419..573M}
}

@ARTICLE{hakobyan_PIC_2019,
    author = {{Hakobyan}, Hayk and {Philippov}, Alexander and {Spitkovsky}, Anatoly},
    title = "{Effects of Synchrotron Cooling and Pair Production on Collisionless Relativistic Reconnection}",
    journal = {\apj},
    year = 2019,
    month = may,
    volume = {877},
    number = {1},
    eid = {53},
    pages = {53},
    doi = {10.3847/1538-4357/ab191b},
    archivePrefix = {arXiv},
    eprint = {1809.10772},
    primaryClass = {astro-ph.HE},
    adsurl = {https://ui.adsabs.harvard.edu/abs/2019ApJ...877...53H}
}

@ARTICLE{schoeffler_PIC_2019,
    author = {{Schoeffler}, K.~M. and {Grismayer}, T. and {Uzdensky}, D. and {Fonseca}, R.~A. and {Silva}, L.~O.},
    title = "{Bright Gamma-Ray Flares Powered by Magnetic Reconnection in QED-strength Magnetic Fields}",
    journal = {\apj},
    year = 2019,
    month = jan,
    volume = {870},
    number = {1},
    eid = {49},
    pages = {49},
    doi = {10.3847/1538-4357/aaf1b9},
    archivePrefix = {arXiv},
    eprint = {1807.09750},
    primaryClass = {astro-ph.HE},
    adsurl = {https://ui.adsabs.harvard.edu/abs/2019ApJ...870...49S}
}

@ARTICLE{nattila_turbulence_2024,
    author = {{N{\"a}ttil{\"a}}, Joonas},
    title = "{Radiative plasma simulations of black hole accretion flow coronae in the hard and soft states}",
    journal = {Nature Communications},
    year = 2024,
    month = dec,
    volume = {15},
    number = {1},
    eid = {7026},
    pages = {7026},
    doi = {10.1038/s41467-024-51257-1},
    archivePrefix = {arXiv},
    eprint = {2408.08161},
    primaryClass = {astro-ph.HE},
    adsurl = {https://ui.adsabs.harvard.edu/abs/2024NatCo..15.7026N}
}

@ARTICLE{ferriere_GMF_2014,
    author = {{Ferri{\`e}re}, Katia and {Terral}, Philippe},
    title = "{Analytical models of X-shape magnetic fields in galactic halos}",
    journal = {\aap},
    year = 2014,
    month = jan,
    volume = {561},
    eid = {A100},
    pages = {A100},
    doi = {10.1051/0004-6361/201322966},
    archivePrefix = {arXiv},
    eprint = {1312.1974},
    primaryClass = {astro-ph.GA},
    adsurl = {https://ui.adsabs.harvard.edu/abs/2014A&A...561A.100F}
}

@ARTICLE{guenduez_GMF_2020,
    author = {{Guenduez}, M. and {Becker Tjus}, J. and {Ferri{\`e}re}, K. and {Dettmar}, R.-J.},
    title = "{A novel analytical model of the magnetic field configuration in the Galactic center}",
    journal = {\aap},
    year = 2020,
    month = dec,
    volume = {644},
    eid = {A71},
    pages = {A71},
    doi = {10.1051/0004-6361/201936081},
    archivePrefix = {arXiv},
    eprint = {1906.05211},
    primaryClass = {astro-ph.GA},
    adsurl = {https://ui.adsabs.harvard.edu/abs/2020A&A...644A..71G}
}

@ARTICLE{winkler_INTEGRAL_2003,
    author = {{Winkler}, C. and {Courvoisier}, T.~J.-L. and {Di Cocco}, G. and {Gehrels}, N. and {Gim{\'e}nez}, A. and {Grebenev}, S. and {Hermsen}, W. and {Mas-Hesse}, J.~M. and {Lebrun}, F. and {Lund}, N. and {Palumbo}, G.~G.~C. and {Paul}, J. and {Roques}, J.-P. and {Schnopper}, H. and {Sch{\"o}nfelder}, V. and {Sunyaev}, R. and {Teegarden}, B. and {Ubertini}, P. and {Vedrenne}, G. and {Dean}, A.~J.},
    title = "{The INTEGRAL mission}",
    journal = {\aap},
    year = 2003,
    month = nov,
    volume = {411},
    pages = {L1-L6},
    doi = {10.1051/0004-6361:20031288},
    adsurl = {https://ui.adsabs.harvard.edu/abs/2003A&A...411L...1W}
}

@ARTICLE{bowyer_Cyg_X-1_1965,
    author = {{Bowyer}, S. and {Byram}, E.~T. and {Chubb}, T.~A. and {Friedman}, H.},
    title = "{Observational results of X-ray astronomy}",
    journal = {Annales d'Astrophysique},
    year = 1965,
    month = feb,
    volume = {28},
    pages = {791},
    adsurl = {https://ui.adsabs.harvard.edu/abs/1965AnAp...28..791B}
}

@ARTICLE{pooley_GRS_1915_1997,
    author = {{Pooley}, G.~G. and {Fender}, R.~P.},
    title = "{The variable radio emission from GRS 1915+=105}",
    journal = {\mnras},
    year = 1997,
    month = dec,
    volume = {292},
    number = {4},
    pages = {925-933},
    doi = {10.1093/mnras/292.4.925},
    archivePrefix = {arXiv},
    eprint = {astro-ph/9708171},
    primaryClass = {astro-ph},
    adsurl = {https://ui.adsabs.harvard.edu/abs/1997MNRAS.292..925P}
}

@ARTICLE{miller-jones_Cyg_X-1_2021,
    author = {{Miller-Jones}, James C.~A. and {Bahramian}, Arash and {Orosz}, Jerome A. and {Mandel}, Ilya and {Gou}, Lijun and {Maccarone}, Thomas J. and {Neijssel}, Coenraad J. and {Zhao}, Xueshan and {Zi{\'o}{\l}kowski}, Janusz and {Reid}, Mark J. and {Uttley}, Phil and {Zheng}, Xueying and {Byun}, Do-Young and {Dodson}, Richard and {Grinberg}, Victoria and {Jung}, Taehyun and {Kim}, Jeong-Sook and {Marcote}, Benito and {Markoff}, Sera and {Rioja}, Mar{\'\i}a J. and {Rushton}, Anthony P. and {Russell}, David M. and {Sivakoff}, Gregory R. and {Tetarenko}, Alexandra J. and {Tudose}, Valeriu and {Wilms}, Joern},
    title = "{Cygnus X-1 contains a 21-solar mass black hole{\textemdash}Implications for massive star winds}",
    journal = {Science},
    year = 2021,
    month = mar,
    volume = {371},
    number = {6533},
    pages = {1046-1049},
    doi = {10.1126/science.abb3363},
    archivePrefix = {arXiv},
    eprint = {2102.09091},
    primaryClass = {astro-ph.HE},
    adsurl = {https://ui.adsabs.harvard.edu/abs/2021Sci...371.1046M}
}

@ARTICLE{hasinger_GX_13+1_1989,
    author = {{Hasinger}, G. and {van der Klis}, M.},
    title = "{Two patterns of correlated X-ray timing and spectral behaviour in low-mass X-ray binaries.}",
    journal = {\aap},
    year = 1989,
    month = nov,
    volume = {225},
    pages = {79-96},
    adsurl = {https://ui.adsabs.harvard.edu/abs/1989A&A...225...79H}
}

@ARTICLE{fridriksson_NSLMXBs_2015,
    author = {{Fridriksson}, Joel K. and {Homan}, Jeroen and {Remillard}, Ronald A.},
    title = "{Common Patterns in the Evolution between the Luminous Neutron Star Low-Mass X-ray Binary Subclasses}",
    journal = {\apj},
    year = 2015,
    month = aug,
    volume = {809},
    number = {1},
    eid = {52},
    pages = {52},
    doi = {10.1088/0004-637X/809/1/52},
    archivePrefix = {arXiv},
    eprint = {1504.00022},
    primaryClass = {astro-ph.HE},
    adsurl = {https://ui.adsabs.harvard.edu/abs/2015ApJ...809...52F}
}

@ARTICLE{kuulkers_4U_1630-47_1998,
    author = {{Kuulkers}, Erik and {Wijnands}, Rudy and {Belloni}, Tomaso and {M{\'e}ndez}, Mariano and {van der Klis}, Michiel and {van Paradijs}, Jan},
    title = "{Absorption Dips in the Light Curves of GRO J1655-40 and 4U 1630-47 during Outburst}",
    journal = {\apj},
    year = 1998,
    month = feb,
    volume = {494},
    number = {2},
    pages = {753-758},
    doi = {10.1086/305248},
    archivePrefix = {arXiv},
    eprint = {astro-ph/9710024},
    primaryClass = {astro-ph},
    adsurl = {https://ui.adsabs.harvard.edu/abs/1998ApJ...494..753K}
}

@ARTICLE{bandyopadhyay_GX_13+1_2002,
    author = {{Bandyopadhyay}, Reba M. and {Charles}, Philip A. and {Shahbaz}, Tariq and {Wagner}, R. Mark},
    title = "{Infrared Photometric Variability of GX 13+1 and GX 17+2}",
    journal = {\apj},
    year = 2002,
    month = may,
    volume = {570},
    number = {2},
    pages = {793-798},
    doi = {10.1086/339776},
    archivePrefix = {arXiv},
    eprint = {astro-ph/0201390},
    primaryClass = {astro-ph},
    adsurl = {https://ui.adsabs.harvard.edu/abs/2002ApJ...570..793B}
}

@ARTICLE{reid_GRS_1915+105_2014,
    author = {{Reid}, M.~J. and {McClintock}, J.~E. and {Steiner}, J.~F. and {Steeghs}, D. and {Remillard}, R.~A. and {Dhawan}, V. and {Narayan}, R.},
    title = "{A Parallax Distance to the Microquasar GRS 1915+105 and a Revised Estimate of its Black Hole Mass}",
    journal = {\apj},
    year = 2014,
    month = nov,
    volume = {796},
    number = {1},
    eid = {2},
    pages = {2},
    doi = {10.1088/0004-637X/796/1/2},
    archivePrefix = {arXiv},
    eprint = {1409.2453},
    primaryClass = {astro-ph.GA},
    adsurl = {https://ui.adsabs.harvard.edu/abs/2014ApJ...796....2R}
}

@ARTICLE{harlaftis_GRS_1915+105_2004,
    author = {{Harlaftis}, E.~T. and {Greiner}, J.},
    title = "{The rotational broadening and the mass of the donor star  of GRS 1915+105}",
    journal = {\aap},
    year = 2004,
    month = jan,
    volume = {414},
    pages = {L13-L16},
    doi = {10.1051/0004-6361:20031754},
    archivePrefix = {arXiv},
    eprint = {astro-ph/0312373},
    primaryClass = {astro-ph},
    adsurl = {https://ui.adsabs.harvard.edu/abs/2004A&A...414L..13H}
}

@ARTICLE{jourdain_cyg_X-1_2012,
    author = {{Jourdain}, E. and {Roques}, J.~P. and {Malzac}, J.},
    title = "{The Emission of Cygnus X-1: Observations with INTEGRAL SPI from 20 keV to 2 MeV}",
    journal = {\apj},
    year = 2012,
    month = jan,
    volume = {744},
    number = {1},
    eid = {64},
    pages = {64},
    doi = {10.1088/0004-637X/744/1/64},
    archivePrefix = {arXiv},
    eprint = {1109.2053},
    primaryClass = {astro-ph.HE},
    adsurl = {https://ui.adsabs.harvard.edu/abs/2012ApJ...744...64J}
}

@ARTICLE{higdon_SNe_2009,
       author = {{Higdon}, J.~C. and {Lingenfelter}, R.~E. and {Rothschild}, R.~E.},
        title = "{The Galactic Positron Annihilation Radiation and the Propagation of Positrons in the Interstellar Medium}",
      journal = {\apj},
         year = 2009,
        month = jun,
       volume = {698},
       number = {1},
        pages = {350-379},
          doi = {10.1088/0004-637X/698/1/350},
archivePrefix = {arXiv},
       eprint = {0711.3008},
 primaryClass = {astro-ph},
       adsurl = {https://ui.adsabs.harvard.edu/abs/2009ApJ...698..350H}
}

@ARTICLE{beacom_in_flight_2006,
       author = {{Beacom}, John F. and {Y{\"u}ksel}, Hasan},
        title = "{Stringent Constraint on Galactic Positron Production}",
      journal = {\prl},
         year = 2006,
        month = aug,
       volume = {97},
       number = {7},
          eid = {071102},
        pages = {071102},
          doi = {10.1103/PhysRevLett.97.071102},
archivePrefix = {arXiv},
       eprint = {astro-ph/0512411},
 primaryClass = {astro-ph},
       adsurl = {https://ui.adsabs.harvard.edu/abs/2006PhRvL..97g1102B}
}

@ARTICLE{knodlseder_in_flight_2025,
    author = {{Kn{\"o}dlseder}, J. and {Sabri}, K. and {Jean}, P. and {von Ballmoos}, P. and {Skinner}, G. and {Collmar}, W.},
    title = "{Detection of positron in-flight annihilation from the Galaxy}",
    journal = {\aap},
    year = 2025,
    month = aug,
    volume = {700},
    eid = {A257},
    pages = {A257},
    doi = {10.1051/0004-6361/202556046},
    archivePrefix = {arXiv},
    eprint = {2506.17427},
    primaryClass = {astro-ph.HE},
    adsurl = {https://ui.adsabs.harvard.edu/abs/2025A&A...700A.257K}
}

@ARTICLE{gryzinski_collisions_1965,
    author = {{Gryzi{\'n}ski}, Michal},
    title = "{Classical Theory of Atomic Collisions. I. Theory of Inelastic Collisions}",
    journal = {Physical Review},
    year = 1965,
    month = apr,
    volume = {138},
    number = {2A},
    pages = {336-358},
    doi = {10.1103/PhysRev.138.A336},
    adsurl = {https://ui.adsabs.harvard.edu/abs/1965PhRv..138..336G}
}

@ARTICLE{strong_GALPROP_1998,
    author = {{Strong}, Andrew W. and {Moskalenko}, Igor V.},
    title = "{Propagation of Cosmic-Ray Nucleons in the Galaxy}",
    journal = {\apj},
    year = 1998,
    month = dec,
    volume = {509},
    number = {1},
    pages = {212-228},
    doi = {10.1086/306470},
    archivePrefix = {arXiv},
    eprint = {astro-ph/9807150},
    primaryClass = {astro-ph},
    adsurl = {https://ui.adsabs.harvard.edu/abs/1998ApJ...509..212S}
}

@ARTICLE{elwert_bremsstrahlung_1939,
    author = {{Elwert}, Gerhard},
    title = "{Versch{\"a}rfte Berechnung von Intensit{\"a}t und Polarisation im kontinuierlichen R{\"o}ntgenspektrum1}",
    journal = {Annalen der Physik},
    year = 1939,
    month = jan,
    volume = {426},
    number = {2},
    pages = {178-208},
    doi = {10.1002/andp.19394260206},
    adsurl = {https://ui.adsabs.harvard.edu/abs/1939AnP...426..178E}
}

@ARTICLE{gould_bremsstrahlung_1990,
    author = {{Gould}, Robert J.},
    title = "{Multipole Radiation in Charged-Particle Scattering}",
    journal = {\apj},
    year = 1990,
    month = oct,
    volume = {362},
    pages = {284},
    doi = {10.1086/169265},
    adsurl = {https://ui.adsabs.harvard.edu/abs/1990ApJ...362..284G}
}

@ARTICLE{carbone_DC_2019,
    author = {{Carbone}, D. and {Wijnands}, R.},
    title = "{Constraining the duty cycle of transient low-mass X-ray binaries through simulations}",
    journal = {\mnras},
    year = 2019,
    month = sep,
    volume = {488},
    number = {2},
    pages = {2767-2779},
    doi = {10.1093/mnras/stz1645},
    archivePrefix = {arXiv},
    eprint = {1904.06713},
    primaryClass = {astro-ph.HE},
    adsurl = {https://ui.adsabs.harvard.edu/abs/2019MNRAS.488.2767C}
}

@INCOLLECTION{belloni_states_2010,
    author = {{Belloni}, T.~M.},
    title = "{States and Transitions in Black Hole Binaries}",
    booktitle = {Lecture Notes in Physics, Berlin Springer Verlag},
    publisher={Springer Science and Business Media LLC},
    year = 2010,
    editor = {{Belloni}, Tomaso},
    volume = {794},
    pages = {53},
    doi = {10.1007/978-3-540-76937-8_3},
    adsurl = {https://ui.adsabs.harvard.edu/abs/2010LNP...794...53B}
}

@ARTICLE{wang_26Al_2009,
    author = {{Wang}, W. and {Lang}, M.~G. and {Diehl}, R. and {Halloin}, H. and {Jean}, P. and {Kn{\"o}dlseder}, J. and {Kretschmer}, K. and {Martin}, P. and {Roques}, J.~P. and {Strong}, A.~W. and {Winkler}, C. and {Zhang}, X.~L.},
    title = "{Spectral and intensity variations of Galactic $^{26}$Al emission}",
    journal = {\aap},
    year = 2009,
    month = mar,
    volume = {496},
    number = {3},
    pages = {713-724},
    doi = {10.1051/0004-6361/200811175},
    archivePrefix = {arXiv},
    eprint = {0902.0211},
    primaryClass = {astro-ph.HE},
    adsurl = {https://ui.adsabs.harvard.edu/abs/2009A&A...496..713W}
}

@ARTICLE{martin_26Al_2009,
    author = {{Martin}, P. and {Kn{\"o}dlseder}, J. and {Diehl}, R. and {Meynet}, G.},
    title = "{New estimates of the gamma-ray line emission of the Cygnus region from INTEGRAL/SPI observations}",
    journal = {\aap},
    year = 2009,
    month = nov,
    volume = {506},
    number = {2},
    pages = {703-710},
    doi = {10.1051/0004-6361/200912178},
    archivePrefix = {arXiv},
    eprint = {1001.1521},
    primaryClass = {astro-ph.HE},
    adsurl = {https://ui.adsabs.harvard.edu/abs/2009A&A...506..703M}
}

@ARTICLE{diehl_26Al_2006,
    author = {{Diehl}, Roland and {Halloin}, Hubert and {Kretschmer}, Karsten and {Lichti}, Giselher G. and {Sch{\"o}nfelder}, Volker and {Strong}, Andrew W. and {von Kienlin}, Andreas and {Wang}, Wei and {Jean}, Pierre and {Kn{\"o}dlseder}, J{\"u}rgen and {Roques}, Jean-Pierre and {Weidenspointner}, Georg and {Schanne}, Stephane and {Hartmann}, Dieter H. and {Winkler}, Christoph and {Wunderer}, Cornelia},
    title = "{Radioactive $^{26}$Al from massive stars in the Galaxy}",
    journal = {\nat},
    year = 2006,
    month = jan,
    volume = {439},
    number = {7072},
    pages = {45-47},
    doi = {10.1038/nature04364},
    archivePrefix = {arXiv},
    eprint = {astro-ph/0601015},
    primaryClass = {astro-ph},
    adsurl = {https://ui.adsabs.harvard.edu/abs/2006Natur.439...45D}
}

@INPROCEEDINGS{prantzos_XRBs_2004,
       author = {{Prantzos}, N.},
        title = "{Astrophysical Gamma-Ray Lines: A Probe of Stellar Nucleosynthesis and Star Formation}",
    booktitle = {5th INTEGRAL Workshop on the INTEGRAL Universe},
         year = 2004,
       editor = {{Schoenfelder}, V. and {Lichti}, G. and {Winkler}, C.},
       series = {ESA Special Publication},
       volume = {552},
        month = oct,
        pages = {15},
          doi = {10.48550/arXiv.astro-ph/0404501},
archivePrefix = {arXiv},
       eprint = {astro-ph/0404501},
 primaryClass = {astro-ph},
       adsurl = {https://ui.adsabs.harvard.edu/abs/2004ESASP.552...15P}
}

@ARTICLE{grimm_XRBs_2002,
    author = {{Grimm}, H.-J. and {Gilfanov}, M. and {Sunyaev}, R.},
    title = "{The Milky Way in X-rays for an outside observer. Log(N)-Log(S) and luminosity function of X-ray binaries from RXTE/ASM data}",
    journal = {\aap},
    year = 2002,
    month = sep,
    volume = {391},
    pages = {923-944},
    doi = {10.1051/0004-6361:20020826},
    archivePrefix = {arXiv},
    eprint = {astro-ph/0109239},
    primaryClass = {astro-ph},
    adsurl = {https://ui.adsabs.harvard.edu/abs/2002A&A...391..923G}
}

@ARTICLE{bouchet_26Al_2015,
    author = {{Bouchet}, Laurent and {Jourdain}, Elisabeth and {Roques}, Jean-Pierre},
    title = "{The Galactic $^{26}$Al Emission Map as Revealed by INTEGRAL SPI}",
    journal = {\apj},
    year = 2015,
    month = mar,
    volume = {801},
    number = {2},
    eid = {142},
    pages = {142},
    doi = {10.1088/0004-637X/801/2/142},
    archivePrefix = {arXiv},
    eprint = {1501.05247},
    primaryClass = {astro-ph.HE},
    adsurl = {https://ui.adsabs.harvard.edu/abs/2015ApJ...801..142B}
}



\end{document}